\documentclass[reprint, aps, prb, superscriptaddress, longbibliography
]{revtex4-2}

\usepackage{amsthm}
\usepackage{mathtools}
\usepackage{xcolor}
\usepackage{graphicx}
\usepackage[T1]{fontenc}
\usepackage[utf8]{inputenc}
\usepackage[
separate-uncertainty=true,
list-units=single,
range-units=single,
retain-explicit-plus,
group-minimum-digits=3,
]{siunitx}
\usepackage{multirow}
\DeclareSIUnit\angstrom{\protect \text {Å}}
\usepackage[colorlinks, allcolors=blue]{hyperref}

\begin{document}

\title{Atomistic modeling of bulk and grain boundary diffusion in solid electrolyte \texorpdfstring{Li\textsubscript{6}PS\textsubscript{5}Cl}{Li6PS5Cl} using machine-learning interatomic potentials}

\author{Yongliang Ou}
\email{yongliang.ou@imw.uni-stuttgart.de}
\affiliation{Institute for Materials Science, University of Stuttgart, Pfaffenwaldring 55, 70569 Stuttgart, Germany}

\author{Yuji Ikeda}
\email{yuji.ikeda@imw.uni-stuttgart.de}
\affiliation{Institute for Materials Science, University of Stuttgart, Pfaffenwaldring 55, 70569 Stuttgart, Germany}

\author{Lena Scholz}
\affiliation{Institute of Applied Mechanics, University of Stuttgart, Universitätsstr.~32, 70569 Stuttgart, Germany}

\author{Sergiy Divinski}
\affiliation{Institute of Materials Physics, University of Münster, Wilhelm-Klemm-Str.~10, 48149 Münster, Germany}

\author{Felix Fritzen}
\affiliation{Institute of Applied Mechanics, University of Stuttgart, Universitätsstr.~32, 70569 Stuttgart, Germany}

\author{Blazej Grabowski}
\affiliation{Institute for Materials Science, University of Stuttgart, Pfaffenwaldring 55, 70569 Stuttgart, Germany}

\date{\today}

\begin{abstract}

Li\textsubscript{6}PS\textsubscript{5}Cl is a promising candidate for the solid electrolyte in all-solid-state Li-ion batteries due to its high ionic conductivity. In applications, this material is in a polycrystalline state with grain boundaries (GBs) that can affect ionic conductivity. While atomistic modeling provides valuable information on the impact of GBs on Li diffusion, such studies face either high computational cost (when using \textit{ab initio} methods) or accuracy limitations (when using classical potentials) as challenges. Here, we develop a quality-level-based active learning scheme for efficient and systematic development of \textit{ab initio}-based machine-learning interatomic potentials, specifically moment tensor potentials (MTPs), for large-scale, long-time, and high-accuracy simulations of complex atomic structures and diffusion mechanisms as encountered in solid electrolytes. Based on this scheme, we obtain MTPs for Li\textsubscript{6}PS\textsubscript{5}Cl and investigate two tilt GBs, $\Sigma3(1\bar{1}2)[110]$, $\Sigma3(\bar{1}11)[110]$, and one twist GB, $\Sigma5(001)[001]$. All three GBs exhibit low formation energies of less than \SI{20}{meV/\angstrom\textsuperscript{2}}, indicating their high stability in polycrystalline Li\textsubscript{6}PS\textsubscript{5}Cl. Using the MTPs, diffusion coefficients of the anion-ordered and anion-disordered bulk, as well as the three GBs, are obtained from molecular dynamics simulations of atomistic models with more than \SI{16000}{atoms} for \SI{5}{\nano\second}. At \SI{300}{\kelvin}, the GB diffusion coefficients fall between the ones of the anion-ordered bulk structure (\SI{1.2e-9}{cm^2/s}, corresponding ionic conductivity about \SI{0.2}{mS/cm}) and the anion-disordered bulk structure (\SI{50}{\percent} Cl/S-anion disorder; \SI{2.2e-7}{cm^2/s}, about \SI{29.8}{mS/cm}) of Li\textsubscript{6}PS\textsubscript{5}Cl. Experimental data fall between the Arrhenius-extrapolated diffusion coefficients of the investigated atomic structures, supporting our quantitative \textit{in silico} predictions.

\end{abstract}

\maketitle

\section{Introduction} \label{sec:intr}

All-solid-state Li-ion batteries have attracted attention for their improved safety compared to conventional Li-ion batteries by virtue of their solid instead of a flammable liquid electrolyte~\cite{Famprikis2019}. However, finding a suitable material for the solid electrolyte is not trivial~\cite{Zhao2020}. In 2008, the argyrodite-type Li$_6$PS$_5$Cl was first reported, featuring an unusually high Li-ion mobility~\cite{Deiseroth2008}. Later, intensive experimental investigations~\cite{Boulineau2013,Kraft2017,Ruhl2021} affirmed the superior ionic conductivity, and ever since Li$_6$PS$_5$Cl is considered a candidate for the solid electrolyte.

Diffusion of Li ions in Li$_6$PS$_5$Cl was measured using nuclear magnetic resonance~\cite{Adeli2019,Schlenker2020}. The room-temperature diffusion coefficients reported in the two studies are of the same order of magnitude (\SI{3.87e-8}{cm^2/s}~\cite{Adeli2019} vs.~\SI{2.5e-8}{cm^2/s}~\cite{Schlenker2020}), and the derived activation energies differ by about 20\%, \SI{0.35\pm0.01}{eV}~\cite{Adeli2019} vs.~\SI{0.28\pm0.01}{eV}~\cite{Schlenker2020}. Differences in the activation energies can come from the relatively narrow temperature intervals of the diffusion measurements and difficulties in reproducibly synthesizing such materials. Depending on the synthesis conditions, polycrystalline materials with Cl/S-anion disorder~\cite{Minafra2020} and potentially different grain boundary (GB) distributions~\cite{Ganapathy2019,Milan2023} are obtained. Further, the contributions of crystalline bulk and GBs to the Li diffusion can hardly be separated in experiments~\cite{Ganapathy2019}, which might affect the measured coefficients. 

Atomistic simulations provide an important complementary tool for an improved understanding of diffusion mechanisms and the impact of GBs. Table~\ref{tab:history} summarizes molecular dynamics (MD) simulations for Li self-diffusion in Li$_6$PS$_5$Cl. The diffusion coefficient for the anion-ordered bulk structure was computed in several previous studies~\cite{Deng2017,Stamminger2019,Jiang2022,Jeon2024}, all of which gave values that are orders of magnitude smaller than those in experiments. Further investigations (both simulations~\cite{Klerk2016,Morgan2021} and  experiments~\cite{Minafra2020}) showed that the origin of superionic Li diffusion is most likely related to the Cl/S-anion disorder in the bulk structure of Li$_6$PS$_5$Cl. This anion disorder triggers Li inter-cage diffusion. Recently, quantitative investigations of Li diffusivity in the anion-disordered structure were also performed by \textit{ab initio} MD simulations~\cite{Lee2022,Sadowski2023,Jeon2024}. With \SI{50}{\percent} Cl/S-anion disorder in the bulk structure, the Arrhenius-extrapolated Li diffusion coefficients at room temperature are consistent with experiments~\cite{Lee2022,Jeon2024}. The simulated activation energies of Li diffusion are lower (\SIrange{0.20}{0.26}{eV}~\cite{Lee2022} and \SI{0.25}{eV}~\cite{Jeon2024}) than the above-listed experimental data. Due to the typical high computational requirements of \textit{ab initio} MD simulations~\cite{He2018}, the sampling time was relatively short in these previous studies, and defects (e.g., GBs) were often neglected. 

\begin{table*}[tbp]
\centering
    \caption{Theoretical studies which provided calculated diffusion coefficients for Li self-diffusion in solid electrolyte Li$_{6}$PS$_{5}$Cl. A closely related Li$_6$PS$_5$Br study~\cite{Sadowski2023} including grain boundaries (GBs) is also listed. Previous studies used \textit{ab initio} molecular dynamics (MD) simulations, while in the present work, a machine-learning interatomic potential (MLIP), specifically moment tensor potential (MTP), fitted to \textit{ab initio} data is used to accelerate MD. The number of atoms in the simulation cell is denoted by $N_{\textrm{at}}$. ``Yes'' or ``no'' means that the corresponding structures have or have not been considered in the given study. }
	\begin{ruledtabular}
	\begin{tabular}{ccS[table-format=3.0]S[table-format=2.2]cccc}
		Year  &   Sampling  & {Size ($N_{\mathrm{at}}$)} & {Time (ns)} & Anion-ordered bulk & Anion-disordered bulk & GBs  & Reference \\
        \hline\\[-0.3cm]
            2017  &   \textit{ab initio} MD & 52 & 0.1 & yes  & no &   no   & \cite{Deng2017}\\
            2019  &   \textit{ab initio} MD & 52 & 0.3 & yes  & partially\makebox[0pt][l]{\footnote{Only one configuration with a Cl/S antisite defect was considered in Ref.~\cite{Stamminger2019}.}} &   no   & \cite{Stamminger2019}\\ 
            2022  &   \textit{ab initio} MD & 52 & 0.12 & yes  & no &   no   & \cite{Jiang2022}\\ 
            2022  &   \textit{ab initio} MD & 52 & 0.15 & yes  & yes &   no   & \cite{Lee2022}\\ 
            2022  &   classical force field & 416 & 20 & no  & yes &   no   & \cite{Das2022}\\ 
            2023\makebox[0pt][l]{\footnote{A structurally similar material, Li$_6$PS$_5$Br, was investigated in Ref.~\cite{Sadowski2023}.}}  & \textit{ab initio} MD & 312 & 0.04 & yes  & yes &   yes   & \cite{Sadowski2023}\\ 
            2024  &   \textit{ab initio} MD & 52 & 0.3 & yes  & yes &   no   & \cite{Jeon2024}\\ 
            2024  &  \textit{ab initio} $\rightarrow$ MLIP MD  &  \SI{>16000}{} & 5  & yes  & yes &   yes  & this work \\
	\end{tabular}
	\end{ruledtabular}
	\label{tab:history}
\end{table*}

Interatomic potentials can be utilized instead of \textit{ab initio} simulations to reduce the computational cost. This enables large-scale and long-time MD simulations, equipping us with tools for a statistically reliable prediction of diffusion properties. For example, classical force fields parameterized by \textit{ab initio} data were utilized for simulating solid electrolytes~\cite{Stegmaier2021,Stegmaier2022,Das2022}. Due to the complex structure of these materials and the limited number of fitting parameters in the force fields, simulation results with lower accuracy have to be expected. A good alternative is given by machine-learning interatomic potentials calibrated to \textit{ab initio} data, which have recently emerged as a powerful tool to accelerate MD simulations while preserving near \textit{ab initio} accuracy~\cite{Gubaev2023,Erhard2024}. Machine-learning interatomic potentials were shown to accurately describe diffusion~\cite{Novoselov2019, Winter2023} and work well even for complex electrolytes~\cite{Musaelian2023}. Li diffusion in Li$_6$PS$_5$Cl has not been investigated systematically with machine-learning interatomic potentials to our knowledge. 

The present study investigates the impact of three GBs with different structural characteristics ($\Sigma3(1\bar{1}2)[110]$, $\Sigma3(\bar{1}11)[110]$, and $\Sigma5(001)[001]$) on Li diffusion in the solid electrolyte Li$_6$PS$_5$Cl, aiming to provide accurate GB diffusion coefficients and thereby enhance the understanding of experimentally measured data. To this end, we propose and apply an active learning scheme that systematically exploits the quality levels of machine-learning interatomic potentials, specifically of moment tensor potentials (MTPs)~\cite{Shapeev2016}. Based on the scheme, we fit machine-learning potentials to \textit{ab initio} data and systematically analyze their performance using the target quantity, i.e., the diffusion coefficient, as a measure. Accelerated by the thus obtained and validated machine-learning interatomic potentials, large-scale and long-time MD simulations are performed to optimize the GB structures and analyze the difference in GB diffusion mechanisms compared to the bulk. From these simulations, we extract the diffusion coefficients for the GBs as well as for the anion-ordered and anion-disordered bulk structures. The results provide a theoretically admissible range of Li diffusivity in Li$_6$PS$_5$Cl. 

\section{Methodology} \label{sec:meth}

\subsection{Grain boundary construction}\label{sec:gbcon}

\begin{figure*}[tbp]
    \centering
    \includegraphics{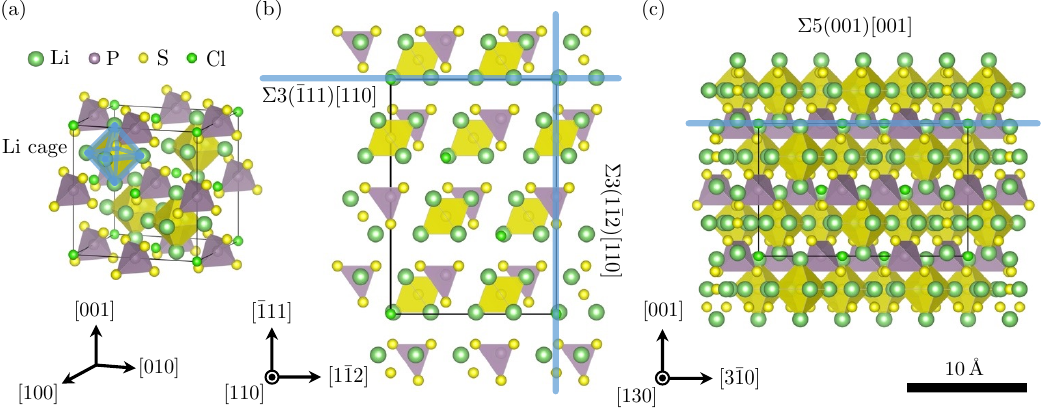}
    \caption{(a) Conventional unit cell of anion-ordered bulk Li$_6$PS$_5$Cl with 52 atoms (four formula units). One octahedral cage consisting of six Li atoms is highlighted in blue. (b) Coincidence-site lattice (CSL) unit cell with 78 atoms (six formula units). The planes for constructing the two $\Sigma3$ GBs are highlighted in blue. (c) CSL unit cell with 130 atoms (ten formula units). The plane for constructing the $\Sigma5$ GB is highlighted in blue. The CSL unit cells are used to construct the GB simulation models (Table~\ref{tab:gb_info}). Visualization performed using \textsc{vesta}~\cite{Momma2011}.} 
    \label{fig:gb_con}
\end{figure*}

Li$_{6}$PS$_{5}$Cl exhibits an argyrodite-type structure and belongs to the cubic crystal system with the space group $F\bar{4}3m$ (No.~216)~\cite{Deiseroth2008, Kong2010}. Figure~\ref{fig:gb_con}(a) shows the conventional unit cell of the bulk structure, which contains four formula units (52 atoms). The shown bulk structure is anion-ordered, i.e., the Wyckoff sites $4a$ [the sites symmetrically equivalent to (0, 0, 0)] and $4c$ [equivalent to (0.25, 0.25, 0.25)] are fully occupied by Cl and S atoms, respectively. Further, PS$_{4}$ tetrahedral units are present with the P and S atoms located at $4b$ [equivalent to (0.5, 0.5, 0.5)] and $16e$ sites, respectively.  The Cl and the P atoms each form a face-centered cubic sublattice. These relatively heavy atoms (P, S, and Cl) comprise a three-dimensional backbone that provides multiple interstitial sites to accommodate the lightweight Li atoms. The Li atoms are positioned at $24g$ sites in the ideal model with high symmetry. As highlighted in Fig.~\ref{fig:gb_con}(a), six Li atoms octahedrally coordinate an S atom at a $4c$ Wyckoff site, forming a so-called Li cage. 

It was reported both experimentally~\cite{Kong2010, Deiseroth2011} and theoretically~\cite{Wang2017,Golov2021,DAmore2022} that the high-symmetry $F\bar{4}3m$ structure is not stable at low temperatures and that neighboring interstitial sites, e.g., $48h$ Wyckoff sites, offer lower-energy states for the Li atoms. This results in symmetry-breaking displacements of the Li atoms, referred to as tilting of the Li cages. The dynamic stabilization of the Li atoms at the high-symmetry positions (phase transition) with increasing temperature was investigated in Ref.~\cite{Deiseroth2011}.
Such a temperature-induced phase transition is also observed in many other materials. For example, the high-symmetry structure of BaFeO\textsubscript{3} perovskite is dynamically unstable at lower temperatures and becomes dynamically stabilized at elevated temperatures~\cite{Ou2022}.

Based on the cubic symmetry of Li$_6$PS$_5$Cl and the coincidence-site lattice theory, periodic and commensurate GB structures can be generated~\cite{Cheng2018}. Considering the complexity of the bulk structure, three relatively simple GBs, yet featuring different geometrical arrangements, namely $\Sigma 3(1\bar{1}2)[110]$, $\Sigma3(\bar{1}11)[110]$, and $\Sigma5(001)[001]$, were chosen for the present study. Table~\ref{tab:gb_info} lists information about the simulation models for the GBs and also the bulk. $\Sigma 3(1\bar{1}2)[110]$ and $\Sigma3(\bar{1}11)[110]$ are tilt GBs symmetric for the Cl and P atoms with misorientation angles of \SI{70.53}{\degree} and \SI{109.47}{\degree}, respectively. The tilt axis is $[110]$, and the GB planes are $(1\bar{1}2)$ and $(\bar{1}11)$. $\Sigma5(001)[001]$ is a twist GB with a rotational axis of $[001]$ and a misorientation angle of \SI{36.87}{\degree}. The GB plane is $(001)$ and lies perpendicular to the rotational axis. 

\begin{table*}[tbp]
\centering
    \caption{Information on the anion-ordered bulk and the GB models. The $x$, $y$, and $z$ entrees give the relation of the Cartesian axes of each model with the crystallographic directions of the conventional unit cell of the bulk structure [Fig.~\ref{fig:gb_con}(a)]. For all GB simulation cells, $x$ and $y$ are parallel ($\parallel$) and $z$ is perpendicular ($\perp$) to the GB plane. The tilt axis of the $\Sigma 3$ tilt GBs is along the $y$ direction, and the rotational axis of the $\Sigma5$ twist GB is along the $z$ direction. The total number of atoms is denoted by $N_{\textrm{at}}$. The dimensions of the optimized structures, the GB formation energies [$\gamma$, Eq.~\eqref{eq:gbe}] calculated at \SI{0}{\kelvin} (see Sec.~\ref{sec:struc}), and the GB widths ($\delta$) evaluated at \SI{600}{\kelvin} (see Secs.~\ref{sec:struc} and~\ref{sec:dis}) are shown. The excess volumes per unit GB area of all three GBs are less than \SI{1}{\angstrom}.
    }
	\begin{ruledtabular}
	\begin{tabular}{ccccccS[table-format=2.2]S[table-format=2.1]}
		Model & $x$ $(\parallel)$ & $y$ $(\parallel)$ & $z$ $(\perp)$  & $N_{\mathrm{at}}$ & Dimension (\SI{}{\angstrom}) & {$\gamma$ (meV/\AA$^2$)}  & {$\delta$ (\AA)} \\
		\hline\\[-0.3cm]
            anion-ordered bulk & $[100]$ & $[010]$ & $[001]$ & \SI{17836}{} & $70\times70\times70$ & n/a & n/a \\
		GB $\Sigma3(1\bar{1}2)[110]$ tilt \SI{70.53}{\degree} & $[\bar{1}11]$ & $[110]$ & $[1\bar{1}2]$ & \SI{19656}{} & $53\times51\times148$ & 10.44 & 28.8 \\
		GB $\Sigma3(\bar{1}11)[110]$ tilt \SI{109.47}{\degree} & $[1\bar{1}2]$ & $[110]$ & $[\bar{1}11]$ & \SI{17472}{} & $50\times51\times133$ & 7.12 & 16.7 \\
        GB $\Sigma5(001)[001]$ twist \SI{36.87}{\degree} & $[130]$ & $[3\bar{1}0]$ & $[001]$ & \SI{16380}{} & $49\times49\times139$ & 18.78 & 26.5 
	\end{tabular}
	\end{ruledtabular}
	\label{tab:gb_info}
\end{table*}

To obtain the three GB supercells, two coincidence-site lattice unit cells [Fig.~\ref{fig:gb_con}(b) for the two $\Sigma3$ GBs, and Fig.~\ref{fig:gb_con}(c) for the $\Sigma5$ GB] were constructed by coordinate transformations of the conventional unit cell [Fig.~\ref{fig:gb_con}(a)]. The GB planes used for constructing the GBs are highlighted in blue. For $\Sigma 3(1\bar{1}2)[110]$, the GB plane cuts through some PS$_{4}$ units and some Li cages. For $\Sigma3(\bar{1}11)[110]$, in contrast, the integrity of all Li cages and PS$_{4}$ units is preserved by setting the GB plane to the position of the Cl atomic layers. The integrity of all Li cages is also preserved for $\Sigma5(001)[001]$, but some PS$_{4}$ units are cut through by the GB plane. With these distinct structural features of the GBs, different diffusion behaviors of Li is expected. 

The GB simulation models were constructed as periodic bicrystals, where two grains of different crystallographic orientations are stacked in the $z$ direction. The two grains were slightly separated to ensure that the smallest atomic distance in the GB simulation cells is larger than \SI{1.5}{\angstrom}. In each GB simulation cell, $x$ and $y$ lie within the GB plane ($\parallel$), and $z$ is perpendicular ($\perp$) to the GB plane. The tilt axis of the $\Sigma 3$ GBs is in the $y$ direction, and the rotational axis of the $\Sigma5$ GB is in the $z$ direction. There are two GBs of the same type per simulation cell but with inverted orientations due to the coincidence-site lattice construction. To avoid the interaction between these GBs, supercells were made so that the distances between the GBs in the direction perpendicular to the GB planes ($z$) are larger than \SI{65}{\angstrom}. Table~\ref{tab:gb_info} lists the total number of atoms for each GB simulation cell. All simulation cells maintain the stoichiometry of Li$_{6}$PS$_{5}$Cl, ensuring formal charge neutrality. 

\subsection{Anion disorder in bulk structure}~\label{sec:dis_const}

Anion disorder in Li$_6$PS$_5$Cl refers to the mixing of Cl and S atoms at the $4a$ and the $4c$ Wyckoff sites of the bulk structure, as observed in experiments~\cite{Deiseroth2008, Kong2010}.
In the present study, \SI{50}{\percent} mixing of Cl and S was considered. The atomistic model with the anion-disordered bulk structure was constructed based on that with the anion-ordered bulk structure (listed in Table~\ref{tab:gb_info}). For each anion-disordered bulk model, half of the Cl atoms at the $4a$ sites and half of the S atoms at the $4c$ sites were randomly selected and exchanged.

\subsection{Self-diffusion coefficient calculation}\label{sec:diffco}

The self-diffusion coefficients were computed from the mean square displacements (MSDs) of Li atoms in MD simulations. The MSD at simulation time $t$ and temperature $T$ reads,
\begin{equation}
    \textrm{MSD}(t, T) = \langle |\boldsymbol{r}(t) - \boldsymbol{r}_0|^2\rangle_T,
    \label{eq:msd}
\end{equation}
where $\boldsymbol{r}(t)$ is the position vector of a Li atom at time~$t$, $\boldsymbol{r}_0$ is its position vector at the reference time $t=0$, and $\langle\cdots\rangle_T$ denotes the average at $T$ over all Li atoms in the simulation cell. Generally, the MSD is related to the simulation time $t$ by
\begin{equation}
    \textrm{MSD}(t, T)\propto t^{\alpha},
    \label{eq:msdfit}
\end{equation}
where the parameter $\alpha$ captures the type of diffusion regime~\cite{Metzler2000}. For $\alpha\neq1$, diffusion is considered to be anomalous. Specifically, the $\alpha<1$ regime is referred to as subdiffusion and the $\alpha>1$ regime as superdiffusion~\cite{Metzler2000}. For $\alpha=1$, i.e., when the MSD depends linearly on $t$, a normal diffusion regime is identified, and the diffusion coefficient $D$ can be calculated by the Einstein equation~\cite{Einstein1905},
\begin{equation}
    D(T) = \frac{1}{2n}\lim_{t\to\infty} \left(\frac{\mathrm{d}}{\mathrm{d}t}\mathrm{MSD}(t,T)\right),
    \label{eq:diffcoecal}
\end{equation}
where $n$ is the number of dimensions. In simulations, the $t\to\infty$ limit in Eq.~\eqref{eq:diffcoecal} cannot be reached and has to be replaced by a finite period. In the present work, the MSD from the simulation time interval \SIrange{3}{5}{\nano\second} was used to fit based on Eq.~\eqref{eq:msdfit} and to determine $\alpha$. Considering statistics, normal diffusion was assumed when $\alpha \in [0.9, 1.1]$. 

Further, within the normal diffusion regime, the diffusion coefficient typically follows an Arrhenius relation, 
\begin{equation}
    D(T) = D_{0} \exp\left(-\frac{E_{\textrm{a}}}{k_{\textrm{B}}T}\right),
    \label{eq:diffcoearr}
\end{equation}
where $D_{0}$ is a constant, $E_{\mathrm{a}}$ is the activation energy for diffusion, and $k_{\textrm{B}}$ is the Boltzmann constant. In the present study, the bulk and GB diffusion coefficients of Li$_6$PS$_5$Cl in the normal diffusion regime were assumed to obey the Arrhenius relation (see Sec.~\ref{sec:separa} for the confirmation of this assumption for the bulk).

Previous studies extracted GB self-diffusion coefficients by considering the MSD only within the rather small GB region (typical widths of about \SI{10}{\angstrom})~\cite{Dawson2018,Lee2023,Jalem2023}.
However, the limited number of jumping events captured in the corresponding MD simulations can lead to significant uncertainties in the calculated GB diffusion coefficients. Therefore, we applied a different approach to improve the accuracy of the GB diffusion coefficients. Specifically, we first computed the \emph{effective} diffusion coefficient $D^{\textrm{eff}}$ for a bicrystal supercell made of two symmetrically equivalent GBs and the bulk separating them (see Sec.~\ref{sec:gbcon}). To extract from $D^{\textrm{eff}}$ the GB diffusion coefficient, $D^{\textrm{GB}}$, we assumed that $D^{\textrm{eff}}$ is linearly composed from $D^{\textrm{GB}}$ and a bulk diffusion contribution, $D^{\textrm{bulk}}$. The procedure was applied separately to the GB diffusion component perpendicular to the GB plane, $D^{\textrm{GB}}_\perp$, and the component within the GB plane, $D^{\textrm{GB}}_\parallel$. To separate $D^{\textrm{GB}}$ into $D^{\textrm{GB}}_\perp$ and $D^{\textrm{GB}}_\parallel$, the vectors $\boldsymbol{r}$ entering Eq.~\eqref{eq:msd} were projected onto the $z$ direction or the $x$-$y$ plane, yielding the respective projected diffusion coefficients (subscripts $\perp$ and $\parallel$, respectively). For the diffusion direction within the GB plane, the so-called Hart equation~\cite{Hart1957,Belova2003}, 
\begin{equation}
    D^{\textrm{eff}}_\parallel(T) = (1-\tau)D^{\textrm{bulk}}_\parallel(T) + \tau D^{\textrm{GB}}_\parallel(T), \quad n=2,
    \label{eq:para}
\end{equation}
where $\tau$ is the fraction of time spent by the diffusing atoms in the GB region~\cite{Kaur1995} and $n$ indicates the dimension as used in Eq.~\eqref{eq:diffcoecal}, manifests the linear decomposition of the effective bicrystal diffusion coefficient. For the diffusion direction perpendicular to the GB plane, long-range diffusion of Li atoms crossing the bulk and the GB area may become relevant. However, the present MD simulations indicate that most Li atoms diffuse around a local area, i.e., there is no significant long-range mass transport of Li (see Sec.~\ref{sec:bulk}). Thus, approximately, the Hart equation can also be used to extract the perpendicular GB diffusion coefficient, 
\begin{equation}
    D^{\textrm{eff}}_\perp(T) \approx (1-\tau)D^{\textrm{bulk}}_\perp(T) + \tau D^{\textrm{GB}}_\perp(T), \quad n=1.
    \label{eq:perp}
\end{equation}
For consistency, the projected diffusion coefficients for the bulk, $D^{\textrm{bulk}}_{\parallel} (T)$ and $D^{\textrm{bulk}}_{\perp} (T)$, were obtained from the anion-ordered bulk models with supercell geometries corresponding to the respective GB models (Table~\ref{tab:gb_info}). 

The linear mixing parameter $\tau$ can be approximated by the volume fraction of the GB region in the GB model, i.e., 
\begin{equation}
    \tau \approx \frac{\delta}{l},
    \label{eq:fracapprox}
\end{equation}
where $\delta$ is the GB width, and $l$ is the distance between the GBs. Equation~\eqref{eq:fracapprox} applies to a steady state with negligible segregation (see Sec.~\ref{sec:struc}) and when GBs are parallel with each other in the periodic simulation model, as in the present case (see Sec.~\ref{sec:gbcon}). The GB width $\delta$ was determined based on the GB profiles (see Sec.~\ref{sec:separa}).

\subsection{Computational details}\label{sec:com}

\textit{Ab initio} simulations were carried out under the density-functional theory (DFT) framework using the projector augmented wave method~\cite{Bloechl1994} and the generalized gradient approximation in the PBE parametrization~\cite{Perdew1996}, as implemented in \textsc{vasp}~\cite{Kresse1995, Kresse1996, Kresse1999}. The $1s^22s^1$, $3s^23p^3$, $3s^23p^4$, and $3s^23p^5$ electrons were treated as valence electrons for Li, P, S, and Cl, respectively. The plane-wave cutoff was set to \SI{500}{\electronvolt}. For calculations of the conventional unit cell [Fig.~\ref{fig:gb_con}(a)], the reciprocal space was sampled using the tetrahedron method with Blöchl corrections~\cite{Bloechl1994a} and a $\Gamma$-centered $2\times2\times2$ $\boldsymbol{k}$-point mesh (\SI{416}{kp\cdot atom}). For calculations of the GB structures, the reciprocal space was sampled at the $\Gamma$-point using the Gaussian smearing with a width of \SI{0.03}{\electronvolt}. For structural optimization, the energy and the maximum residual force were converged to better than \SI{e-5}{\electronvolt} per simulation cell and \SI{e-2}{eV.\angstrom\textsuperscript{\textminus1}}, respectively. 

Large-scale MD simulations were performed using the validated MTP (see Sec.~\ref{sec:mtp}) within \textsc{lammps}~\cite{Thompson2022} for the models listed in Table~\ref{tab:gb_info}. The MD time step was set to \SI{2}{\femto\second}. To fully relax the structures, the so-called annealing-and-quenching approach (a$+$q), which was used previously for polycrystal relaxation in Ref.~\cite{Wagih2023Dec}, was employed in the present study, i.e., the structures were first equilibrated at \SI{600}{\kelvin} for \SI{0.2}{\nano\second} and then cooled to \SI{1}{\kelvin} within \SI{1}{\nano\second}. Thermal vibrations at the annealing stage enable Li diffusion, and the slow quenching stage ensures that the system is maintained close to a thermodynamically favorable state. The a$+$q approach was performed in the \textit{NPT} ensemble. Lastly, geometry optimization was performed using the conjugate gradient algorithm implemented in \textsc{lammps}. Only the lattice variation along the $z$ direction was allowed during relaxation. The GB formation energy $\gamma$ at \SI{0}{\kelvin} was calculated by,
\begin{equation}
    \gamma = \frac{E_{\textrm{GB}} - E_{\textrm{bulk}}}{A_\textrm{tot}},
    \label{eq:gbe}
\end{equation}
where $E_{\textrm{GB}}$ is the total energy of the respective GB models (Table~\ref{tab:gb_info}) after relaxation, and $E_{\textrm{bulk}}$ is the total energy of the reference anion-ordered bulk model after relaxation. Further, $A_\textrm{tot}$ is the total GB area in the simulation cell, which is twice as large as the $x$--$y$ cross-sectional area because of two GB planes in the present simulation cells. For consistency, anion-ordered bulk models with a supercell geometry corresponding to the GB models were used in Eq.~\eqref{eq:gbe}. For the diffusion investigations, the structures were first relaxed and equilibrated using the Nosé--Hoover thermostat for \SI{0.2}{\nano\second} at the target temperature. The \textit{NPT} ensemble with a temperature rescaling every 100 time-steps at zero pressure was used. Next, to avoid any spurious effects of the thermostat on diffusion, a microcanonical (\textit{NVE}) ensemble was used, and the system was sampled for \SI{5}{\nano\second}. The GB profiles were created by binning Li atoms into slices of a width of about \SI{5}{\angstrom} in the $z$ direction of the simulation cell and averaging over the entire \textit{NVE} MD sampling period.

\section{Results}\label{sec:result}

\subsection{Training scheme and validation of MTP}\label{sec:mtp}

\begin{figure}[tbp]
    \centering
    \includegraphics{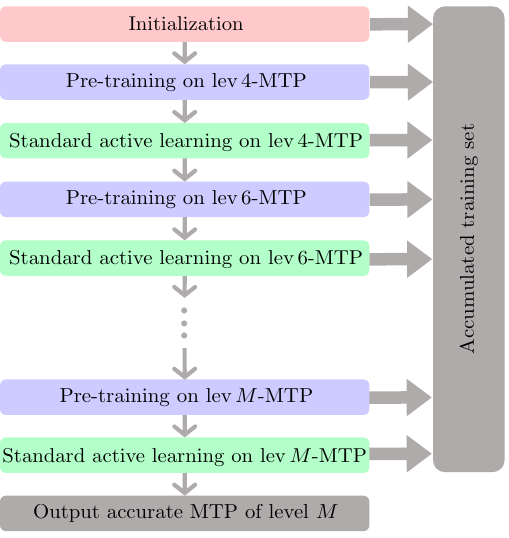}
    \caption{Proposed quality-level-based active learning scheme for the MTP. After initialization, pre-training and standard active learning are performed iteratively for MTPs with increasing levels, and the generated configurations are accumulated in the training set. When the targeted level $M$ is reached, the final MTP with high accuracy is outputted for subsequent simulations. } 
    \label{fig:al_scheme}
\end{figure}

A special training and active learning scheme for the MTP, as summarized in Fig.~\ref{fig:al_scheme} and outlined in the following, is proposed in the present study. The initialization involves \textit{ab initio} MD simulations for the anion-ordered bulk structure. The trajectories obtained from these initial runs are used as the basis for the training set. Next, the first pre-training and active learning cycle is performed for the target training structure (e.g., a GB structure) for an MTP of level 4. The ``level'' indicates the quality, i.e., the number of fitting parameters, in the MTP~\cite{Novikov2020}. The new configurations obtained during standard active learning are labeled by DFT calculations (i.e., energies, forces, and stresses are computed)~\cite{Gubaev2021} and then added to and accumulated in the training set. In the following steps, MTPs of higher levels (6, 8, 10, and so on) are subsequently used for pre-training and standard active learning. Configurations obtained after each pre-training or standard active learning cycle are labeled and added to the training set. The accuracy of the trained MTPs at different levels is estimated by the fitting errors with respect to DFT. Finally, when a high-accuracy MTP is obtained (as quantified below), the active-learning scheme finishes. Further details on the proposed scheme are given in Appendix~\ref{sec:al_scheme}.

In the first few cycles of the proposed active learning scheme, a relatively large number of new configurations are being sampled and added to the training set. The corresponding information (energies, forces, stresses) is required to provide a larger-scale estimate of the potential energy surface of the target structure. Consequently, multiple iterations within each standard active learning cycle of the initial MTPs are required. It is, therefore, beneficial to use low-level MTPs at the beginning of the active learning scheme since they can be generally constructed within a smaller number of iterations than high-level MTPs due to their fewer fitting parameters. The low-level MTPs are also more robust in phase-space extrapolation, requiring less computational cost for refitting. Starting the whole active learning scheme with low-level MTPs thus generates, in a computationally efficient manner, an ample training set for the high-level MTPs at the later stages of the active learning scheme. In this way, the number of standard active learning iterations is kept small for the computationally expensive high-level MTPs. Overall, the proposed training and active learning scheme systematizes and accelerates the construction of precise MTPs for chemically and structurally complex materials. 

\begin{figure}
    \centering
    \includegraphics{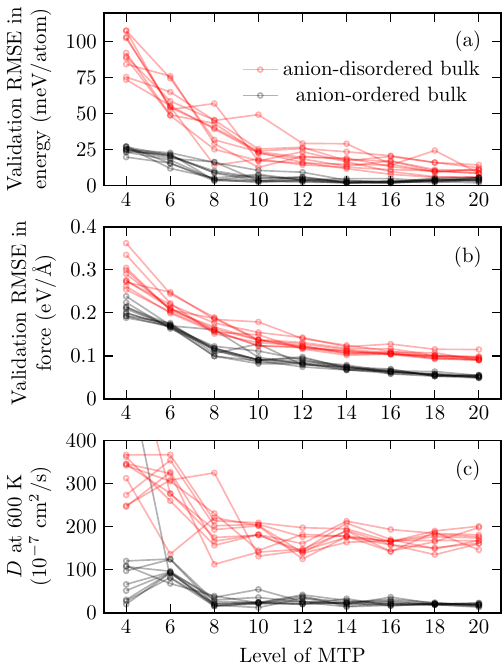}
    \caption{Validation root-mean-square error (RMSE) in (a) energy and (b) force of the MTPs obtained for the anion-ordered and the anion-disordered bulk structures from level 4 to 20 with the proposed scheme (Fig.~\ref{fig:al_scheme}). (c) Diffusion coefficients $D$ at \SI{600}{\kelvin} calculated with the obtained MTPs. Each line represents the result of an independently fitted MTP. Multiple MTPs were fitted for each structure to show the statistical variation. Simulations were performed with supercells with \num{3328} atoms [equivalent to a $4\times4\times4$ expansion of the unit cell shown in Fig.~\ref{fig:gb_con}(a)]. }
    \label{fig:vali_bulk}
\end{figure}

In the present study, various MTPs were fitted from level 4 up to level 20 for the anion-ordered and the anion-disordered bulk structures. Based on the bulk results, single MTPs were fitted for each GB from level 4 up to level 18. Figures~\ref{fig:vali_bulk}(a) and (b) show the validation root-mean-square errors (RMSEs) in energies and forces for the anion-ordered and anion-disordered bulk, respectively. Additionally, Fig.~\ref{fig:vali_bulk}(c) shows the dependence of the target quantity, i.e., the diffusion coefficient, on the MTP level. For the anion-ordered bulk structure, the RMSE in energy is less than \SI{10}{meV/atom} for MTPs from level 14 on. The RMSE in force slowly decreases with the MTP level and reaches a small final value of about \SI{0.05}{eV/\angstrom} for level 20. Larger RMSEs in both energy and force are observed for the anion-disordered bulk structure, likely due to its complex anionic arrangement. With MTPs from level 10 on, the calculated diffusion coefficients for the ordered and the disordered bulk structures converge to final values of about \SIlist{20e-7;180e-7}{cm^2/s}, respectively. The ordered bulk structure also shows less variation among multiple MTPs for both the RMSEs and the diffusion coefficients. 

\begin{figure}[tbp]
    \centering
    \includegraphics{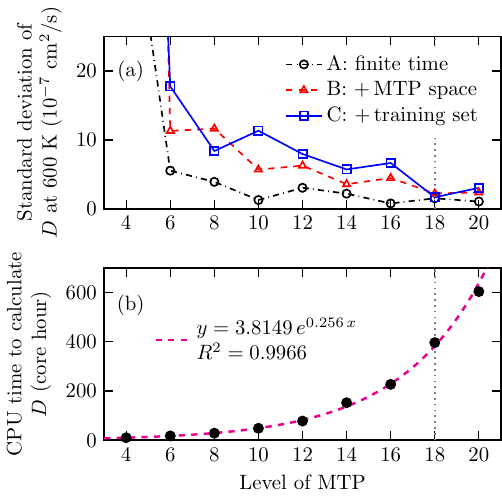}
    \caption{(a) Standard deviation of the diffusion coefficients $D$ of the anion-ordered bulk structure at \SI{600}{\kelvin} for the three different sets of simulations A, B, and C as described in the main text. As a reference, the diffusion coefficient of the anion-ordered bulk structure calculated with a level-20 MTP at \SI{600}{\kelvin} is about \SI{20e-7}{cm^2/s} [Fig.~\ref{fig:vali_bulk}(c)]. (b) CPU time required to calculate $D$ for the ordered bulk structure. An exponential function was fitted to show the CPU time and the MTP level relation. To balance the accuracy and simulation cost, MTPs of level 18 (Table~\ref{tab:mtp_fit_validate}) were used in the subsequent simulations. Simulations were performed with supercells with \num{3328} atoms. } 
    \label{fig:std}
\end{figure}

Variation in the calculated diffusion coefficients may originate from the following sources: 1.~Limited diffusion sampling time; 2.~Different local minima within the MTP parameter space; 3.~Different accumulated training sets. To systematically investigate these sources, the diffusion coefficients of the anion-ordered bulk structure were calculated at \SI{600}{\kelvin} for three different sets of simulations:
\begin{enumerate}
\renewcommand{\theenumi}{\Alph{enumi}}
\item The complete active learning scheme was run once, thereby generating one MTP at each level (from 4 to 20). For each MTP, five independent runs for \SI{5}{\nano\second} were performed, and the corresponding standard deviation for the diffusion coefficient was determined. The standard deviation for set A represents the contribution to the variation due to the finite sampling time of the diffusion coefficient.
\item The final training set obtained after the active learning in A was used to fit ten MTPs for each level. Different MTPs (in terms of the basis set coefficients) can be obtained even for the same training set and for the same MTP level because of multiple local minima in the MTP parameter space visited due to randomized initial starting conditions. At each level, the 10 MTPs were used to run diffusion coefficient simulations for \SI{5}{\nano\second}. The corresponding standard deviation in the diffusion coefficient was determined for each level. The standard deviation for set B contains the contribution from the finite sampling time (set A) and additionally the variation from the different MTP parameters. 
\item The whole active learning scheme was run ten times, thereby generating 10 different MTPs at each level. These MTPs vary not only in the MTP parameters but, importantly, also in the training set used for the fitting. Thus, set C includes, in addition to the previous variation contributions (as for set B), the variation due to different training sets.
\end{enumerate}
The standard deviations of the diffusion coefficients obtained for these three sets of simulations are shown in Fig.~\ref{fig:std}(a). The results allow us to draw a very important conclusion. All curves decrease with an increasing MTP level and reach very small values for the highest investigated levels. This means that the final diffusion-coefficient values for the highest MTP levels do not depend (beyond the remaining statistical variation) on the initial starting conditions of the active learning scheme. Thus, it is possible to systematically increase the precision in the predicted diffusion coefficient values of the MTPs. However, one should be aware of the accompanying strongly increasing computational cost with MTP level as shown in Fig.~\ref{fig:std}(b). To balance the accuracy and simulation cost, MTPs of level 18 with 807 fitting parameters were used in the present study. 

Table~\ref{tab:mtp_fit_validate} lists the information on the final level-18 MTPs trained for different target structures. Larger training RMSEs in both energies and forces are observed for the GBs compared to the bulk structures (more than \SI{1}{meV/atom} in energy). 
Despite applying a different anion disorder in constructing the validation set for the MTP of the disordered bulk structure, the validation RMSE remains consistent with the training RMSE. Details of the construction of the validation set are given in Appendix~\ref{sec:al_scheme}. 

\begin{table*}[tbp]
\centering
\caption{Information of the final MTPs. The number of atoms in the simulation cell of the training structures is denoted by $N_{\textrm{at}}$, and the number of configurations in the final accumulated training set is denoted by $N_{\textrm{conf}}$. Different MTPs were independently trained to different structures according to the proposed scheme up to level 18 (i.e., $M=18$ in Fig.~\ref{fig:al_scheme}; cf.~Fig.~\ref{fig:std}) with 807 fitting parameters. Training and validation root-mean-square error (RMSE) is shown for the bulk structures, and training RMSE is shown for the GB structures. }
	\begin{ruledtabular}
	\begin{tabular}{ccccccc}
	Training structure & $N_{\textrm{at}}$ &  $N_{\textrm{conf}}$ & \multicolumn{2}{c}{Training RMSE} & \multicolumn{2}{c}{Validation RMSE} \\ 
 \cline{4-7} \\[-0.25cm]
       & & & Energy (meV/atom) & Force (eV/\AA) & Energy (meV/atom) & Force (eV/\AA) \\[+0.05cm]
        \hline \\[-0.25cm]
anion-ordered bulk &  104  &  2219  &  4.7  &  0.132    &  3.1 & 0.056  \\  
anion-disordered bulk & 208 &  2210  &  5.1  &  0.153   & 6.0 &  0.095\\  
GB $\Sigma3(1\bar{1}2)[110]$ & 312 & 2350 & 6.6  & 0.164 & n/a & n/a \\ 
GB $\Sigma3(\bar{1}11)[110]$ & 156 & 2293 & 6.9  &  0.157 & n/a & n/a \\ 
GB $\Sigma5(001)[001]$ & 260 & 2241 & 6.1 & 0.157 & n/a & n/a \\ 
	\end{tabular}
	\end{ruledtabular}
	\label{tab:mtp_fit_validate}
\end{table*}

\begin{figure}[tbp]
    \centering
    \includegraphics{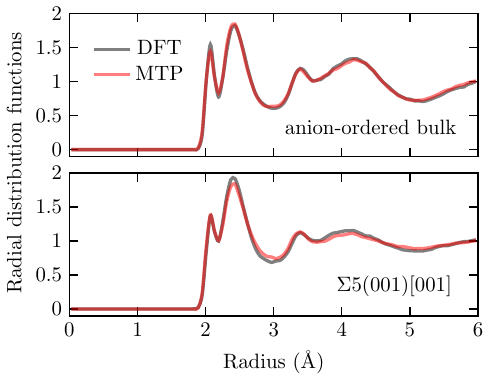}
    \caption{Comparison of radial distribution functions obtained by DFT and MTP MD at \SI{600}{\kelvin} for the anion-ordered bulk and the $\Sigma5(001)[001]$ GB averaged over 4000 and 2000 MD steps, respectively. Analysis was performed using \textsc{ovito}~\cite{Stukowski2009}. } 
    \label{fig:pbemtp}
\end{figure}

Figure~\ref{fig:pbemtp} shows the radial distribution functions of the anion-ordered bulk structure in the conventional unit cell [Fig.~\ref{fig:gb_con}(a)] and the $\Sigma5(001)[001]$ GB obtained from MD simulations at \SI{600}{\kelvin}. The accuracy of the MTPs is validated by the small difference in the radial distribution functions obtained by MTP and DFT. 

\subsection{Structures and energies}\label{sec:struc}

Table~\ref{tab:opt} shows structural and energetic information on the anion-ordered bulk Li$_6$PS$_5$Cl at \SI{0}{\kelvin} as obtained with various optimization approaches. For the optimization with symmetry constraint (i.e., preserving the space group $F\bar{4}3m$), DFT and MTP show similar lattice constants of about \SI{10.25}{\angstrom}. The symmetry-constrained structure is dynamically unstable at \SI{0}{\kelvin} as confirmed by additional phonon calculations. The related double-well potentials of two imaginary modes were analyzed previously in Ref.~\cite{DAmore2022}. The dynamical instability is also consistent with previous experimental~\cite{Kong2010,Deiseroth2011} and computational~\cite{Golov2021,Wang2017} studies reporting that the high symmetry Li$_6$PS$_5$Cl structure [Fig.~\ref{fig:gb_con}(a)] is not the most stable phase at low temperatures and \SI{0}{\kelvin}, respectively.

\begin{table*}[tbp]
\centering
	\caption{Structural and energetic data for the anion-ordered Li$_6$PS$_5$Cl bulk at \SI{0}{\kelvin} based on different methods.\footnote{A lattice constant of \SI{9.818}{\angstrom} was obtained in experiments at \SI{150}{\kelvin} and \SI{1}{atm} for a sample with about \SI{56.2}{\percent} of Cl/S-anion disorder showing the space group $F\bar{4}3m$~\cite{Minafra2020}. Note that this experimental value is not directly comparable with the simulation results given in the table because of the differences in the space group, temperature, and anion ordering.} The total number of atoms in the simulation cell is denoted by $N_{\textrm{at}}$. The structure before optimization [Fig.~\ref{fig:gb_con}(a)] has a cubic symmetry with the space group $F\bar{4}3m$. The cubic shape of the simulation cell is constrained during the a$+$q optimization, and $a$ refers to the cubic lattice constant. The energy $\Delta E$ is the energy of the optimized structure referenced with respect to the symmetry-constrained structure.}
	\begin{ruledtabular}
	\begin{tabular}{cS[table-format=5.0]cccccc}
		Source &  {$N_{\mathrm{at}}$} & Method & {$a$ (\AA)} & Space group  & Stability\footnote{The Stability column indicates the dynamical stability as determined by the presence or absence of imaginary phonons at \SI{0}{\kelvin}.} & {$\Delta E$ (meV/atom)} & Reference \\
		\hline\\[-0.3cm]
  DFT &  52 & CG\footnote{The conjugate gradient (CG) algorithm implemented in \textsc{vasp} was used. } with sym.~constraint & \tablenum[table-format=2.3]{10.246} & $F\bar{4}3m$ & unstable & \tablenum[table-format=4.1]{0.0} & this work \\
  MTP &  52 & BFGS\footnote{The Broyden–Fletcher–Goldfarb–Shanno (BFGS) algorithm implemented in \textsc{ase}~\cite{ase-paper} was used. } with sym.~constraint & \tablenum[table-format=2.3]{10.264}  & $F\bar{4}3m$ & unstable &  \tablenum[table-format=4.1]{0.0} & this work \\
  MTP   & 52 &  a$+$q\footnote{The annealing-and-quenching (a$+$q) approach described in Sec.~\ref{sec:com} was used.} & \tablenum[table-format=2.3]{9.950} & $P1$ & stable &  \tablenum[table-format=4.1]{-55.3} & this work \\
  DFT   & 52 &  MTP a$+$q $\rightarrow$ DFT CG\footnote{The optimized structure with MTP using the a$+$q\footnotemark[4] method was further optimized with DFT using the CG\footnotemark[2] method.} & \tablenum[table-format=2.3]{9.940}\rlap{\footnote{The cubic cell shape was changed after this optimization, so the effective cubic lattice constant is shown.}} & $P1$ & n/a &  \tablenum[table-format=4.1]{-58.9} & this work \\
  DFT   & 13 & USPEX\footnote{The universal structure predictor: evolutionary xtallography (USPEX) described in Refs.~\cite{Glass2006,Wang2017} was used. } & \tablenum[table-format=2.3]{10.209}\rlap{\footnotemark[7]} & $P1$ & stable &  \tablenum[table-format=4.1]{-50.0} & \cite{Wang2017}\\
MTP   & 17836 &  a$+$q & \tablenum[table-format=2.3]{9.954} & $P1$ & n/a & \tablenum[table-format=4.1]{-57.7}  & this work \\
	\end{tabular}
	\end{ruledtabular}
	\label{tab:opt}
\end{table*}

Due to the complicated potential energy surface of Li$_6$PS$_5$Cl with multiple local minima, it is difficult to find the fully optimized structure based on direct optimization methods (e.g., the conjugate gradient algorithm). To enhance the search for the global energy minimum, an MD-based annealing$+$quenching (a$+$q) optimization approach with volume relaxation was carried out with the MTP (see Sec.~\ref{sec:com}). The MTP-optimized structure was further relaxed with DFT utilizing the conjugate gradient algorithm to validate the MTP result. The resulting structure has an energy of \SI{-58.9}{meV/atom} with respect to the $F\bar{4}3m$ high-symmetry state, which is energetically more favorable than the structure obtained previously based on optimization with DFT and the USPEX method (\SI{-50.0}{meV/atom})~\cite{Wang2017}. The corresponding MTP energy of our optimized structure is \SI{-55.3}{meV/atom} and thus in agreement with the corresponding DFT value with an error of \SI{+3.6}{meV/atom}, consistent with the validation RMSE in energy [Fig.~\ref{fig:vali_bulk}(a) and Table~\ref{tab:mtp_fit_validate}]. For a larger simulation cell with \num{17836} atoms (the anion-ordered bulk model listed in Table~\ref{tab:gb_info}), the optimization with the MTP yields an energy that is further reduced by \SI{2.4}{meV/atom}. This result indicates that there are favorable arrangements of the Li atoms that cannot be represented with the small simulation cell. Tests show that repeating relaxations using the a$+$q approach may lead to different arrangements of Li atoms but have a small impact on the total energy of the relaxed structure (less than \SI{0.1}{meV/atom}). 

For the GB models, an additional step, the so-called $\gamma$-surface search~\cite{Guo2017}, should be considered to optimize the interface between the two grains before using the a$+$q approach. Specifically, in the $\gamma$-surface search, different rigid shifts of the two grains in directions parallel to the GB plane ($x$-$y$) are investigated in order to probe different initial GB structures for further optimization. For each GB model, we investigated a mesh of $20\times 20$ points on the respective $\gamma$-surface, by performing direct optimization for each such point. The structure of the lowest-energy point in the $\gamma$-surface was selected for further relaxation with the a$+$q approach. Additionally, the original GB structures without a shift were also used for relaxation with the a$+$q approach. Full relaxation of the structures with or without the $\gamma$-surface search leads to similar final optimized structures. This finding indicates that the shifting of two grains is implicitly included in the a$+$q optimization process, at least for the three GBs considered in the present study.

\begin{figure*}[tbp]
    \centering
    \includegraphics{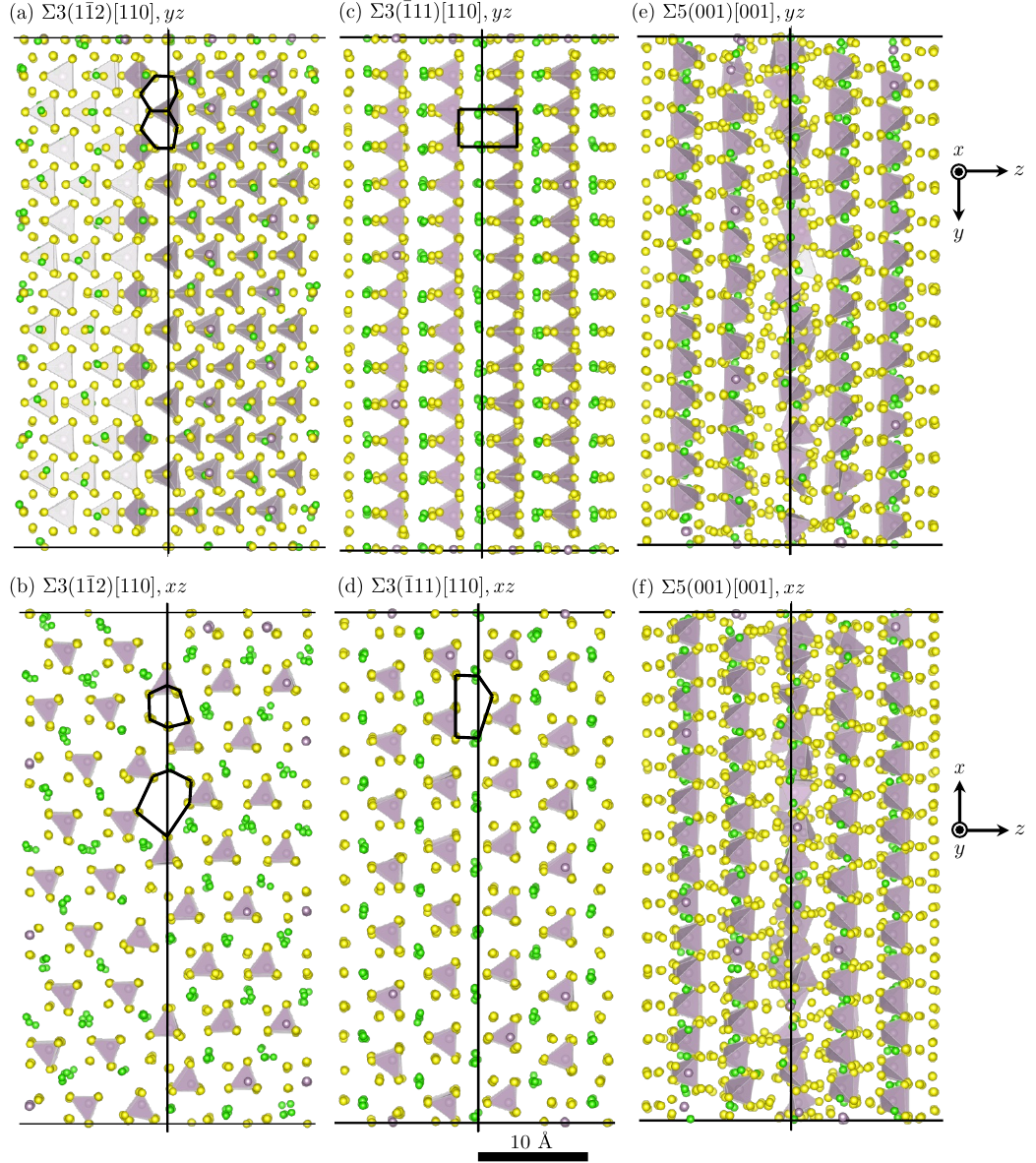}
    \caption{Atomic structures of the three GB models (Table~\ref{tab:gb_info}) after optimization. Each GB model is visualized in two views (top and bottom row), with the GB normal lying in the paper plane for both views. The GB plane is indicated by the vertical black line in the middle of each structure. For visibility, Li atoms and cages are not visualized, and only the atoms and PS$_4$ units close to the GB plane (within about $\pm\SI{10}{\angstrom}$) are shown. The smallest repeating GB structural units are highlighted for the $\Sigma3$ GBs. }
    \label{fig:gb_rel}
\end{figure*}

The atomic structures of the three GB models after optimization are shown in Fig.~\ref{fig:gb_rel}. To better emphasize the characteristic features, Li atoms are not visualized, and only the atoms close to the GB plane are shown. The dimensions of the optimized structures, and GB energies [$\gamma$, Eq.~\eqref{eq:gbe}] 
calculated at \SI{0}{\kelvin} are shown in Table~\ref{tab:gb_info}. For $\Sigma3 (1\bar{1}2)[110]$ [Figs.~\ref{fig:gb_rel}(a) and (b)], voids (as large as \SI{6}{\angstrom} in diameter) form at the GB plane. The formation of these voids requires the reordering of a range of atoms, which in turn requires enough thermal energy that is available through the a$+$q process. In contrast, direct optimization could not reproduce the voids, emphasizing again the importance of the a$+$q process. For the $\Sigma3 (\bar{1}11)[110]$ GB [Figs.~\ref{fig:gb_rel}(c) and (d)], for which the cage structure is preserved during the GB construction, only a small change of the cage arrangement is visible at the GB plane after optimization. In particular, no void formation is observed. Correspondingly, $\Sigma3(\bar{1}11)[110]$ shows a GB energy of only \SI{7.12}{meV/\angstrom\textsuperscript{2}}, which is the smallest among the three investigated GBs. Such a small value indicates that the GB structure does not differ much from the bulk structure. For the $\Sigma5(001)[001]$ GB [Figs.~\ref{fig:gb_rel}(e) and (f)], for which the cage structure is preserved and the GB plane is densely filled with atoms, yet other results are observed than for the two $\Sigma3$ GBs. For the $\Sigma5$ GB, optimization leads to an amorphous-like area at and near the GB plane. The amorphous area extends from the Cl atoms at the GB plane to the next layer of the Cl atoms away from the GB plane. The atoms inside the affected area appear to be structurally disordered. Since the bonds at the GB plane are severely modified compared to the bulk, the largest GB energy (\SI{18.78}{meV/\angstrom\textsuperscript{2}}) is found for the $\Sigma5(001)[001]$ GB. Results show that the excess volumes per unit GB area of all GBs are small, with values less than \SI{1}{\angstrom}. 

\begin{figure*}[tbp]
    \centering
    \includegraphics{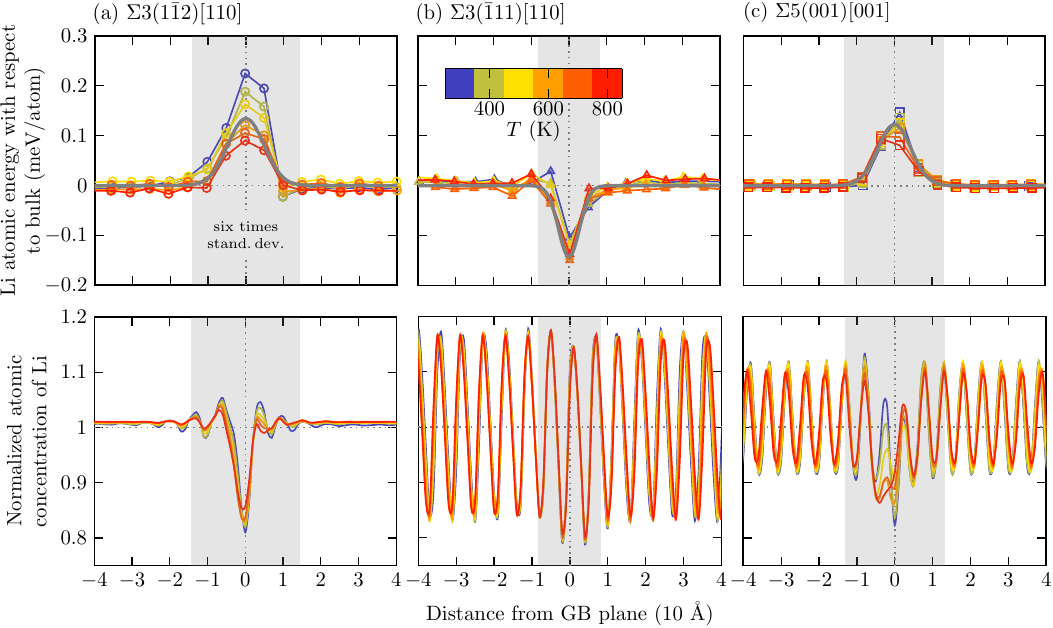}
    \caption{Temperature-dependent GB profiles of the three GB models (Table~\ref{tab:gb_info}) averaged over \SI{5}{\nano\second} of MD simulation time. The upper panel shows the atomic energy of Li atoms referenced with respect to the bulk. A Gaussian fit (gray line) of the data from \SI{600}{\kelvin} was used for GB width analysis. The lower panel shows the atomic concentration of Li atoms normalized with the average Li concentration of the GB simulation cells. The kernel-density estimation was used for smoothing. The gray area indicates the GB regions, which were defined to be six times the standard deviations (stand.~dev.) of the Gaussian fits in the upper panel. The Li concentration averaged over the GB region was used to confirm that the segregation of Li in the GB region is negligible. }
    \label{fig:gb_profile}
\end{figure*}

Figure \ref{fig:gb_profile} shows the atomic energy and the atomic concentration profiles of the Li atoms for the three GB models. To emphasize the impact of the GBs, the energy profiles are referenced with respect to their bulk counterparts, and the concentration profiles are normalized with the average Li concentration of the GB simulation cells. According to the energy profiles (upper panel), the atomic energy of the Li atoms increases in the GB region for the $\Sigma 3(1\bar{1}2)[110]$ (up to \SI{0.2}{meV/atom}) and the $\Sigma 5(001)[001]$ (up to \SI{0.1}{meV/atom}) GBs. For the $\Sigma3(\bar{1}11)[110]$ GB, the change in atomic energy of Li in the GB region is negative (about \SI{-0.1}{meV/atom}). For the $\Sigma 3(1\bar{1}2)[110]$ and $\Sigma 5(001)[001]$ GBs, the impact of the GB structure on the Li atomic energy is slightly reduced as temperature increases. As regards the concentration profiles (lower panel), we observe layers with increased and decreased Li concentrations around the GB plane. Especially for $\Sigma5(001)[001]$, Li atoms tend to segregate from the center to the edge of the amorphous-like GB region. With higher temperatures, smaller Li concentration fluctuations are found. 

The atomic energy profiles in Fig.~\ref{fig:gb_profile} were utilized to determine the GB widths as needed for the GB diffusion coefficient calculations (see Sec.~\ref{sec:separa}). Specifically, a Gaussian function was fitted to the peaks in the Li atomic energy, and the GB width was defined to be six times the standard deviation of the fitted Gaussian function. A similar approach to determine the GB width was used in Ref.~\cite{Koju2020}. Since the temperature dependence of the standard deviation of the fitted Gaussian function is very small, the GB widths obtained from the \SI{600}{\kelvin} profiles (shown in Table~\ref{tab:gb_info}) were used throughout the GB diffusion coefficient calculations. The estimated GB widths are similar between $\Sigma3(1\bar{1}2)[110]$ (\SI{28.8}{\angstrom}) and $\Sigma5(001)[001]$ (\SI{26.5}{\angstrom}) and smaller for $\Sigma3(\bar{1}11)[110]$ (\SI{16.7}{\angstrom}), consistent with visual inspection of the Li diffusion trajectories (see Sec.~\ref{sec:bulk}). The volume fractions [$\tau$ in Eq.~\eqref{eq:fracapprox}] of the three GBs in the GB models are between \SIrange{25}{39}{\percent}. 

The determined GB regions for the three GBs are indicated in the atomic concentration profiles in Fig.~\ref{fig:gb_profile}. Analysis shows that the Li concentration averaged within the GB region is comparable to the average Li concentration of the reference bulk structure. It indicates that the segregation of Li in the GB region is negligible, supporting the approximation in Eq.~\eqref{eq:fracapprox}. Li segregation in the present simulations may be hindered by the restriction of all simulation cells to the stoichiometric composition of Li$_6$PS$_5$Cl (see Sec.~\ref{sec:gbcon}). 

\subsection{Bulk and GB diffusion}\label{sec:bulk}

\begin{figure*}[tbp]
    \centering
    \includegraphics{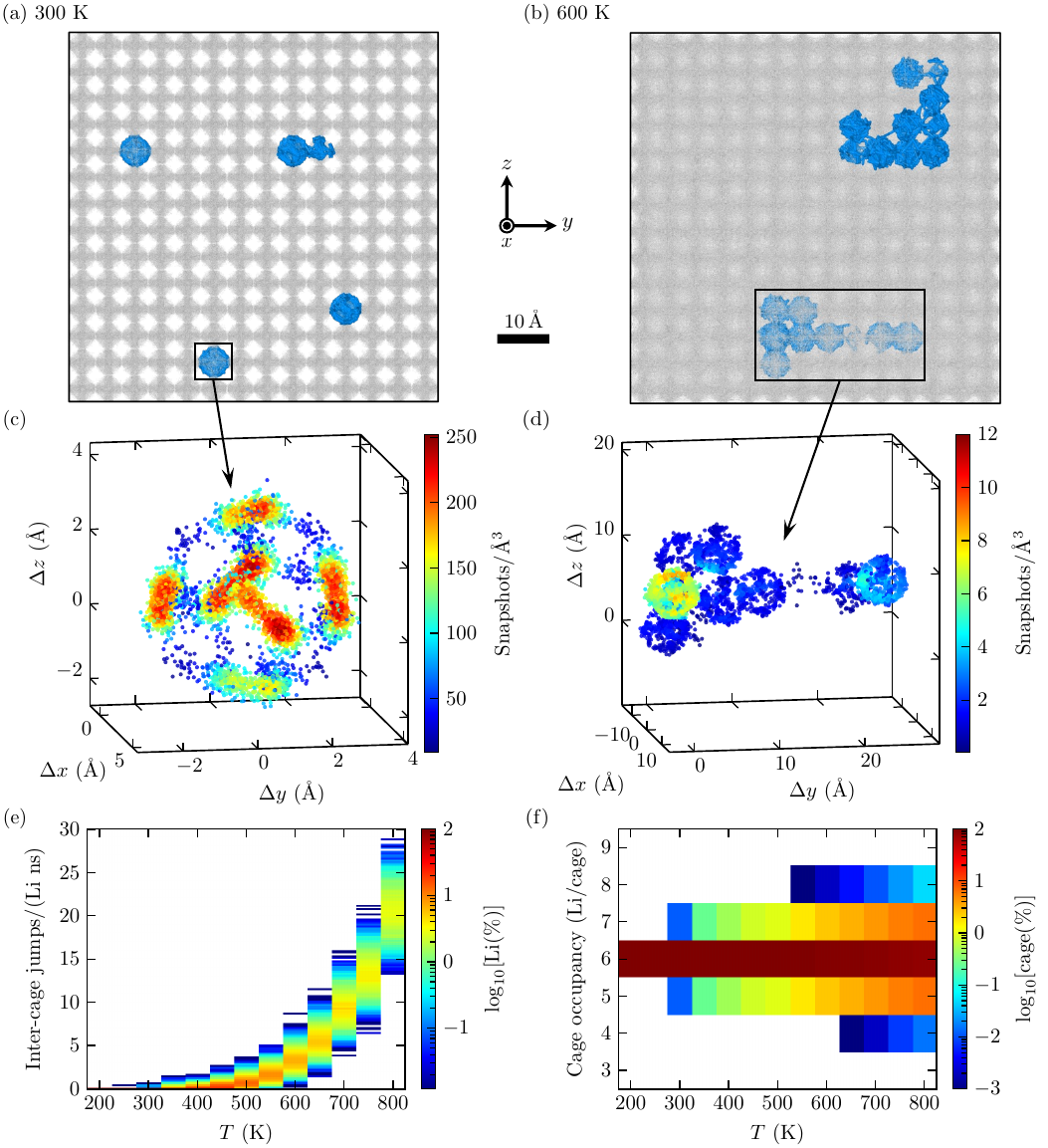}
    \caption{Trajectories of all Li atoms in the ordered bulk structure (Table~\ref{tab:gb_info}) at (a) \SI{300}{\kelvin} and (b) \SI{600}{\kelvin} after \SI{5}{\nano\second} of MD simulations. The trajectories of a few selected Li atoms are highlighted in blue, and the trajectories of the other Li atoms are shown in gray. Magnified views of one of the highlighted Li trajectories at (c) \SI{300}{\kelvin} and (d) \SI{600}{\kelvin} relative to the initial position are also shown. The color mapping indicates the probability density smoothed by kernel-density estimation. The elapsed time between the snapshots is \SI{1}{\pico\second}. (e) Inter-cage jumping rate and (f) cage occupancy of Li at different temperatures. The color mapping indicates the percentage of Li atoms that exhibit the corresponding jumping rate or cage occupancy. The result of each temperature is averaged over \SI{5}{\nano\second} of MD simulation. Trajectories in (a) and (b) are visualized by \textsc{ovito}~\cite{Stukowski2009}. 
    }
    \label{fig:bulk_diff}
\end{figure*}

Trajectories of all Li atoms obtained from \SI{5}{ns} MD runs at \SIlist{300;600}{\kelvin} for the anion-ordered bulk structure are shown in Figs.~\ref{fig:bulk_diff}(a) and (b), respectively. The trajectories of a few selected Li atoms are highlighted in blue. At \SI{300}{\kelvin}, the Li trajectories are mostly localized within the cages, and only very few inter-cage jumps occur during the simulation time. At \SI{600}{\kelvin}, the Li trajectories are distributed much more homogeneously in the simulation cell due to the enhanced inter-cage (long-range) diffusion. Clearly, temperature strongly affects Li diffusion, especially inter-cage diffusion. It is worth emphasizing that the macroscopic conductivity measured on the experimental scale mainly results from inter-cage diffusion.

Figures~\ref{fig:bulk_diff}(c) and (d) are zoom-ins of the trajectories of a single Li atom at \SIlist{300;600}{\kelvin}, respectively. At \SI{300}{\kelvin}, the selected Li atom shows frequent \emph{intra}-cage diffusion visiting multiple interstitial sites of the same cage within the \SI{5}{ns} of simulation time but no \emph{inter}-cage diffusion. At \SI{600}{\kelvin}, in contrast, significant intra- and inter-cage diffusion is observed for the selected Li atom. After jumping to a neighboring cage, the Li atom typically resides in the new cage for some time, during which various interstitial sites are visited via intra-cage diffusion. Occasionally, the residence time is shorter, and the Li atom diffuses quickly to another cage. Short residence times in the cages are more frequently observed at higher temperatures, and they further enhance long-range diffusion and conductivity.

To quantify the frequency of inter-cage jumps,  Fig.~\ref{fig:bulk_diff}(e) shows the inter-cage jump rates per Li atom and per nanosecond as a function of temperature. At the lowest investigated temperature (\SI{200}{\kelvin}), almost no inter-cage jumps are observed within the \SI{5}{ns} simulation time. The average inter-cage jumping rate increases strongly with increasing temperature. Consequently, the macroscopic conductivity likewise increases strongly with temperature. 

In the anion-ordered bulk structure, each cage consists of six Li atoms around one S atom. When inter-cage jumps of Li occur, it can be anticipated that there will be an imbalanced distribution of Li in the cages, i.e., single cages may show a Li occupancy of more or less than six. Figure~\ref{fig:bulk_diff}(f) shows the cage occupancy averaged over the entire MD simulation time at different temperatures. The color mapping in the plot indicates the percentage of the cages in the simulation cell that exhibit the corresponding occupancy. As expected, when very few inter-cage jumps are observed, e.g., at \SI{200}{\kelvin}, almost all cages are occupied by six Li atoms, i.e., neutral occupation. With increasing temperature and more inter-cage jumps [Fig.~\ref{fig:bulk_diff}(e)], the cage occupancy distribution broadens, i.e., cages with more or less than six Li atoms are observed. Taking the highest investigated temperature of \SI{800}{\kelvin}, about 18\% of cages have an imbalance of $\pm1$ Li atoms (i.e., cages with five or seven Li atoms). The majority (82\%) of the cages are still occupied with six Li atoms.

\begin{figure*}[tbp]
    \centering \includegraphics{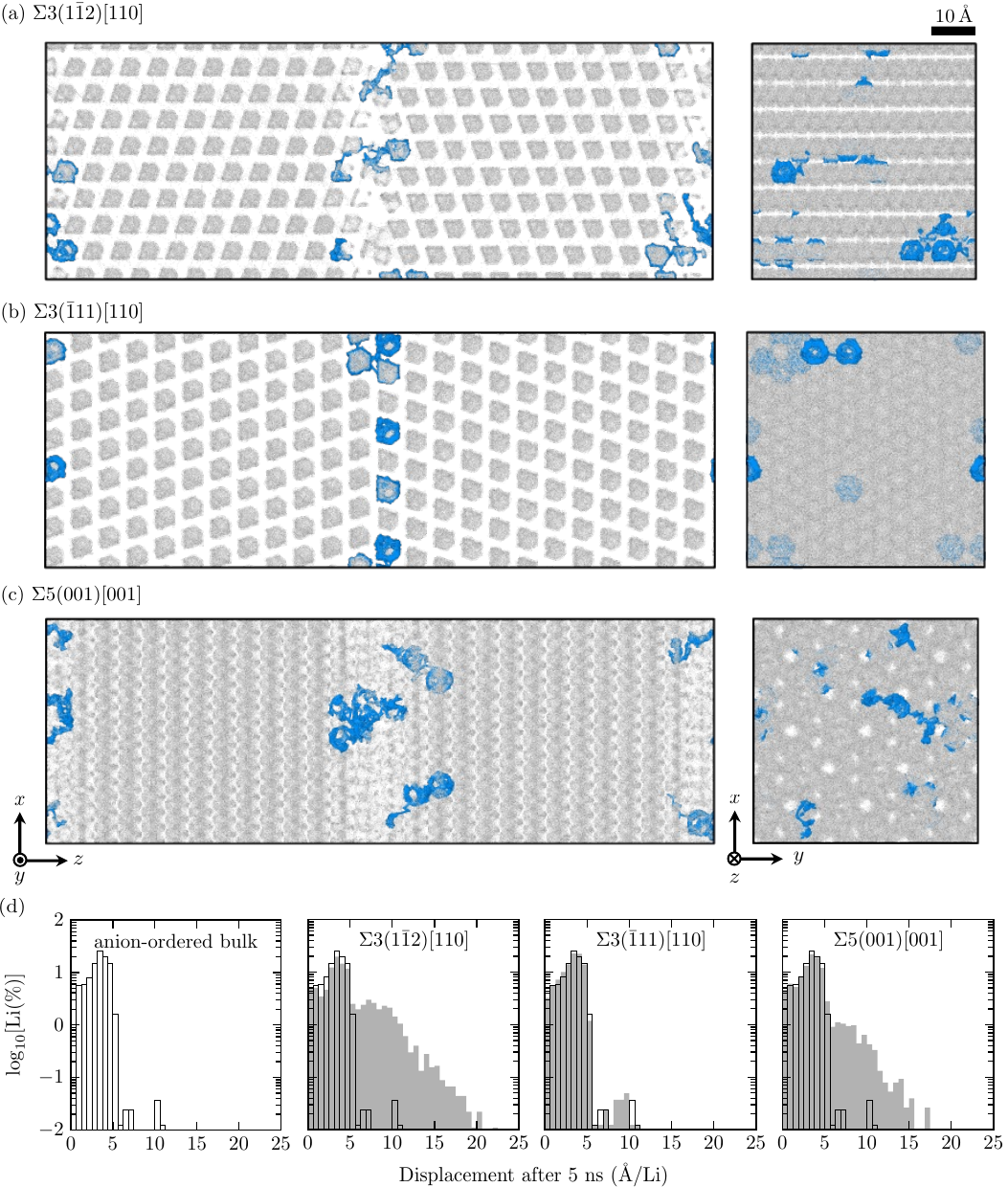}
    \caption{Analysis of Li diffusion for the three GB models (Table~\ref{tab:gb_info}) at \SI{300}{\kelvin} and after \SI{5}{\nano\second} MD simulations. (a), (b), and (c) Trajectories of all Li atoms in the simulation cells. The trajectories of a few selected Li atoms in the GB region are highlighted in blue, and the trajectories of the other Li atoms are shown in gray. For each GB model, trajectories are visualized in two views. GB planes are located in the middle and at the edge of the GB simulation cells in the $z$-direction. (d) Distribution of the Li displacements for the GB models (gray-shaded) in comparison with that for the anion-ordered bulk model (black-edged). }
    \label{fig:gb_tra}
\end{figure*}

Figure~\ref{fig:gb_tra} shows an analysis of Li diffusion for the three GB models, with a focus on \SI{300}{\kelvin}. Similar to the bulk analysis [Figs.~\ref{fig:bulk_diff}(a) and (b)], trajectories of all Li atoms obtained from \SI{5}{ns} MD simulations for the three GB models are shown in Figs.~\ref{fig:gb_tra}(a)--(c). Trajectories far from the GB regions are similar to the bulk results discussed above. In the GB regions, differences can be seen depending on the GB type. For $\Sigma3(1\bar{1}2)[110]$ [Fig.~\ref{fig:gb_tra}(a)], broken cages are connected by Li trajectories across the GB plane, leading to multiple complex diffusion paths. Additionally, small voids (regions not covered by Li trajectories) in diameter of about \SI{5}{\angstrom} are found at the GB plane. These voids result from the intersection of rows of cages from the two grains, oriented with a rotational angle of \SI{70.53}{\degree} (the tilt angle). The impact on Li diffusion is substantially different when the GB planes do not break any cages: For $\Sigma3(\bar{1}11)[110]$ [Fig.~\ref{fig:gb_tra}(b)], the GB plane is located between two rows of cages from each grain. Most Li atoms show only intra-cage diffusion, even in the GB region. Only a few trajectories cross the GB planes. For $\Sigma5(001)[001]$ [Fig.~\ref{fig:gb_tra}(d)], the analysis of the Li trajectories is complicated by the projection plane. Nevertheless, one can clearly distinguish different trajectories within the GB region. The trajectories appear homogeneously distributed, with no clear voids visible. 

The trajectories of a few selected Li atoms in the GB regions are highlighted in Figs.~\ref{fig:gb_tra}(a), (b), and (c). Relatively long trajectories are found, which indicates that the GB structure increases the probability of long-range diffusion compared to the anion-ordered bulk. For $\Sigma3(1\bar{1}2)[110]$, the long-range trajectories connect the broken cages at the GB plane. Trajectories combined with both intra- and inter-cage jumps are observed for the $\Sigma3(\bar{1}11)[110]$ GB. Similar trajectories are also found for the anion-ordered bulk but at a higher temperature, e.g., \SI{600}{\kelvin} [Figs.~\ref{fig:bulk_diff}(c) and (d)]. For $\Sigma5(001)[001]$, trajectories reveal straighter diffusion paths for Li crossing the GB plane, suggesting a weaker trapping effect of the cages for Li. Trajectories within the GB plane [right column of Figs.~\ref{fig:gb_tra}(a)--(c)] suggest negligible in-plane anisotropy in the GB diffusion for the three GBs. 

In Fig.~\ref{fig:gb_tra}(d), distributions of Li according to displacements from the GB simulations are compared to those derived from simulations of the anion-ordered bulk structure. For the bulk structure, most Li atoms show displacements up to about \SI{6}{\angstrom}. Considering that the maximum atomic distance within one cage at \SI{0}{\kelvin} is about \SI{4.6}{\angstrom}, it can be concluded that most Li atoms at \SI{300}{\kelvin} are restricted to intra-cage diffusion, consistent with the previous analysis (Fig.~\ref{fig:bulk_diff}). At the same temperature and simulation time, longer Li displacements and a higher probability of long-distance displacements are observed for the GB models compared to the bulk. Importantly, a different impact on Li mobility is observed for the different GB types. A strong enhancement in Li diffusion is shown for $\Sigma3(1\bar{1}2)[110]$ and $\Sigma5(001)[001]$ GBs. A relatively small enhancement in Li diffusion is seen for $\Sigma3(\bar{1}11)[110]$, which mainly results from a higher probability of Li atoms that show displacements between 7 and \SI{11}{\angstrom}. 

\subsection{Diffusion coefficients}\label{sec:separa}

\begin{figure*}[tbp]
    \centering
    \includegraphics{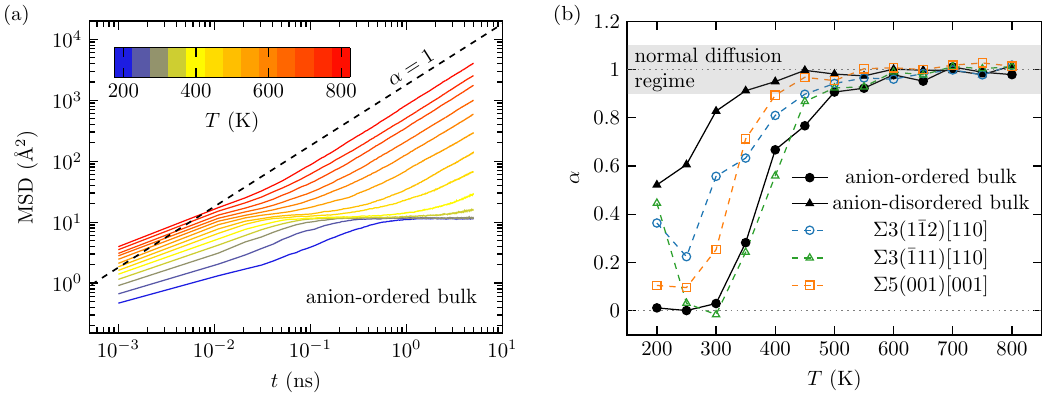}
    \caption{(a) Mean square displacements (MSD) of all Li atoms in MD simulations of the anion-ordered bulk models (Table~\ref{tab:gb_info}) as a function of simulation time up to \SI{5}{ns}.  The color of the MSD curves indicates the simulation temperature. The slope corresponding to $\alpha=1$ [Eq.~\eqref{eq:msdfit}, normal diffusion] is shown by dashed black lines for reference. (b) $\alpha$ values as a function of temperature obtained by fitting the MSD within the simulation time interval \SIrange{3}{5}{ns} according to Eq.~\eqref{eq:msdfit}. The normal diffusion regime, defined by the criterion $\alpha \in [0.9, 1.1]$ (Sce.~\ref{sec:diffco}), is indicated. Note that the MSDs from the anion-disordered bulk and the GB models are not shown here, and the GB models include both the bulk and the GB regions (see Sec.~\ref{sec:gbcon}).
    }
    \label{fig:msd}
\end{figure*}

Diffusion coefficients were calculated from the MSDs according to Eq.~\eqref{eq:diffcoecal} for normal diffusion. Figure~\ref{fig:msd}(a) shows the MSDs of the Li atoms in the anion-ordered bulk as a function of time and for different temperatures at intervals of $\SI{50}{\kelvin}$ as a representative example. Note that a double-logarithmic scale is used to reveal the diffusion type. At low temperatures (up to \SI{300}{\kelvin}), a plateau in the MSD is visible after a certain simulation time. The plateau indicates that most Li atoms in the simulation cell exhibit only intra-cage (short-range) diffusion within the simulation time, i.e., Li atoms are trapped in the cages. This is fully consistent with the above analysis of the Li trajectories [Figs.~\ref{fig:bulk_diff}(a) and (b)]. At elevated temperatures (\SIrange{300}{600}{\kelvin}), the MSD curves show a characteristic increase--plateau--increase shape. This shape indicates a combination of intra- and inter-cage jumps (short- and long-range diffusion) of Li atoms in the simulation cell, as also implied by the increased inter-cage jump rates at higher temperatures shown in Fig.~\ref{fig:bulk_diff}(e). At temperatures exceeding \SI{600}{\kelvin}, the plateau in the MSD disappears because the long-range diffusion of Li sets in already in the initial stages of the simulation. Similar MSD curves are found for the anion-disordered bulk and the three GB models.

A linear dependence of the MSD with a slope of $\alpha=1$ [see Eq.~\eqref{eq:msdfit}] identifies normal diffusion, while the respective offset corresponds to the diffusion–coefficient value. The dotted line in Fig.~\ref{fig:msd}(a) indicates a slope of one as a visual guide. According to the MSDs, subdiffusion ($\alpha < 1$) is found for simulations at low temperatures in the whole simulation time frame up to \SI{5}{ns}. At high temperatures, subdiffusion is also observed for short observation times as evidenced by the initial slope below one (e.g., at \SI{800}{\kelvin} in the time interval up to \SI{10}{\pico\second}). During these short observation times, there are not enough inter–cage jumps in the simulation to trigger long-range diffusion. These results emphasize the importance of carefully checking the slope of the MSD curve to ensure a correct application of the Einstein equation [Eq.~\eqref{eq:diffcoecal}]. Figure~\ref{fig:msd}(b) shows the $\alpha$ values as a function of temperature obtained by fitting the MSD within the simulation time interval of \SIrange{3}{5}{ns} according to Eq.~\eqref{eq:msdfit}. We define normal diffusion by the criterion $\alpha \in [0.9, 1.1]$ (see Sec.~\ref{sec:diffco}). Based on this definition, normal diffusion occurs for the anion-ordered bulk at temperatures higher than \SI{500}{\kelvin}. For the GBs that enhance Li diffusion, normal diffusion is already observed at lower temperatures. The anion-disordered bulk structure with \SI{50}{\percent} anion disorder shows normal diffusion at the lowest temperatures, i.e., from \SI{350}{\kelvin} on. Diffusion coefficients were calculated for temperatures within the defined normal–diffusion regime.

\begin{figure*}[tbp]
    \centering
    \includegraphics{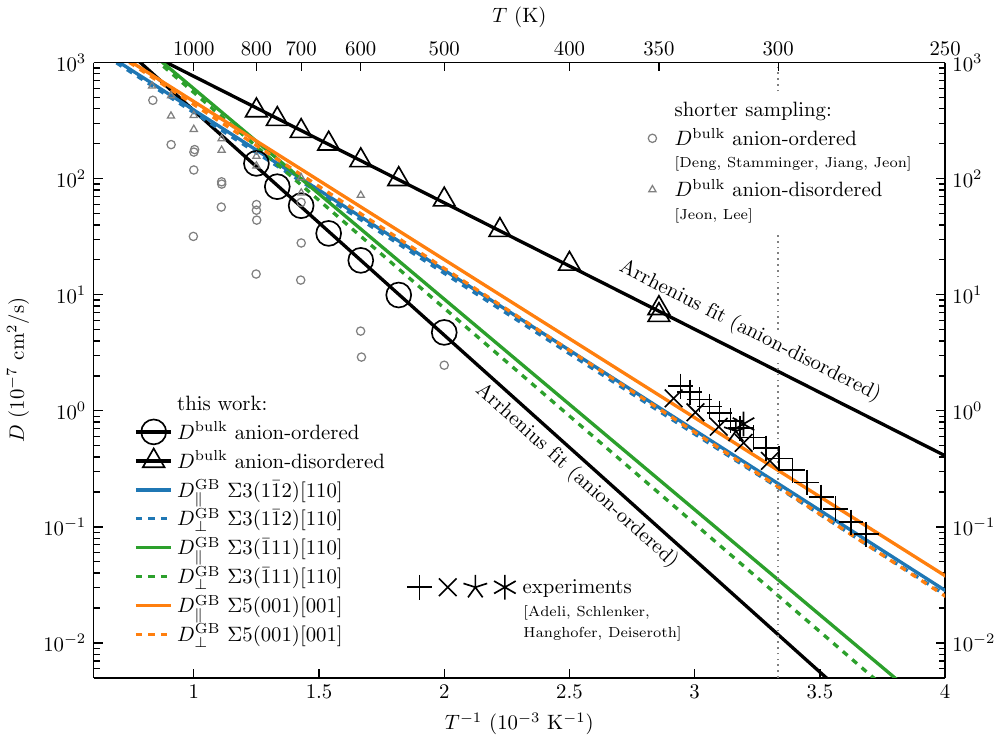}
    \caption{Arrhenius diagram for diffusion coefficients $D^\mathrm{bulk}$, $D^\mathrm{GB}_\parallel$, and $D^\mathrm{GB}_\perp$ obtained by atomistic simulations. Anion-ordered and anion-disordered (\SI{50}{\percent} Cl/S-anion disorder) bulk results are included. 
    The fitted Arrhenius lines for the three GBs and directions parallel ($\parallel$, solid lines) and perpendicular ($\perp$, dashed lines) to the GB plane are shown. 
    For comparison, bulk diffusion coefficients obtained by \textit{ab initio} MD [shorter sampling, up to \SI{300}{ps} vs.~\SI{5}{\nano\second} in the present work (Table~\ref{tab:history})] from Refs.~\cite{Deng2017,Stamminger2019,Jiang2022,Jeon2024} (ordered) and~\cite{Lee2022,Jeon2024} (disordered, \SI{50}{\percent} Cl/S-anion disorder) are shown.
    The experimental data marked by $+$~(\citet{Adeli2019}), \texttimes~(\citet{Schlenker2020}), and $\star$~(\citet{Hanghofer2019}) are based on nuclear magnetic resonance measurements, while $\ast$~(\citet{Deiseroth2011}) are based on polarization measurements. The samples corresponding to $+$ and \texttimes \,showed \SI{61.5}{\percent}~(\citet{Adeli2019}) and \SI{53.8}{\percent}~(\citet{Schlenker2020}) Cl/S-anion disorder, respectively. }
    \label{fig:dis_coe}
\end{figure*}

The calculated diffusion coefficients and the resulting Arrhenius fits are shown in Fig.~\ref{fig:dis_coe} for the anion-ordered and anion-disordered bulk structures (open symbols and black lines). For the three GBs, only the Arrhenius fits are shown, separated into diffusion coefficients perpendicular to and within the GB plane. The corresponding Arrhenius fitting parameters are listed in Table~\ref{tab:diff_coe_fit}. Clear linear Arrhenius relations are observed for the diffusion coefficients of both the anion-ordered and the anion-disordered bulk structures with respect to the inverse temperature. The ordered bulk structure shows a high diffusion activation energy of \SI{384}{meV}, which is lowered to \SI{216}{meV} by the random \SI{50}{\percent} anion disorder of the Cl and S anions. This is in good agreement with previous experiments~\cite{Morgan2021} and simulations~\cite{Minafra2020} focusing on the effect of anion disorder. For comparison, Fig.~\ref{fig:dis_coe} includes results of anion-ordered~\cite{Deng2017,Stamminger2019,Jiang2022} and anion-disordered~\cite{Lee2022,Jeon2024} bulk structures obtained by \textit{ab initio} MD simulations from previous studies, all of which show significant deviations compared to the values obtained in the present work. This is likely due to smaller simulation cells \{\SI{52}{atoms}~\cite{Deng2017,Stamminger2019,Jiang2022,Lee2022,Jeon2024} vs.~more than~\SI{16000}{atoms} in the present work (Table~\ref{tab:history})\} and shorter simulation times \{up to \SI{300}{ps}~\cite{Deng2017,Stamminger2019,Jiang2022,Lee2022,Sadowski2023,Jeon2024} vs.~\SI{5}{\nano\second} in the present work (Table~\ref{tab:history})\} due to the expensive \textit{ab initio} MD simulations utilized in the previous studies. 

\begin{table*}[tbp]
\centering
	\caption{Fitted Arrhenius parameters [$D_0$, $E_\textrm{a}$, see Eq.~\eqref{eq:diffcoearr}] and self-diffusion coefficients at \SI{300}{\kelvin} of bulk and pure GBs, i.e., $D^\mathrm{bulk}$, $D^\mathrm{GB}_\parallel$, and $D^\mathrm{GB}_\perp$ obtained in the present work. Note that $D^\mathrm{GB}_\parallel$ and $D^\mathrm{GB}_\perp$ in Eqs.~\eqref{eq:para} and~\eqref{eq:perp}, respectively, are evaluated with the estimated GB widths listed in Table~\ref{tab:gb_info} (for the impact of GB widths, see Sec.~\ref{sec:dis}). The experimental results obtained by nuclear magnetic resonance measurements in previous studies~\cite{Adeli2019,Schlenker2020} are shown for comparison. The ionic conductivities, $\sigma$, are converted from the calculated diffusion coefficients, $D$, based on a relation extracted from the experimental data in Ref.~\cite{Schlenker2020} (see Appendix~\ref{sec:conductivity}).}
	\begin{ruledtabular}
	\begin{tabular}{cccS[table-format=3.0(2)] S[table-format=1.3] S[table-format=2.1]}
		Structure & Direction &  {$D_{0}$ (\SI{e-7}{cm^2/s})} & {$E_{\mathrm{a}}$ (meV)} & {$D$ at $\SI{300}{\kelvin}$ (\SI{e-7}{cm^2/s})} & {$\sigma$ at $\SI{300}{\kelvin}$ (\SI{}{mS/cm})}\\
		\hline\\[-0.3cm]
	anion-ordered bulk & all & \SI{3.3e4}{} & 384  &  0.012 &  0.2\\
	  anion-disordered bulk\footnote{The structure has \SI{50}{\percent} of Cl/S-anion disorder.} & all & \SI{9.4e3}{} & 216 &  2.203 & 29.8 \\
        GB $\Sigma3(1\bar{1}2)[110]$ & $\parallel$ & \SI{9.4e3}{} & 274  &  0.236 &  3.2\\
         & $\perp$ & \SI{9.0e3}{} & 275 & 0.218  &  2.9\\
	GB $\Sigma3(\bar{1}11)[110]$ & $\parallel$ & \SI{3.9e4}{} & 360  & 0.035  &  0.5\\
         & $\perp$ & \SI{4.0e4}{} &  369  & 0.026  &  0.3\\
        GB $\Sigma5(001)[001]$ & $\parallel$ & \SI{1.1e4}{} & 270  & 0.307  &  4.1\\
         & $\perp$ & \SI{1.1e4}{} & 280  &  0.222 & 3.0 \\
         polycrystal in experiment\footnote{The sample showed \SI{61.5}{\percent} of Cl/S-anion disorder~\cite{Adeli2019}.} & n/a & n/a & 350\pm10 & 0.387 & 2.5 \\
         polycrystal in experiment\footnote{The sample showed \SI{53.8}{\percent} of Cl/S-anion disorder~\cite{Schlenker2020}.} & n/a & n/a & 280\pm10 & 0.25 & 3.4\\
	\end{tabular}
	\end{ruledtabular}
	\label{tab:diff_coe_fit}
\end{table*}

Enhanced diffusion is observed for all three GBs as compared to the anion-ordered bulk (\SIrange{0.026e-7}{0.307e-7}{cm^2/s} for GBs vs.~\SI{0.012e-7}{cm^2/s} for the anion-ordered bulk at \SI{300}{\kelvin}). The enhancement is sensitive to the GB width (details in Sec.~\ref{sec:dis}). Based on the current width estimates [Figs.~\ref{fig:gb_profile} and Table~\ref{tab:gb_info}], varying increases are shown at \SI{300}{\kelvin}: from $\Sigma3(\bar{1}11)[110]$ (about two to three times) to $\Sigma3(1\bar{1}2)[110]$ or $\Sigma5(001)[001]$ (more than an order of magnitude). For all GBs, higher diffusivity is shown for Li diffusing along rather than across the GB plane at \SI{300}{\kelvin}, i.e., $D^\mathrm{GB}_\parallel>D^\mathrm{GB}_\perp$. The calculated GB diffusion coefficients are within the interval set by the anion-ordered and anion-disordered bulk diffusion coefficients for temperatures below \SI{1000}{\kelvin}. Table~\ref{tab:diff_coe_fit} shows that the diffusion activation energies of all three GBs are also within the activation energy range of the anion-ordered and anion-disordered bulk structures (\SIrange{216}{384}{\meV}). 

Experimental data obtained by nuclear magnetic resonance measurements from Refs.~\cite{Adeli2019, Hanghofer2019, Schlenker2020} and obtained by polarization measurements from Ref.~\cite{Deiseroth2011} are shown in Fig.~\ref{fig:dis_coe} and Table~\ref{tab:diff_coe_fit} for comparison. For reference, ionic conductivities, $\sigma$, which are converted from the calculated diffusion coefficients from atomistic simulations (see Appendix~\ref{sec:conductivity}), are also shown in Table~\ref{tab:diff_coe_fit}. The available experimental data, including all the activation energies, the diffusion coefficients, and the conductivities, fall between or together with the computed and Arrhenius-extrapolated values of the investigated atomic structures.

\section{Discussion}\label{sec:dis}

We have proposed an active learning scheme based on MTPs of progressively increasing quality. A systematic validation for the thus-obtained MTPs has also been conducted. Our results demonstrate that the proposed active learning scheme is able to automatically and efficiently sample configurations for constructing the training set of complex atomic structures in Li$_6$PS$_5$Cl, including GBs, starting from \textit{ab initio} MD simulations of only the anion-ordered bulk structure. Importantly, diffusion coefficients obtained from the final high-level MTPs show small variations with respect to the training set and MTP parameter space, guaranteeing the reproducibility of atomistic simulation results using machine-learning interatomic potentials. We note that active learning ideas akin to our current scheme were investigated recently for structurally and chemically simpler systems, i.e., a unary metallic system (Zr) with defects~\cite{Luo2023} and a defect-free ternary system (Ti$_{0.5}$Al$_{0.5}$N)~\cite{Bock2024}. With the verified applicability to a structurally and chemically complex material system, Li$_6$PS$_5$Cl, the present work systematically brings active learning to the next development stage. 

The high-level MTPs trained with the proposed scheme show energy RMSEs from \SIrange{6}{7}{meV/atom}, which are higher than that found for metallic systems, either with defects (up to \SI{4}{meV/atom})~\cite{Xu2023,Xu2024,Zotov2024} or without defects (\SIrange{1}{3}{meV/atom})~\cite{Grabowski2019,Jung2023,Jung2023a,Forslund2023,Zhou2022}. This comparison indicates the inherent complexity of the electronic interactions in Li$_6$PS$_5$Cl. Since the computational time grows quickly for high MTP levels without substantial reduction of the RMSEs, a further increase of the MTP level does not seem to be a viable option. In the future, the proposed active learning scheme could be coupled with other recent machine-learning potentials, e.g., NequIP~\cite{Batzner2022} or MACE~\cite{NEURIPS2022_4a36c3c5}, to further optimize the balance of accuracy and efficiency.

The GB energies obtained with the MTPs and the annealing-and-quenching (a$+$q) approach for three structurally distinct GBs in Li$_6$PS$_5$Cl are within \SIrange{7}{19}{meV/\angstrom\textsuperscript{2}}, which translates to \SIrange{0.1}{0.3}{J/m^2}. 
The range is consistent with that calculated by DFT for GBs in Li$_6$PS$_5$Cl with smaller simulation cells (\SIrange{0.09}{0.54}{J/m^2})~\cite{Xie2024}. A close value of \SI{0.26}{J/m^2} was reported in Ref.~\cite{Sadowski2023} for Li$_6$PS$_5$Br, a similar material system. The calculated GB energies for Li$_6$PS$_5$Cl are mostly lower than those for other recently explored solid electrolytes, e.g., Li$_3$PS$_4$ (\SIrange{0.2}{1.2}{J/m^2})~\cite{Jalem2023}, Li$_{0.16}$La$_{0.62}$TiO$_3$ (\SIrange{0.3}{1.3}{J/m^2})~\cite{Symington2021} or Li$_3$OCl (\SIrange{0.3}{1.1}{J/m^2})~\cite{Dawson2018,Quirk2023}. 

\begin{figure}[tbp]
    \centering
    \includegraphics{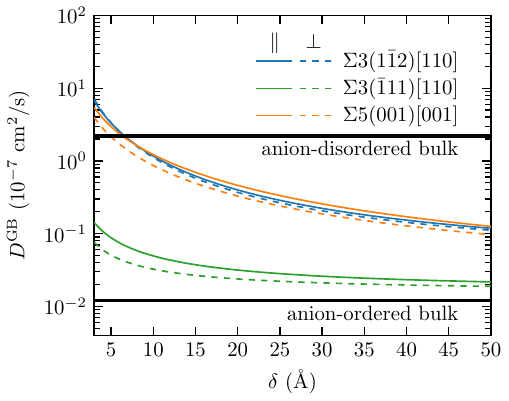}
    \caption{Calculated GB diffusion coefficients at \SI{300}{\kelvin} as a function of the utilized GB width. Diffusion coefficients of the anion-ordered and the anion-disordered bulk structures at \SI{300}{\kelvin} are shown for comparison. In the present study, the GB widths of the three investigated GBs were determined to be between \SIrange{15}{30}{\angstrom}.}
    \label{fig:width}
\end{figure}

In the present study, we have defined the GB width to be six times the standard deviation (covering 99.7\%) of the Gaussian fitted to the relative atomic energy profiles of Li. In this way, the GB regions cover most of the Li layers showing GB-induced concentration deviations with respect to the bulk, resulting in a negligible Li segregation. The GB widths determined in this way are within \SIrange{15}{30}{\angstrom}, and thus larger than GB width estimates for other solid electrolytes in previous computational studies, e.g., $\beta$-Li$_{3}$PS$_{4}$ (\SIrange{7}{10}{\angstrom})~\cite{Dawson2018} or Li$_{0.375}$Sr$_{0.4375}$Ta$_{0.75}$Zr$_{0.25}$O$_{3}$ (\SIrange{5}{10}{\angstrom})~\cite{Lee2023}. Our results indicate that GBs have a relatively large impact on the local structural modification of complex solid electrolytes. The computed GB diffusion coefficients show sensitivity to the GB width employed in the calculations, as demonstrated in Fig.~\ref{fig:width}. For Li$_6$PS$_5$Cl, where GBs enhance Li diffusion compared to the anion-ordered bulk, smaller estimates of GB widths result in larger GB diffusion coefficients and vice versa. With a GB width larger than \SI{7}{\angstrom}, the diffusion coefficients at \SI{300}{\kelvin} computed for all three GBs are between that computed for the anion-ordered and the anion-disordered bulk structures. 

Our simulations show that all three GBs enhance the Li self-diffusion in Li$_6$PS$_5$Cl compared to the anion-ordered bulk, which is consistent with a previous study for Li$_6$PS$_5$Br~\cite{Sadowski2023}. The effect of GBs on Li diffusion depends on the characteristic atomic arrangements induced by the GBs. The $\Sigma3(1\bar{1}2)[110]$ GB breaks the Li-coordinated cages and thus enhances Li diffusion, which we refer to as the ``cage-opening effect''. A similar enhancement of Li diffusion was found due to the effect of anion disorder~\cite{Minafra2020,Morgan2021}. For $\Sigma3(\bar{1}11)[110]$, fewer structural changes (distortions) are seen in the GB regions, and correspondingly, less long-range Li diffusion is observed in the MD simulations compared to $\Sigma3(1\bar{1}2)[110]$. The amorphous-like structure formed in the GB $\Sigma5(001)[001]$ increases the Li diffusion. Our findings are in line with the Li-diffusion enhancement due to amorphization reported for another sulfide-type solid electrolyte, i.e., Li$_3$PS$_4$~\cite{Hayamizu2016,Jalem2023}. 

At temperatures below \SI{1000}{\kelvin}, the calculated GB diffusion coefficients fall within the interval set by the Arrhenius fits for the diffusion coefficients of the anion-ordered and anion-disordered bulk structures. Similar phenomena were also seen for other solid electrolytes: For Li$_6$PS$_5$Br the MSDs of Li atoms in the GB regions are in between those for the anion-ordered and the anion-disordered structures~\cite{Sadowski2023}, and for $\beta$-Li$_3$PS$_4$ GB diffusion coefficients are in between those for the bulk and the amorphous structures~\cite{Jalem2023}. We expect that Li diffusion coefficients of other (high-angle) GBs in Li$_6$PS$_5$Cl will also fall into the here-established Arrhenius interval. For the anion-disordered structures of argyrodites Li$_6$PS$_5X$ ($X$ $\in \{\text{Cl, Br, I}\}$), both experiments~\cite{Kraft2017, Zhou2021} and simulations~\cite{Morgan2021,Sadowski2023} indicate that maximum diffusivity is likely achieved at about \SI{50}{\percent} $X$/S-anion disorder. Therefore, we expect that the Li diffusion coefficient of the bulk structure with a different anion-disorder ratio may also fall into the obtained order-disorder interval. 

The Borisov relation~\cite{Borisov1964, Pelleg1966} and its derivatives~\cite{Page2021} are often applied to establish correlations between GB energy and diffusivity for metallic materials~\cite{Divinski2010,Prokoshkina2013,Page2021,Li2023}. For Li$_6$PS$_5$Cl, the \SI{0}{\kelvin} GB energy of $\Sigma3(1\bar{1}2)[110]$ is larger than that of $\Sigma3(\bar{1}11)[110]$. The diffusion coefficient of $\Sigma3(1\bar{1}2)[110]$ is also larger than that of $\Sigma3(\bar{1}11)[110]$ over a wide temperature range. These results indicate that the Borisov relation may be applied for Li$_6$PS$_5$Cl, but further research on more GBs is required to validate the hypothesis.

\begin{figure*}[tbp]
    \centering
    \includegraphics{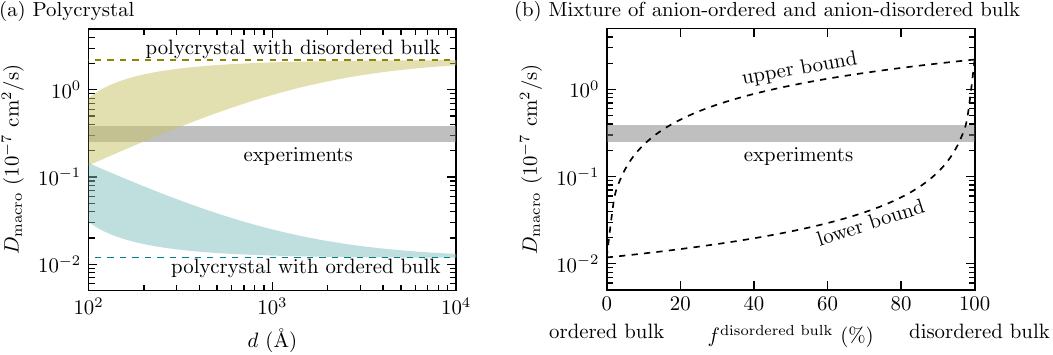}
    \caption{Extension of diffusion coefficients at \SI{300}{\kelvin} from atomistic (Fig.~\ref{fig:dis_coe} and Table~\ref{tab:diff_coe_fit}) to macro-scale based on the Wiener bounds~\cite{Karkkainen2000,Chen2006} [Eq.~\eqref{eq:upper} for the upper bound and Eq.~\eqref{eq:lower} for the lower bound]. (a) Grain size ($d$) effect for a polycrystal with anion-ordered or \SI{50}{\percent} anion-disordered bulk structures. The shaded areas are the intervals set by the Wiener bounds. (b) Effect of mixing the anion-ordered and the \SI{50}{\percent} anion-disordered bulk structures for a single crystal without GBs. The range covering diffusion coefficients measured in experiments at about \SI{300}{\kelvin}~\cite{Adeli2019,Schlenker2020} is shown for comparison. It was evidenced from experiments that the sample had an averaged grain size of less than \SI{1.5}{\micro\meter}~\cite{Schlenker2020}.}
    \label{fig:micro}
\end{figure*}

Since the available experimental diffusivity data for Li$_6$PS$_5$Cl correspond to polycrystalline samples~\cite{Adeli2019,Schlenker2020,Hanghofer2019,Deiseroth2011}, it is meaningful to analyze the diffusion coefficients of polycrystalline Li$_6$PS$_5$Cl. Figure~\ref{fig:micro} shows an extrapolation of the diffusion coefficients obtained from the atomistic simulations at \SI{300}{\kelvin} to the macro-scale based on the Wiener bounds~\cite{Karkkainen2000,Chen2006} (see Appendix~\ref{sec:micro}). The color-shaded regions in Fig.~\ref{fig:micro}(a) indicate the spread of the effective macroscopic diffusion coefficient $D_{\textrm{macro}}$ for a polycrystalline microstructure with GBs, assuming either the anion-ordered or anion-disordered bulk structure. Both color-shaded regions narrow down with increasing grain size, and the impact of GBs becomes nearly negligible above a grain size of \SI{1000}{\nano\meter}. This finding is consistent with experiments that reported a minor influence of GBs on Li transport~\cite{Schlenker2020, Hanghofer2019} for grain sizes of about \SI{1500}{\nano\meter}~\cite{Schlenker2020}.  To analyze a joint contribution of the anion-ordered and the anion-disordered bulk regions in a macroscopic sample, Fig.~\ref{fig:micro}(b) shows the Wiener bounds for a mixture of the anion-ordered and anion-disordered regions (without an additional GB impact). Compared with experimental data, a significant interval of possible $D_{\textrm{macro}}$ values is observed over a large volume fraction range of the two regions (about \SIrange{10}{95}{\percent}). To systematically examine the experimental and simulation results and to investigate the combined effect of different bulk structures and GBs, continuum simulations with explicit consideration of the microstructure are needed~\cite{Heo2021}, which is planned for a subsequent study. Additionally, the impact of other defects, e.g., dislocation~\cite{Faka2024} and micro-pores~\cite{Janek2023}, and strain~\cite{Zguns2022Jul} may also be considered. 

\section{Conclusions}\label{sec:conc}
Diffusion in solid electrolytes is a complex, multiscale phenomenon influenced by microstructure and chemistry, particularly grain boundaries (GBs) and anion disorder. This multiscale challenge pushes simulations to their limits as classical force fields lack accuracy and large-scale \textit{ab initio} methods are computationally out of reach.

We have tackled this challenge and developed a quality-level-based active learning scheme to efficiently and systematically train accurate machine-learning interatomic potentials for complex atomic structures. These potentials enable, for example, the acquisition of diffusivity data for GBs and anion-disordered structures in solid electrolytes through accelerated atomistic simulations with near \textit{ab initio} accuracy.

Utilizing the proposed scheme, we have investigated Li-ion diffusion for three structurally distinct GBs in Li$_6$PS$_5$Cl. These GBs exhibit low formation energies, indicating their high stability in polycrystalline Li$_6$PS$_5$Cl. The GBs enhance Li-ion diffusion compared to the anion-ordered bulk structure, with the degree of enhancement varying according to the specific GB structure. The underlying reason for the enhancement is traced back to the Li-cage opening effect in the GB region. GBs may, thus, generally affect the macroscopic diffusivity of polycrystalline Li$_6$PS$_5$Cl. Based on the present data, the limiting case for diffusion enhancement is the 50\% anion-disordered bulk. Experimental diffusion data for Li$_6$PS$_5$Cl around room temperature fall into the wide Arrhenius-extrapolated interval of diffusion coefficients for the investigated atomic structures.

The proposed scheme facilitates the computation of energies and diffusivities for various complex atomic structures in solid electrolytes. These data can, for example, be subsequently integrated into continuum simulations to model large-scale microstructures comprising multiple complex atomic structures. Consequently, high-precision simulations of macroscale diffusion in solid electrolytes are within reach, with potential applications in microstructure engineering. 

\begin{acknowledgments}
This project is funded by Deutsche Forschungsgemeinschaft (DFG, German Research Foundation) under Germany's Excellence Strategy -- EXC 2075 -- 390740016. We acknowledge the support by the Stuttgart Center for Simulation Science (SimTech), the state of Baden-Württemberg through bwHPC, and the German Research Foundation (DFG) through grant no.~INST 40/575-1 FUGG (JUSTUS 2 cluster) -- 405998092. Yuji Ikeda is funded by the Deutsche Forschungsgemeinschaft (DFG, German Research Foundation) -- \mbox{IK~125/1-1} -- 519607530. Felix Fritzen is funded by the Deutsche Forschungsgemeinschaft (DFG, German Research Foundation) in the Heisenberg program -- \mbox{FR~2702/10} -- 517847245. Sergiy Divinski thanks the Deutsche Forschungsgemeinschaft (DFG, German Research Foundation) for partial support via the grant \mbox{DI~1419/18-1}. Blazej Grabowski is funded by the European Research Council (ERC) under the European Union’s Horizon 2020 Research and Innovation Program (grant agreement no.~865855).
\end{acknowledgments}

\appendix
\section{Details on training and validation of MTP}\label{sec:al_scheme}

\begin{figure*}[tbp]
    \centering
    \includegraphics{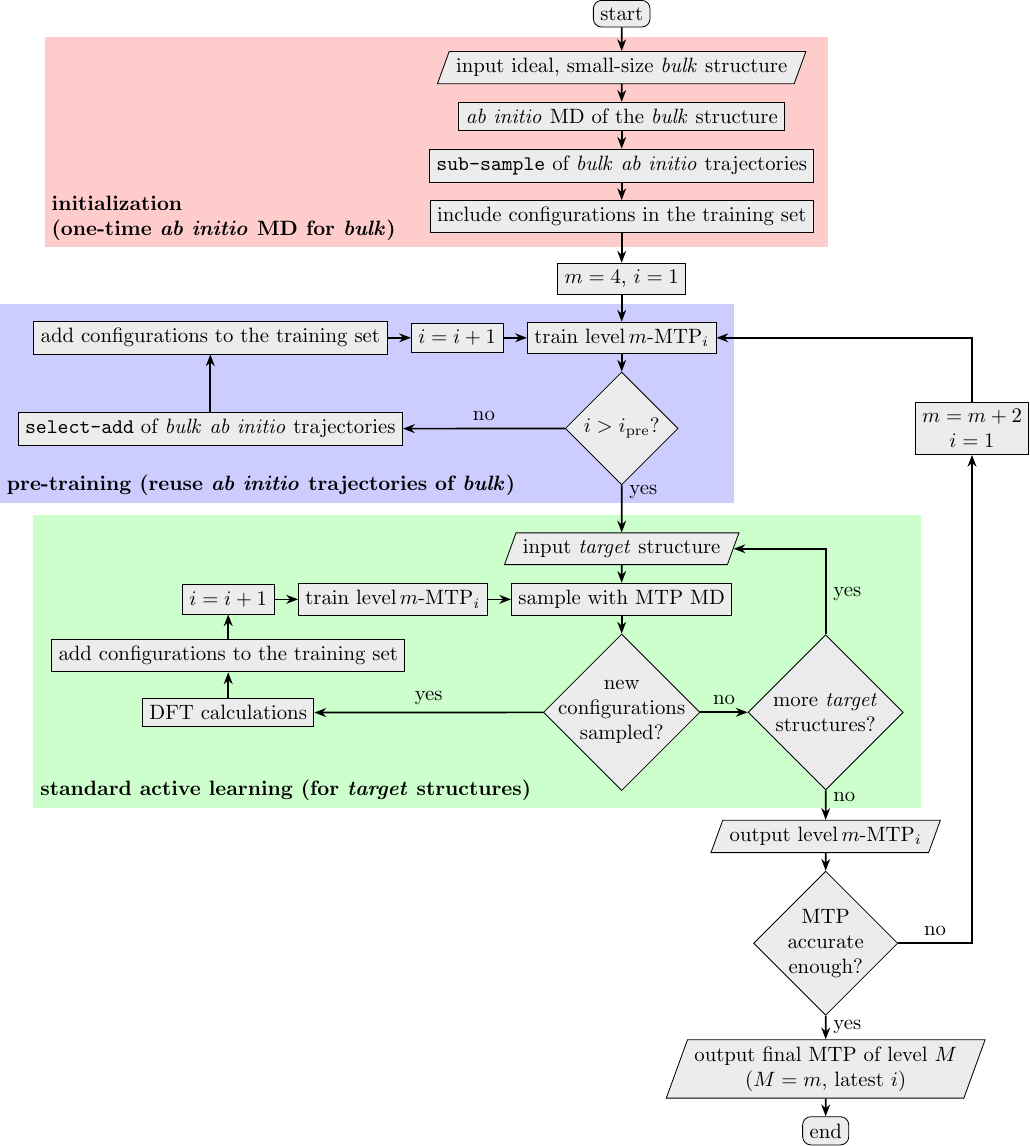}
    \caption{Detailed flowchart of the quality-level-based active learning scheme proposed and utilized in the present study. ``\texttt{Sub-sample}'' and ``\texttt{select-add}'' are commands in the \textsc{mlip} package~\cite{Novikov2020}. The MTP level~\cite{Novikov2020} is denoted by $m$, and $i$ enumerates the different, retrained MTPs of the same level. The ``pre-training'' and 
 ``standard active learning'' stages are performed at each MTP level, as shown in Fig.~\ref{fig:al_scheme}. The number of ``pre-training'' cycles is controlled by the hyperparameter $i_\textrm{pre}$. When the desired accuracy of the MTP is reached, the proposed scheme finishes. In the present study, we used the anion-ordered bulk structure as the input structure in the ``initialization'' and set $i_\textrm{pre}=1$. MTP levels up to 20 ($M=20$) were tested separately for different target structures in Li$_6$PS$_5$Cl (anion-ordered and anion-disordered bulk, and GBs). The large-scale production runs were done with level 18 MTPs to balance the accuracy against the computational costs (Fig.~\ref{fig:std}). }
    \label{fig:al_scheme_large}
\end{figure*}

A detailed flowchart of the proposed quality-level-based active learning scheme is shown in Fig.~\ref{fig:al_scheme_large}. This scheme consists of three core parts: Initialization, pre-training, and standard active learning. 

In the initialization, \textit{ab initio} MD simulations are performed for a small supercell of the ideal (ordered) bulk structure of the target material to provide the basis for the training set. To reduce the size of the training set (which enables faster MTP fitting at the initial stage), a small number of (several hundred) uncorrelated configurations are extracted (the so-called ``\texttt{sub-sample}'') from the obtained \textit{ab initio} MD trajectories. These configurations, together with the DFT-computed energies, forces, and stresses, are then included in the training set. Note that, in the proposed active learning scheme, \textit{ab initio} MD simulations are performed only during the initialization stage and that they are comparably efficient because of the utilized ideal bulk structure and the small supercell size. 

The labeled configurations obtained in the initialization are propagated to the pre-training stage. The key idea behind the pre-training stage is to pre-train the MTP on the ideal bulk structure before entering the standard active learning stage for the target structures. The ideal bulk structure already contains a lot of information about the atomic interactions, and thus, the following standard active learning cycle for the target structure is shortened (i.e., fewer DFT calculations and fewer standard active learning cycles are needed). The pre-training stage can be run once or more times (depending on $i_\mathrm{pre}$; we chose one) to select the most informative configurations (so-called ``\texttt{select-add}'' based on the D-optimality criterion~\cite{Podryabinkin2017,Gubaev2018}) from the initial \textit{ab initio} molecular dynamics runs. The selected configurations are added to the training set, and the MTP is retrained. When the pre-training stage is entered for the first time after initialization, an MTP of level 4 ($m=4$) is chosen for the fitting. In later stages, higher-level MTPs ($m=6, 8, 10,$~etc.) are chosen.

The pre-trained MTP and the accumulated training set (from initialization and pre-training) are propagated to the next stage, which is a standard active learning cycle carried out for the target structure (e.g., a structure with a grain boundary). The pre-trained MTP is used to perform MD simulations for the target structure, and configurations with large extrapolation grades are selected (so-called ``sample''). The energies, forces, and stresses of the sampled configurations are calculated by DFT and added to the training set. The MTP is then retrained using the training set enlarged with the sampled configurations. The standard active learning for one target structure is finished when no more configurations are sampled. In case the MTP is required to describe multiple target structures, e.g., different anion-disordered structures of Li$_6$PS$_5$Cl or different types of GBs, the standard active learning is repeated for each target structure with the MTP of the same level ($i$ increases while $m$ is unchanged). 

After running the standard active learning cycle for the first time, the trained level 4 MTP ($m=4$) is outputted, and the accuracy of the MTP is evaluated according to the fitting RMSEs in energies and forces. If higher accuracy is required, an untrained MTP of the next level (level 6, $m=6$) is utilized for another training round ($i$ reset to 1) with the pre-training and standard active learning stages. The new configurations generated during each stage are added to and accumulated in the training set. The MTP level in the pre-training and standard active learning can continue to increase in this way until the accuracy requirement of the MTP is satisfied. After that, the highest level of the utilized MTPs is denoted by $M$, and the whole active learning scheme finishes by outputting the finally-trained level~$M$-MTP. With this quality-level-based scheme, the configuration space of the target structures is explored efficiently and systematically. It is important to note that the computational cost of high-level MTPs increases rapidly, necessitating a balance between accuracy and efficiency (see Sec.~\ref{sec:mtp}). An automatic code to perform the proposed active learning scheme is available from the corresponding author upon reasonable request.

In the present study, conventional unit cells of the anion-ordered bulk structure of Li$_6$PS$_5$Cl [Fig.~\ref{fig:gb_con}(a)] in three volumes with atomic densities \qtylist{17.73; 20.68; 23.94}{\angstrom\textsuperscript{3}/atom} were used in the initialization stage. \textit{Ab initio} MD simulations were performed at \SI{1500}{\kelvin} for \SI{8}{\pico\second} with the Nosé--Hoover thermostat and the canonical ($NVT$) ensemble implemented in \textsc{vasp}. A time step of \SI{2}{\femto\second} and a Nosé mass of \SI{3}{u.\angstrom\textsuperscript{2}} were utilized. At each ionic step, the energy was converged to within \SI{e-3}{\electronvolt} per simulation cell. The MD trajectories were ``\texttt{sub-sampled}'' every \SI{1}{\pico\second}. A set of untrained MTPs configured at different levels~\cite{Novikov2020} and with 8 radial basis functions was utilized. The minimum and maximum cutoffs were set to \qtylist{1.5;5}{\angstrom}, respectively. The fitting weights for the energies, the atomic forces, and the stresses were set to 1, \SI{0.1}{\angstrom\textsuperscript{2}}, and \SI{0.001}{\angstrom\textsuperscript{6}}, respectively. In the standard active learning stage, ``sample'' was performed by MTP MD simulations with \textsc{lammps} at a temperature of \SI{1000}{\kelvin} and for a maximum of \SI{0.4}{\nano\second}. The \textit{NPT} ensemble with the same parameters as described in Sec.~\ref{sec:com} was used. The extrapolation grade values of the selecting and breaking thresholds were chosen to be 2 and 5, as suggested in Ref.~\cite{Gubaev2021}. 

Target structures for the anion-ordered and anion-disordered bulk, as well as the three GBs of Li$_6$PS$_5$Cl were constructed. A $1\times 1\times 2$ supercell was used for the ordered bulk structure. The disordered bulk structure was constructed based on a $1\times 2\times 2$ supercell of the ordered bulk structure (see Sec.~\ref{sec:dis_const}). Three different anion-disordered structures with random anion disorder were used to train MTPs of the anion-disordered bulk at each level. In total, 2210 configurations were accumulated in the training set up to level 18 (Table~\ref{tab:mtp_fit_validate}). GB target structures were constructed following Sec.~\ref{sec:gbcon} and with supercells made of the coincidence-site lattice unit cell. The validation sets were constructed separately for the anion-ordered and the anion-disordered bulk structures. For each target structure, the final MTP trained up to level 20 was used to perform an MD simulation at \SI{600}{\kelvin}. The validation set of each structure contained 500 configurations ``\texttt{sub-sampled}'' every \SI{1}{\pico\second} from the MD simulation. The energies, forces, and stresses of those configurations were calculated with DFT. Note that for the anion-disordered bulk structure, the anion disorder of the configurations in the validation set differed from those in the training set. 

\section{Conversion of self-diffusion coefficients to ionic conductivities}\label{sec:conductivity}

Based on the Nernst--Einstein equation~\cite{Uitz2017,Kizilyalli1999}, a relation between ionic conductivity, $\sigma$, and the self-diffusion coefficient, $D$, at temperature, $T$, can be written as,
\begin{equation}
    \sigma(T) = \Lambda\frac{D(T) e^2}{k_{\mathrm{B}}T},
    \label{eq:nernst}
\end{equation}
where $e$ is the elementary charge. The coefficient $\Lambda$ is related to the concentration of mobile ions, i.e., the Li-ions in the present case, and to the Haven ratio~\cite{Uitz2017,Kizilyalli1999}. Table~\ref{tab:exp_conduct} lists the experimentally measured values for $\sigma$ and $D$ at about \SI{300}{\kelvin} from different studies, and the resulting $\Lambda$'s [Eq.~\eqref{eq:nernst}].  As reflected by the strong scatter in the $\Lambda$ values from the different studies, the determination of $\Lambda$ is not trivial~\cite{Marcolongo2017,Adeli2019}. We utilized $\Lambda=\SI{21.808e-3}{Li/\angstrom\textsuperscript{3}}$ from the most recent experimental study~\cite{Schlenker2020} to convert the calculated self-diffusion coefficients to ionic conductivity (Table~\ref{tab:diff_coe_fit}). 

\begin{table}[tbp]
\centering
	\caption{Comparison of experimentally measured ionic conductivity $\sigma$ and self-diffusion coefficient $D$ at about \SI{300}{\kelvin}. The coefficient $\Lambda$ is calculated based on the measured values according to Eq.~\eqref{eq:nernst}. }
	\begin{ruledtabular}
	\begin{tabular}{ccS[table-format=1.4]S[table-format=1.3]S[table-format=2.3]c}
	Year & $T$  & {$\sigma$} & {$D$} & {$\Lambda$}  & Ref. \\
    & (\SI{}{K}) & {(\SI{}{mS/cm})} & {(\SI{e-7}{cm^2/s})} & {(\SI{e-3}{Li/\angstrom\textsuperscript{3}})}\\
		\hline
    2011 & 313 &  0.0013  &   0.77 &  0.003 & \cite{Deiseroth2011} \\
    2019 & 313 &  9 &  0.67 & 22.625 & \cite{Hanghofer2019} \\
    2019 & 298 &  2.5 & 0.387 & 10.359  & \cite{Adeli2019}\\
    2020 & 298 &  3.4  &  0.25 & 21.808 & \cite{Schlenker2020} \\
	\end{tabular}
	\end{ruledtabular}
	\label{tab:exp_conduct}
\end{table}

\section{Extension from atomistic to macro-scale}\label{sec:micro}

Models with different geometrical assumptions of grains have been used to obtain effective diffusion coefficients of polycrystals, e.g., the Maxwell--Garnet model~\cite{Garnett1997} and its modification~\cite{Kalnin2002}, the Belova--Murch~\cite{Belova2003} model, or the Chen--Schuh~\cite{Chen2007} model. In continuum mechanics, the Voigt--Reuss bounds~\cite{Voigt1889,Reuss1929,Hill1963} are commonly used to give an admissible range of effective properties rather than a geometry-dependent guess. They are equivalent to the Wiener bounds~\cite{Karkkainen2000,Chen2006} used for permittivity. In the present study, we used the Wiener bounds to obtain diffusion coefficients at the macro-scale ($D_{\textrm{macro}}$) from the atomistic diffusion coefficients (Table~\ref{tab:diff_coe_fit}) at an exemplary temperature of \SI{300}{\kelvin}.

For a polycrystal with a set of components $J$, e.g., the ordered bulk, the disordered bulk, and the GBs, etc., Chen et al.~\cite{Chen2006} have shown that, in the absence of segregation, if diffusivity ratios of any two components are less than \SI{e2}{} (low-contrast system), the Wiener bounds remain valid for evaluating the effective diffusion coefficients. Specifically, the upper and lower bounds of the diffusion coefficients of the polycrystal are given by the weighted arithmetic mean,
\begin{equation}
    D^{\textrm{upper}}_{\textrm{macro}} = \sum\limits_{j\in J} f^{j} D^{j},
    \label{eq:upper}
\end{equation}
and the weighted harmonic mean,
\begin{equation}
    D^{\textrm{lower}}_{\textrm{macro}} = \left(\sum\limits_{j\in J} \frac{f^{j}}{D^{j}}\right)^{-1},
    \label{eq:lower}
\end{equation}
respectively. Here, $f^{j}$ is the volume fraction of the structure $s$ in the polycrystal which satisfies,
\begin{equation}
    \sum\limits_{j\in J}f^{j} = 1. 
\end{equation}
Equations~\eqref{eq:upper} and~\eqref{eq:lower} are equivalent to the equations of the Hart~\cite{Hart1957,Belova2003} and the one-dimensional Maxwell--Garnet models~\cite{Garnett1997}, respectively. As was validated in Ref.~\cite{Yoon2023}, effective diffusion coefficients of a polycrystal calculated from the above-mentioned classical models fall within the Wiener bounds [Eqs.~\eqref{eq:upper} and~\eqref{eq:lower}]. 

Assuming for simplicity an isotropic distribution of GB surface orientations, the averaged GB diffusion coefficients can be estimated as an arithmetic mean of the diffusion coefficients at the atomistic scale in three dimensions (Fig.~\ref{fig:dis_coe} and Table~\ref{tab:diff_coe_fit}),
\begin{equation}
    D^{\textrm{GB}_k} \approx  \frac{2D_{\parallel}^{\mathrm{GB}_k} + D_{\perp}^{\mathrm{GB}_k}}{3},
\end{equation}
where $k$ indicates the GB type. With an average grain size of $d$, the volume fraction $f^{\mathrm{GB}_k}$ of the type-$k$ GB can be approximated~\cite{Thorvaldsen1997,Chen2007}, 
\begin{equation}
    f^{\textrm{GB}_k} \approx p^{\mathrm{GB}_k}H_\mathrm{GB}\frac{\delta^{\mathrm{GB}_k}}{d},
\end{equation}
where $\delta^{\textrm{GB}_k}$ is the width of the type-$k$ GB (Table~\ref{tab:gb_info}). Further, $H_{\textrm{GB}}$ is a dimensionless numerical factor accounting for the shape and size distribution of the grains in a polycrystal. Here, $H_{\textrm{GB}}=\SI{2.9105}{}$ was used, corresponding to the grain shape of Voronoi polyhedra and a nearly log-normal grain-size distribution~\cite{Thorvaldsen1997,Chen2007}. The formation probability of the type-$k$ GB is reflected by $p^{\mathrm{GB}_k}$. Assuming for simplicity all GB types to appear with equal probability gives, for every $k$, $p^{\mathrm{GB}_k}\approx 1/n^{\mathrm{GB}}$, where $n^{\mathrm{GB}}$ is the total number of GB types. The analysis has to be modified if Li ions are not equipartitionally distributed between different components, i.e., in the case of segregation~\cite{Zhang2012-partitioning}.

We note that the analysis above corresponds to the so-called A-type kinetic regime of GB diffusion~\cite{Harrison1961}, which holds if the diffusion length, $\sqrt{D_\textrm{macro} t}$, is larger than the characteristic size of all microstructure elements in the model~\cite{Paul2014}. In the absence of segregation, the ratio of $\sqrt{D_\textrm{macro} t}$ to the characteristic size of any microstructure element is typically required to exceed three~\cite{Paul2014}. Otherwise, one has to deal with diffusion in a complex microstructure with a hierarchy of the kinetic properties from different constituting elements (e.g., GBs of different types, dislocations, pores, etc., and their specific geometric arrangement), which also results in a hierarchy of diffusion regimes~\cite{Divinski2004a, DIVINSKI2004b}.


\begin{thebibliography}{108}%
\makeatletter
\providecommand \@ifxundefined [1]{%
 \@ifx{#1\undefined}
}%
\providecommand \@ifnum [1]{%
 \ifnum #1\expandafter \@firstoftwo
 \else \expandafter \@secondoftwo
 \fi
}%
\providecommand \@ifx [1]{%
 \ifx #1\expandafter \@firstoftwo
 \else \expandafter \@secondoftwo
 \fi
}%
\providecommand \natexlab [1]{#1}%
\providecommand \enquote  [1]{``#1''}%
\providecommand \bibnamefont  [1]{#1}%
\providecommand \bibfnamefont [1]{#1}%
\providecommand \citenamefont [1]{#1}%
\providecommand \href@noop [0]{\@secondoftwo}%
\providecommand \href [0]{\begingroup \@sanitize@url \@href}%
\providecommand \@href[1]{\@@startlink{#1}\@@href}%
\providecommand \@@href[1]{\endgroup#1\@@endlink}%
\providecommand \@sanitize@url [0]{\catcode `\\12\catcode `\$12\catcode
  `\&12\catcode `\#12\catcode `\^12\catcode `\_12\catcode `\%12\relax}%
\providecommand \@@startlink[1]{}%
\providecommand \@@endlink[0]{}%
\providecommand \url  [0]{\begingroup\@sanitize@url \@url }%
\providecommand \@url [1]{\endgroup\@href {#1}{\urlprefix }}%
\providecommand \urlprefix  [0]{URL }%
\providecommand \Eprint [0]{\href }%
\providecommand \doibase [0]{https://doi.org/}%
\providecommand \selectlanguage [0]{\@gobble}%
\providecommand \bibinfo  [0]{\@secondoftwo}%
\providecommand \bibfield  [0]{\@secondoftwo}%
\providecommand \translation [1]{[#1]}%
\providecommand \BibitemOpen [0]{}%
\providecommand \bibitemStop [0]{}%
\providecommand \bibitemNoStop [0]{.\EOS\space}%
\providecommand \EOS [0]{\spacefactor3000\relax}%
\providecommand \BibitemShut  [1]{\csname bibitem#1\endcsname}%
\let\auto@bib@innerbib\@empty
\bibitem [{\citenamefont {Famprikis}\ \emph {et~al.}(2019)\citenamefont
  {Famprikis}, \citenamefont {Canepa}, \citenamefont {Dawson}, \citenamefont
  {Islam},\ and\ \citenamefont {Masquelier}}]{Famprikis2019}%
  \BibitemOpen
  \bibfield  {author} {\bibinfo {author} {\bibfnamefont {T.}~\bibnamefont
  {Famprikis}}, \bibinfo {author} {\bibfnamefont {P.}~\bibnamefont {Canepa}},
  \bibinfo {author} {\bibfnamefont {J.~A.}\ \bibnamefont {Dawson}}, \bibinfo
  {author} {\bibfnamefont {M.~S.}\ \bibnamefont {Islam}},\ and\ \bibinfo
  {author} {\bibfnamefont {C.}~\bibnamefont {Masquelier}},\ }\bibfield  {title}
  {\bibinfo {title} {Fundamentals of inorganic solid-state electrolytes for
  batteries},\ }\href {https://doi.org/10.1038/s41563-019-0431-3} {\bibfield
  {journal} {\bibinfo  {journal} {Nat. Mater.}\ }\textbf {\bibinfo {volume}
  {18}},\ \bibinfo {pages} {1278} (\bibinfo {year} {2019})}\BibitemShut
  {NoStop}%
\bibitem [{\citenamefont {Zhao}\ \emph {et~al.}(2020)\citenamefont {Zhao},
  \citenamefont {Stalin}, \citenamefont {Zhao},\ and\ \citenamefont
  {Archer}}]{Zhao2020}%
  \BibitemOpen
  \bibfield  {author} {\bibinfo {author} {\bibfnamefont {Q.}~\bibnamefont
  {Zhao}}, \bibinfo {author} {\bibfnamefont {S.}~\bibnamefont {Stalin}},
  \bibinfo {author} {\bibfnamefont {C.-Z.}\ \bibnamefont {Zhao}},\ and\
  \bibinfo {author} {\bibfnamefont {L.~A.}\ \bibnamefont {Archer}},\ }\bibfield
   {title} {\bibinfo {title} {Designing solid-state electrolytes for safe,
  energy-dense batteries},\ }\href {https://doi.org/10.1038/s41578-019-0165-5}
  {\bibfield  {journal} {\bibinfo  {journal} {Nat. Rev. Mater.}\ }\textbf
  {\bibinfo {volume} {5}},\ \bibinfo {pages} {229} (\bibinfo {year}
  {2020})}\BibitemShut {NoStop}%
\bibitem [{\citenamefont {Deiseroth}\ \emph {et~al.}(2008)\citenamefont
  {Deiseroth}, \citenamefont {Kong}, \citenamefont {Eckert}, \citenamefont
  {Vannahme}, \citenamefont {Reiner}, \citenamefont {Zaiß},\ and\
  \citenamefont {Schlosser}}]{Deiseroth2008}%
  \BibitemOpen
  \bibfield  {author} {\bibinfo {author} {\bibfnamefont {H.-J.}\ \bibnamefont
  {Deiseroth}}, \bibinfo {author} {\bibfnamefont {S.-T.}\ \bibnamefont {Kong}},
  \bibinfo {author} {\bibfnamefont {H.}~\bibnamefont {Eckert}}, \bibinfo
  {author} {\bibfnamefont {J.}~\bibnamefont {Vannahme}}, \bibinfo {author}
  {\bibfnamefont {C.}~\bibnamefont {Reiner}}, \bibinfo {author} {\bibfnamefont
  {T.}~\bibnamefont {Zaiß}},\ and\ \bibinfo {author} {\bibfnamefont
  {M.}~\bibnamefont {Schlosser}},\ }\bibfield  {title} {\bibinfo {title}
  {{Li$_6$PS$_5X$}: {A} class of crystalline {Li}-rich solids with an unusually
  high {Li}$^{+}$ mobility},\ }\href {https://doi.org/10.1002/anie.200703900}
  {\bibfield  {journal} {\bibinfo  {journal} {Angew. Chem. Int. Ed.}\ }\textbf
  {\bibinfo {volume} {47}},\ \bibinfo {pages} {755} (\bibinfo {year}
  {2008})}\BibitemShut {NoStop}%
\bibitem [{\citenamefont {Boulineau}\ \emph {et~al.}(2013)\citenamefont
  {Boulineau}, \citenamefont {Tarascon}, \citenamefont {Leriche},\ and\
  \citenamefont {Viallet}}]{Boulineau2013}%
  \BibitemOpen
  \bibfield  {author} {\bibinfo {author} {\bibfnamefont {S.}~\bibnamefont
  {Boulineau}}, \bibinfo {author} {\bibfnamefont {J.-M.}\ \bibnamefont
  {Tarascon}}, \bibinfo {author} {\bibfnamefont {J.-B.}\ \bibnamefont
  {Leriche}},\ and\ \bibinfo {author} {\bibfnamefont {V.}~\bibnamefont
  {Viallet}},\ }\bibfield  {title} {\bibinfo {title} {Electrochemical
  properties of all-solid-state lithium secondary batteries using
  {Li}-argyrodite {Li$_6$PS$_5$Cl} as solid electrolyte},\ }\href
  {https://doi.org/10.1016/j.ssi.2013.04.012} {\bibfield  {journal} {\bibinfo
  {journal} {Solid State Ion.}\ }\textbf {\bibinfo {volume} {242}},\ \bibinfo
  {pages} {45} (\bibinfo {year} {2013})}\BibitemShut {NoStop}%
\bibitem [{\citenamefont {Kraft}\ \emph {et~al.}(2017)\citenamefont {Kraft},
  \citenamefont {Culver}, \citenamefont {Calderon}, \citenamefont {Böcher},
  \citenamefont {Krauskopf}, \citenamefont {Senyshyn}, \citenamefont
  {Dietrich}, \citenamefont {Zevalkink}, \citenamefont {Janek},\ and\
  \citenamefont {Zeier}}]{Kraft2017}%
  \BibitemOpen
  \bibfield  {author} {\bibinfo {author} {\bibfnamefont {M.~A.}\ \bibnamefont
  {Kraft}}, \bibinfo {author} {\bibfnamefont {S.~P.}\ \bibnamefont {Culver}},
  \bibinfo {author} {\bibfnamefont {M.}~\bibnamefont {Calderon}}, \bibinfo
  {author} {\bibfnamefont {F.}~\bibnamefont {Böcher}}, \bibinfo {author}
  {\bibfnamefont {T.}~\bibnamefont {Krauskopf}}, \bibinfo {author}
  {\bibfnamefont {A.}~\bibnamefont {Senyshyn}}, \bibinfo {author}
  {\bibfnamefont {C.}~\bibnamefont {Dietrich}}, \bibinfo {author}
  {\bibfnamefont {A.}~\bibnamefont {Zevalkink}}, \bibinfo {author}
  {\bibfnamefont {J.}~\bibnamefont {Janek}},\ and\ \bibinfo {author}
  {\bibfnamefont {W.~G.}\ \bibnamefont {Zeier}},\ }\bibfield  {title} {\bibinfo
  {title} {Influence of lattice polarizability on the ionic conductivity in the
  lithium superionic argyrodites {Li$_6$PS$_5X$} ({$X$} = {Cl}, {Br}, {I})},\
  }\href {https://doi.org/10.1021/jacs.7b06327} {\bibfield  {journal} {\bibinfo
   {journal} {J. Am. Chem. Soc.}\ }\textbf {\bibinfo {volume} {139}},\ \bibinfo
  {pages} {10909} (\bibinfo {year} {2017})}\BibitemShut {NoStop}%
\bibitem [{\citenamefont {Ruhl}\ \emph {et~al.}(2021)\citenamefont {Ruhl},
  \citenamefont {Riegger}, \citenamefont {Ghidiu},\ and\ \citenamefont
  {Zeier}}]{Ruhl2021}%
  \BibitemOpen
  \bibfield  {author} {\bibinfo {author} {\bibfnamefont {J.}~\bibnamefont
  {Ruhl}}, \bibinfo {author} {\bibfnamefont {L.~M.}\ \bibnamefont {Riegger}},
  \bibinfo {author} {\bibfnamefont {M.}~\bibnamefont {Ghidiu}},\ and\ \bibinfo
  {author} {\bibfnamefont {W.~G.}\ \bibnamefont {Zeier}},\ }\bibfield  {title}
  {\bibinfo {title} {Impact of solvent treatment of the superionic argyrodite
  {Li$_6$PS$_5$Cl} on solid-state battery performance},\ }\href
  {https://doi.org/10.1002/aesr.202000077} {\bibfield  {journal} {\bibinfo
  {journal} {Adv. Energy Sustain. Res.}\ }\textbf {\bibinfo {volume} {2}},\
  \bibinfo {pages} {2000077} (\bibinfo {year} {2021})}\BibitemShut {NoStop}%
\bibitem [{\citenamefont {Adeli}\ \emph {et~al.}(2019)\citenamefont {Adeli},
  \citenamefont {Bazak}, \citenamefont {Park}, \citenamefont {Kochetkov},
  \citenamefont {Huq}, \citenamefont {Goward},\ and\ \citenamefont
  {Nazar}}]{Adeli2019}%
  \BibitemOpen
  \bibfield  {author} {\bibinfo {author} {\bibfnamefont {P.}~\bibnamefont
  {Adeli}}, \bibinfo {author} {\bibfnamefont {J.~D.}\ \bibnamefont {Bazak}},
  \bibinfo {author} {\bibfnamefont {K.~H.}\ \bibnamefont {Park}}, \bibinfo
  {author} {\bibfnamefont {I.}~\bibnamefont {Kochetkov}}, \bibinfo {author}
  {\bibfnamefont {A.}~\bibnamefont {Huq}}, \bibinfo {author} {\bibfnamefont
  {G.~R.}\ \bibnamefont {Goward}},\ and\ \bibinfo {author} {\bibfnamefont
  {L.~F.}\ \bibnamefont {Nazar}},\ }\bibfield  {title} {\bibinfo {title}
  {Boosting solid-state diffusivity and conductivity in lithium superionic
  argyrodites by halide substitution},\ }\href
  {https://doi.org/10.1002/anie.201814222} {\bibfield  {journal} {\bibinfo
  {journal} {Angew. Chem. Int. Ed.}\ }\textbf {\bibinfo {volume} {58}},\
  \bibinfo {pages} {8681} (\bibinfo {year} {2019})}\BibitemShut {NoStop}%
\bibitem [{\citenamefont {Schlenker}\ \emph {et~al.}(2020)\citenamefont
  {Schlenker}, \citenamefont {Hansen}, \citenamefont {Senyshyn}, \citenamefont
  {Zinkevich}, \citenamefont {Knapp}, \citenamefont {Hupfer}, \citenamefont
  {Ehrenberg},\ and\ \citenamefont {Indris}}]{Schlenker2020}%
  \BibitemOpen
  \bibfield  {author} {\bibinfo {author} {\bibfnamefont {R.}~\bibnamefont
  {Schlenker}}, \bibinfo {author} {\bibfnamefont {A.-L.}\ \bibnamefont
  {Hansen}}, \bibinfo {author} {\bibfnamefont {A.}~\bibnamefont {Senyshyn}},
  \bibinfo {author} {\bibfnamefont {T.}~\bibnamefont {Zinkevich}}, \bibinfo
  {author} {\bibfnamefont {M.}~\bibnamefont {Knapp}}, \bibinfo {author}
  {\bibfnamefont {T.}~\bibnamefont {Hupfer}}, \bibinfo {author} {\bibfnamefont
  {H.}~\bibnamefont {Ehrenberg}},\ and\ \bibinfo {author} {\bibfnamefont
  {S.}~\bibnamefont {Indris}},\ }\bibfield  {title} {\bibinfo {title}
  {Structure and diffusion pathways in {Li$_6$PS$_5$Cl} argyrodite from neutron
  diffraction, pair-distribution function analysis, and {NMR}},\ }\href
  {https://doi.org/10.1021/acs.chemmater.0c02418} {\bibfield  {journal}
  {\bibinfo  {journal} {Chem. Mater.}\ }\textbf {\bibinfo {volume} {32}},\
  \bibinfo {pages} {8420} (\bibinfo {year} {2020})}\BibitemShut {NoStop}%
\bibitem [{\citenamefont {Minafra}\ \emph {et~al.}(2020)\citenamefont
  {Minafra}, \citenamefont {Kraft}, \citenamefont {Bernges}, \citenamefont
  {Li}, \citenamefont {Schlem}, \citenamefont {Morgan},\ and\ \citenamefont
  {Zeier}}]{Minafra2020}%
  \BibitemOpen
  \bibfield  {author} {\bibinfo {author} {\bibfnamefont {N.}~\bibnamefont
  {Minafra}}, \bibinfo {author} {\bibfnamefont {M.~A.}\ \bibnamefont {Kraft}},
  \bibinfo {author} {\bibfnamefont {T.}~\bibnamefont {Bernges}}, \bibinfo
  {author} {\bibfnamefont {C.}~\bibnamefont {Li}}, \bibinfo {author}
  {\bibfnamefont {R.}~\bibnamefont {Schlem}}, \bibinfo {author} {\bibfnamefont
  {B.~J.}\ \bibnamefont {Morgan}},\ and\ \bibinfo {author} {\bibfnamefont
  {W.~G.}\ \bibnamefont {Zeier}},\ }\bibfield  {title} {\bibinfo {title} {Local
  charge inhomogeneity and lithium distribution in the superionic argyrodites
  {Li$_6$PS$_5X$} ({$X$} = {Cl}, {Br}, {I})},\ }\href
  {https://doi.org/10.1021/acs.inorgchem.0c01504} {\bibfield  {journal}
  {\bibinfo  {journal} {Inorg. Chem.}\ }\textbf {\bibinfo {volume} {59}},\
  \bibinfo {pages} {11009} (\bibinfo {year} {2020})}\BibitemShut {NoStop}%
\bibitem [{\citenamefont {Ganapathy}\ \emph {et~al.}(2019)\citenamefont
  {Ganapathy}, \citenamefont {Yu}, \citenamefont {van Eck},\ and\ \citenamefont
  {Wagemaker}}]{Ganapathy2019}%
  \BibitemOpen
  \bibfield  {author} {\bibinfo {author} {\bibfnamefont {S.}~\bibnamefont
  {Ganapathy}}, \bibinfo {author} {\bibfnamefont {C.}~\bibnamefont {Yu}},
  \bibinfo {author} {\bibfnamefont {E.~R.~H.}\ \bibnamefont {van Eck}},\ and\
  \bibinfo {author} {\bibfnamefont {M.}~\bibnamefont {Wagemaker}},\ }\bibfield
  {title} {\bibinfo {title} {Peeking across grain boundaries in a solid-state
  ionic conductor},\ }\href {https://doi.org/10.1021/acsenergylett.9b00610}
  {\bibfield  {journal} {\bibinfo  {journal} {ACS Energy Lett.}\ }\textbf
  {\bibinfo {volume} {4}},\ \bibinfo {pages} {1092} (\bibinfo {year}
  {2019})}\BibitemShut {NoStop}%
\bibitem [{\citenamefont {Milan}\ and\ \citenamefont
  {Pasta}(2022)}]{Milan2023}%
  \BibitemOpen
  \bibfield  {author} {\bibinfo {author} {\bibfnamefont {E.}~\bibnamefont
  {Milan}}\ and\ \bibinfo {author} {\bibfnamefont {M.}~\bibnamefont {Pasta}},\
  }\bibfield  {title} {\bibinfo {title} {The role of grain boundaries in
  solid-state {Li}-metal batteries},\ }\href
  {https://doi.org/10.1088/2752-5724/aca703} {\bibfield  {journal} {\bibinfo
  {journal} {Mater. Futures}\ }\textbf {\bibinfo {volume} {2}},\ \bibinfo
  {pages} {013501} (\bibinfo {year} {2022})}\BibitemShut {NoStop}%
\bibitem [{\citenamefont {Deng}\ \emph {et~al.}(2017)\citenamefont {Deng},
  \citenamefont {Zhu}, \citenamefont {Chu},\ and\ \citenamefont
  {Ong}}]{Deng2017}%
  \BibitemOpen
  \bibfield  {author} {\bibinfo {author} {\bibfnamefont {Z.}~\bibnamefont
  {Deng}}, \bibinfo {author} {\bibfnamefont {Z.}~\bibnamefont {Zhu}}, \bibinfo
  {author} {\bibfnamefont {I.-H.}\ \bibnamefont {Chu}},\ and\ \bibinfo {author}
  {\bibfnamefont {S.~P.}\ \bibnamefont {Ong}},\ }\bibfield  {title} {\bibinfo
  {title} {Data-driven first-principles methods for the study and design of
  alkali superionic conductors},\ }\href
  {https://doi.org/10.1021/acs.chemmater.6b02648} {\bibfield  {journal}
  {\bibinfo  {journal} {Chem. Mater.}\ }\textbf {\bibinfo {volume} {29}},\
  \bibinfo {pages} {281} (\bibinfo {year} {2017})}\BibitemShut {NoStop}%
\bibitem [{\citenamefont {Stamminger}\ \emph {et~al.}(2019)\citenamefont
  {Stamminger}, \citenamefont {Ziebarth}, \citenamefont {Mrovec}, \citenamefont
  {Hammerschmidt},\ and\ \citenamefont {Drautz}}]{Stamminger2019}%
  \BibitemOpen
  \bibfield  {author} {\bibinfo {author} {\bibfnamefont {A.~R.}\ \bibnamefont
  {Stamminger}}, \bibinfo {author} {\bibfnamefont {B.}~\bibnamefont
  {Ziebarth}}, \bibinfo {author} {\bibfnamefont {M.}~\bibnamefont {Mrovec}},
  \bibinfo {author} {\bibfnamefont {T.}~\bibnamefont {Hammerschmidt}},\ and\
  \bibinfo {author} {\bibfnamefont {R.}~\bibnamefont {Drautz}},\ }\bibfield
  {title} {\bibinfo {title} {Ionic conductivity and its dependence on
  structural disorder in halogenated argyrodites {Li$_6$PS$_5X$} ({$X$} = {Br},
  {Cl}, {I})},\ }\href {https://doi.org/10.1021/acs.chemmater.9b02047}
  {\bibfield  {journal} {\bibinfo  {journal} {Chem. Mater.}\ }\textbf {\bibinfo
  {volume} {31}},\ \bibinfo {pages} {8673} (\bibinfo {year}
  {2019})}\BibitemShut {NoStop}%
\bibitem [{\citenamefont {Jiang}\ \emph {et~al.}(2022)\citenamefont {Jiang},
  \citenamefont {Chen}, \citenamefont {Rao}, \citenamefont {Chen},
  \citenamefont {Zu},\ and\ \citenamefont {Singh}}]{Jiang2022}%
  \BibitemOpen
  \bibfield  {author} {\bibinfo {author} {\bibfnamefont {M.}~\bibnamefont
  {Jiang}}, \bibinfo {author} {\bibfnamefont {Z.-W.}\ \bibnamefont {Chen}},
  \bibinfo {author} {\bibfnamefont {A.}~\bibnamefont {Rao}}, \bibinfo {author}
  {\bibfnamefont {L.-X.}\ \bibnamefont {Chen}}, \bibinfo {author}
  {\bibfnamefont {X.-T.}\ \bibnamefont {Zu}},\ and\ \bibinfo {author}
  {\bibfnamefont {C.~V.}\ \bibnamefont {Singh}},\ }\bibfield  {title} {\bibinfo
  {title} {Se-doped {Li$_6$PS$_5$Cl} and {Li$_{5.5}$PS$_{4.5}$Cl$_{1.5}$} with
  improved ionic conductivity and interfacial compatibility: {A}
  high-throughput {DFT} study},\ }\href {https://doi.org/10.1039/D2TC03696G}
  {\bibfield  {journal} {\bibinfo  {journal} {J. Mater. Chem. C}\ }\textbf
  {\bibinfo {volume} {10}},\ \bibinfo {pages} {18294} (\bibinfo {year}
  {2022})}\BibitemShut {NoStop}%
\bibitem [{\citenamefont {Jeon}\ \emph {et~al.}(2024)\citenamefont {Jeon},
  \citenamefont {Cha},\ and\ \citenamefont {Jung}}]{Jeon2024}%
  \BibitemOpen
  \bibfield  {author} {\bibinfo {author} {\bibfnamefont {T.}~\bibnamefont
  {Jeon}}, \bibinfo {author} {\bibfnamefont {G.~H.}\ \bibnamefont {Cha}},\ and\
  \bibinfo {author} {\bibfnamefont {S.~C.}\ \bibnamefont {Jung}},\ }\bibfield
  {title} {\bibinfo {title} {Understanding the anion disorder governing lithium
  distribution and diffusion in an argyrodite {Li$_6$PS$_5$Cl} solid
  electrolyte},\ }\href {https://doi.org/10.1039/D3TA06069A} {\bibfield
  {journal} {\bibinfo  {journal} {J. Mater. Chem. A}\ }\textbf {\bibinfo
  {volume} {12}},\ \bibinfo {pages} {993} (\bibinfo {year} {2024})}\BibitemShut
  {NoStop}%
\bibitem [{\citenamefont {de~Klerk}\ \emph {et~al.}(2016)\citenamefont
  {de~Klerk}, \citenamefont {Rosłoń},\ and\ \citenamefont
  {Wagemaker}}]{Klerk2016}%
  \BibitemOpen
  \bibfield  {author} {\bibinfo {author} {\bibfnamefont {N.~J.~J.}\
  \bibnamefont {de~Klerk}}, \bibinfo {author} {\bibfnamefont {I.}~\bibnamefont
  {Rosłoń}},\ and\ \bibinfo {author} {\bibfnamefont {M.}~\bibnamefont
  {Wagemaker}},\ }\bibfield  {title} {\bibinfo {title} {Diffusion mechanism of
  {Li} argyrodite solid electrolytes for {Li}-ion batteries and prediction of
  optimized halogen doping: {The} effect of {Li} vacancies, halogens, and
  halogen disorder},\ }\href {https://doi.org/10.1021/acs.chemmater.6b03630}
  {\bibfield  {journal} {\bibinfo  {journal} {Chem. Mater.}\ }\textbf {\bibinfo
  {volume} {28}},\ \bibinfo {pages} {7955} (\bibinfo {year}
  {2016})}\BibitemShut {NoStop}%
\bibitem [{\citenamefont {Morgan}(2021)}]{Morgan2021}%
  \BibitemOpen
  \bibfield  {author} {\bibinfo {author} {\bibfnamefont {B.~J.}\ \bibnamefont
  {Morgan}},\ }\bibfield  {title} {\bibinfo {title} {Mechanistic origin of
  superionic lithium diffusion in anion-disordered {Li$_6$PS$_5X$}
  argyrodites},\ }\href {https://doi.org/10.1021/acs.chemmater.0c03738}
  {\bibfield  {journal} {\bibinfo  {journal} {Chem. Mater.}\ }\textbf {\bibinfo
  {volume} {33}},\ \bibinfo {pages} {2004} (\bibinfo {year}
  {2021})}\BibitemShut {NoStop}%
\bibitem [{\citenamefont {Lee}\ \emph {et~al.}(2022)\citenamefont {Lee},
  \citenamefont {Lee}, \citenamefont {Park}, \citenamefont {Cho}, \citenamefont
  {Park},\ and\ \citenamefont {Sohn}}]{Lee2022}%
  \BibitemOpen
  \bibfield  {author} {\bibinfo {author} {\bibfnamefont {B.~D.}\ \bibnamefont
  {Lee}}, \bibinfo {author} {\bibfnamefont {J.-W.}\ \bibnamefont {Lee}},
  \bibinfo {author} {\bibfnamefont {J.}~\bibnamefont {Park}}, \bibinfo {author}
  {\bibfnamefont {M.~Y.}\ \bibnamefont {Cho}}, \bibinfo {author} {\bibfnamefont
  {W.~B.}\ \bibnamefont {Park}},\ and\ \bibinfo {author} {\bibfnamefont
  {K.-S.}\ \bibnamefont {Sohn}},\ }\bibfield  {title} {\bibinfo {title}
  {Argyrodite configuration determination for {DFT} and {AIMD} calculations
  using an integrated optimization strategy},\ }\href
  {https://doi.org/10.1039/D2RA05889H} {\bibfield  {journal} {\bibinfo
  {journal} {RSC Adv.}\ }\textbf {\bibinfo {volume} {12}},\ \bibinfo {pages}
  {31156} (\bibinfo {year} {2022})}\BibitemShut {NoStop}%
\bibitem [{\citenamefont {Sadowski}(2023)}]{Sadowski2023}%
  \BibitemOpen
  \bibfield  {author} {\bibinfo {author} {\bibfnamefont {M.}~\bibnamefont
  {Sadowski}},\ }\emph {\bibinfo {title} {Properties of sulfide solid
  electrolytes studied by electronic structure calculations}},\ \href
  {https://doi.org/10.26083/tuprints-00023752} {Ph.D. thesis},\ \bibinfo
  {school} {Technische Universität Darmstadt}, \bibinfo {address} {Darmstadt}
  (\bibinfo {year} {2023})\BibitemShut {NoStop}%
\bibitem [{\citenamefont {He}\ \emph {et~al.}(2018)\citenamefont {He},
  \citenamefont {Zhu}, \citenamefont {Epstein},\ and\ \citenamefont
  {Mo}}]{He2018}%
  \BibitemOpen
  \bibfield  {author} {\bibinfo {author} {\bibfnamefont {X.}~\bibnamefont
  {He}}, \bibinfo {author} {\bibfnamefont {Y.}~\bibnamefont {Zhu}}, \bibinfo
  {author} {\bibfnamefont {A.}~\bibnamefont {Epstein}},\ and\ \bibinfo {author}
  {\bibfnamefont {Y.}~\bibnamefont {Mo}},\ }\bibfield  {title} {\bibinfo
  {title} {Statistical variances of diffusional properties from \textit{ab
  initio} molecular dynamics simulations},\ }\href
  {https://doi.org/10.1038/s41524-018-0074-y} {\bibfield  {journal} {\bibinfo
  {journal} {Npj Comput. Mater.}\ }\textbf {\bibinfo {volume} {4}},\ \bibinfo
  {pages} {1} (\bibinfo {year} {2018})}\BibitemShut {NoStop}%
\bibitem [{\citenamefont {Das}\ \emph {et~al.}(2022)\citenamefont {Das},
  \citenamefont {Merinov}, \citenamefont {Yang},\ and\ \citenamefont
  {Iii}}]{Das2022}%
  \BibitemOpen
  \bibfield  {author} {\bibinfo {author} {\bibfnamefont {T.}~\bibnamefont
  {Das}}, \bibinfo {author} {\bibfnamefont {B.~V.}\ \bibnamefont {Merinov}},
  \bibinfo {author} {\bibfnamefont {M.~Y.}\ \bibnamefont {Yang}},\ and\
  \bibinfo {author} {\bibfnamefont {W.~A.~G.}\ \bibnamefont {Iii}},\ }\bibfield
   {title} {\bibinfo {title} {Structural, dynamic, and diffusion properties of
  a {Li$_{6}$}({PS$_4$}){SCl} superionic conductor from molecular dynamics
  simulations; prediction of a dramatically improved conductor},\ }\href
  {https://doi.org/10.1039/D2TA02715A} {\bibfield  {journal} {\bibinfo
  {journal} {J. Mater. Chem. A}\ }\textbf {\bibinfo {volume} {10}},\ \bibinfo
  {pages} {16319} (\bibinfo {year} {2022})}\BibitemShut {NoStop}%
\bibitem [{\citenamefont {Stegmaier}\ \emph {et~al.}(2021)\citenamefont
  {Stegmaier}, \citenamefont {Schierholz}, \citenamefont {Povstugar},
  \citenamefont {Barthel}, \citenamefont {Rittmeyer}, \citenamefont {Yu},
  \citenamefont {Wengert}, \citenamefont {Rostami}, \citenamefont {Kungl},
  \citenamefont {Reuter}, \citenamefont {Eichel},\ and\ \citenamefont
  {Scheurer}}]{Stegmaier2021}%
  \BibitemOpen
  \bibfield  {author} {\bibinfo {author} {\bibfnamefont {S.}~\bibnamefont
  {Stegmaier}}, \bibinfo {author} {\bibfnamefont {R.}~\bibnamefont
  {Schierholz}}, \bibinfo {author} {\bibfnamefont {I.}~\bibnamefont
  {Povstugar}}, \bibinfo {author} {\bibfnamefont {J.}~\bibnamefont {Barthel}},
  \bibinfo {author} {\bibfnamefont {S.~P.}\ \bibnamefont {Rittmeyer}}, \bibinfo
  {author} {\bibfnamefont {S.}~\bibnamefont {Yu}}, \bibinfo {author}
  {\bibfnamefont {S.}~\bibnamefont {Wengert}}, \bibinfo {author} {\bibfnamefont
  {S.}~\bibnamefont {Rostami}}, \bibinfo {author} {\bibfnamefont
  {H.}~\bibnamefont {Kungl}}, \bibinfo {author} {\bibfnamefont
  {K.}~\bibnamefont {Reuter}}, \bibinfo {author} {\bibfnamefont {R.-A.}\
  \bibnamefont {Eichel}},\ and\ \bibinfo {author} {\bibfnamefont
  {C.}~\bibnamefont {Scheurer}},\ }\bibfield  {title} {\bibinfo {title}
  {Nano-scale complexions facilitate {Li} dendrite-free operation in {LATP}
  solid-state electrolyte},\ }\href {https://doi.org/10.1002/aenm.202100707}
  {\bibfield  {journal} {\bibinfo  {journal} {Adv. Energy Mater.}\ }\textbf
  {\bibinfo {volume} {11}},\ \bibinfo {pages} {2100707} (\bibinfo {year}
  {2021})}\BibitemShut {NoStop}%
\bibitem [{\citenamefont {Stegmaier}\ \emph {et~al.}(2022)\citenamefont
  {Stegmaier}, \citenamefont {Reuter},\ and\ \citenamefont
  {Scheurer}}]{Stegmaier2022}%
  \BibitemOpen
  \bibfield  {author} {\bibinfo {author} {\bibfnamefont {S.}~\bibnamefont
  {Stegmaier}}, \bibinfo {author} {\bibfnamefont {K.}~\bibnamefont {Reuter}},\
  and\ \bibinfo {author} {\bibfnamefont {C.}~\bibnamefont {Scheurer}},\
  }\bibfield  {title} {\bibinfo {title} {Exploiting nanoscale complexion in
  {LATP} solid-state electrolyte via interfacial {Mg$^{2+}$} doping},\ }\href
  {https://doi.org/10.3390/nano12172912} {\bibfield  {journal} {\bibinfo
  {journal} {Nanomater.}\ }\textbf {\bibinfo {volume} {12}},\ \bibinfo {pages}
  {2912} (\bibinfo {year} {2022})}\BibitemShut {NoStop}%
\bibitem [{\citenamefont {Gubaev}\ \emph {et~al.}(2023)\citenamefont {Gubaev},
  \citenamefont {Zaverkin}, \citenamefont {Srinivasan}, \citenamefont {Duff},
  \citenamefont {Kästner},\ and\ \citenamefont {Grabowski}}]{Gubaev2023}%
  \BibitemOpen
  \bibfield  {author} {\bibinfo {author} {\bibfnamefont {K.}~\bibnamefont
  {Gubaev}}, \bibinfo {author} {\bibfnamefont {V.}~\bibnamefont {Zaverkin}},
  \bibinfo {author} {\bibfnamefont {P.}~\bibnamefont {Srinivasan}}, \bibinfo
  {author} {\bibfnamefont {A.~I.}\ \bibnamefont {Duff}}, \bibinfo {author}
  {\bibfnamefont {J.}~\bibnamefont {Kästner}},\ and\ \bibinfo {author}
  {\bibfnamefont {B.}~\bibnamefont {Grabowski}},\ }\bibfield  {title} {\bibinfo
  {title} {Performance of two complementary machine-learned potentials in
  modelling chemically complex systems},\ }\href
  {https://doi.org/10.1038/s41524-023-01073-w} {\bibfield  {journal} {\bibinfo
  {journal} {Npj Comput. Mater.}\ }\textbf {\bibinfo {volume} {9}},\ \bibinfo
  {pages} {1} (\bibinfo {year} {2023})}\BibitemShut {NoStop}%
\bibitem [{\citenamefont {Erhard}\ \emph {et~al.}(2024)\citenamefont {Erhard},
  \citenamefont {Rohrer}, \citenamefont {Albe},\ and\ \citenamefont
  {Deringer}}]{Erhard2024}%
  \BibitemOpen
  \bibfield  {author} {\bibinfo {author} {\bibfnamefont {L.~C.}\ \bibnamefont
  {Erhard}}, \bibinfo {author} {\bibfnamefont {J.}~\bibnamefont {Rohrer}},
  \bibinfo {author} {\bibfnamefont {K.}~\bibnamefont {Albe}},\ and\ \bibinfo
  {author} {\bibfnamefont {V.~L.}\ \bibnamefont {Deringer}},\ }\bibfield
  {title} {\bibinfo {title} {Modelling atomic and nanoscale structure in the
  silicon–oxygen system through active machine learning},\ }\href
  {https://doi.org/10.1038/s41467-024-45840-9} {\bibfield  {journal} {\bibinfo
  {journal} {Nat. Commun.}\ }\textbf {\bibinfo {volume} {15}},\ \bibinfo
  {pages} {1927} (\bibinfo {year} {2024})}\BibitemShut {NoStop}%
\bibitem [{\citenamefont {Novoselov}\ \emph {et~al.}(2019)\citenamefont
  {Novoselov}, \citenamefont {Yanilkin}, \citenamefont {Shapeev},\ and\
  \citenamefont {Podryabinkin}}]{Novoselov2019}%
  \BibitemOpen
  \bibfield  {author} {\bibinfo {author} {\bibfnamefont {I.~I.}\ \bibnamefont
  {Novoselov}}, \bibinfo {author} {\bibfnamefont {A.~V.}\ \bibnamefont
  {Yanilkin}}, \bibinfo {author} {\bibfnamefont {A.~V.}\ \bibnamefont
  {Shapeev}},\ and\ \bibinfo {author} {\bibfnamefont {E.~V.}\ \bibnamefont
  {Podryabinkin}},\ }\bibfield  {title} {\bibinfo {title} {Moment tensor
  potentials as a promising tool to study diffusion processes},\ }\href
  {https://doi.org/10.1016/j.commatsci.2019.03.049} {\bibfield  {journal}
  {\bibinfo  {journal} {Comput. Mater. Sci.}\ }\textbf {\bibinfo {volume}
  {164}},\ \bibinfo {pages} {46} (\bibinfo {year} {2019})}\BibitemShut
  {NoStop}%
\bibitem [{\citenamefont {Winter}\ and\ \citenamefont
  {Gómez-Bombarelli}(2023)}]{Winter2023}%
  \BibitemOpen
  \bibfield  {author} {\bibinfo {author} {\bibfnamefont {G.}~\bibnamefont
  {Winter}}\ and\ \bibinfo {author} {\bibfnamefont {R.}~\bibnamefont
  {Gómez-Bombarelli}},\ }\bibfield  {title} {\bibinfo {title} {Simulations
  with machine learning potentials identify the ion conduction mechanism
  mediating non-{Arrhenius} behavior in {LGPS}},\ }\href
  {https://doi.org/10.1088/2515-7655/acbbef} {\bibfield  {journal} {\bibinfo
  {journal} {J. Phys. Energy}\ }\textbf {\bibinfo {volume} {5}},\ \bibinfo
  {pages} {024004} (\bibinfo {year} {2023})}\BibitemShut {NoStop}%
\bibitem [{\citenamefont {Musaelian}\ \emph {et~al.}(2023)\citenamefont
  {Musaelian}, \citenamefont {Batzner}, \citenamefont {Johansson},
  \citenamefont {Sun}, \citenamefont {Owen}, \citenamefont {Kornbluth},\ and\
  \citenamefont {Kozinsky}}]{Musaelian2023}%
  \BibitemOpen
  \bibfield  {author} {\bibinfo {author} {\bibfnamefont {A.}~\bibnamefont
  {Musaelian}}, \bibinfo {author} {\bibfnamefont {S.}~\bibnamefont {Batzner}},
  \bibinfo {author} {\bibfnamefont {A.}~\bibnamefont {Johansson}}, \bibinfo
  {author} {\bibfnamefont {L.}~\bibnamefont {Sun}}, \bibinfo {author}
  {\bibfnamefont {C.~J.}\ \bibnamefont {Owen}}, \bibinfo {author}
  {\bibfnamefont {M.}~\bibnamefont {Kornbluth}},\ and\ \bibinfo {author}
  {\bibfnamefont {B.}~\bibnamefont {Kozinsky}},\ }\bibfield  {title} {\bibinfo
  {title} {Learning local equivariant representations for large-scale atomistic
  dynamics},\ }\href {https://doi.org/10.1038/s41467-023-36329-y} {\bibfield
  {journal} {\bibinfo  {journal} {Nat. Commun.}\ }\textbf {\bibinfo {volume}
  {14}},\ \bibinfo {pages} {579} (\bibinfo {year} {2023})}\BibitemShut
  {NoStop}%
\bibitem [{\citenamefont {Shapeev}(2016)}]{Shapeev2016}%
  \BibitemOpen
  \bibfield  {author} {\bibinfo {author} {\bibfnamefont {A.~V.}\ \bibnamefont
  {Shapeev}},\ }\bibfield  {title} {\bibinfo {title} {Moment tensor potentials:
  {A} class of systematically improvable interatomic potentials},\ }\href
  {https://doi.org/10.1137/15M1054183} {\bibfield  {journal} {\bibinfo
  {journal} {Multiscale Model. Simul.}\ }\textbf {\bibinfo {volume} {14}},\
  \bibinfo {pages} {1153} (\bibinfo {year} {2016})}\BibitemShut {NoStop}%
\bibitem [{\citenamefont {Kong}\ \emph {et~al.}(2010)\citenamefont {Kong},
  \citenamefont {Deiseroth}, \citenamefont {Reiner}, \citenamefont {Gün},
  \citenamefont {Neumann}, \citenamefont {Ritter},\ and\ \citenamefont
  {Zahn}}]{Kong2010}%
  \BibitemOpen
  \bibfield  {author} {\bibinfo {author} {\bibfnamefont {S.-T.}\ \bibnamefont
  {Kong}}, \bibinfo {author} {\bibfnamefont {H.-J.}\ \bibnamefont {Deiseroth}},
  \bibinfo {author} {\bibfnamefont {C.}~\bibnamefont {Reiner}}, \bibinfo
  {author} {\bibfnamefont {{\"O}.}~\bibnamefont {Gün}}, \bibinfo {author}
  {\bibfnamefont {E.}~\bibnamefont {Neumann}}, \bibinfo {author} {\bibfnamefont
  {C.}~\bibnamefont {Ritter}},\ and\ \bibinfo {author} {\bibfnamefont
  {D.}~\bibnamefont {Zahn}},\ }\bibfield  {title} {\bibinfo {title} {Lithium
  argyrodites with phosphorus and arsenic: {Order} and disorder of lithium
  atoms, crystal chemistry, and phase transitions},\ }\href
  {https://doi.org/10.1002/chem.200902470} {\bibfield  {journal} {\bibinfo
  {journal} {Chem. Eur. J.}\ }\textbf {\bibinfo {volume} {16}},\ \bibinfo
  {pages} {2198} (\bibinfo {year} {2010})}\BibitemShut {NoStop}%
\bibitem [{\citenamefont {Deiseroth}\ \emph {et~al.}(2011)\citenamefont
  {Deiseroth}, \citenamefont {Maier}, \citenamefont {Weichert}, \citenamefont
  {Nickel}, \citenamefont {Kong},\ and\ \citenamefont
  {Reiner}}]{Deiseroth2011}%
  \BibitemOpen
  \bibfield  {author} {\bibinfo {author} {\bibfnamefont {H.-J.}\ \bibnamefont
  {Deiseroth}}, \bibinfo {author} {\bibfnamefont {J.}~\bibnamefont {Maier}},
  \bibinfo {author} {\bibfnamefont {K.}~\bibnamefont {Weichert}}, \bibinfo
  {author} {\bibfnamefont {V.}~\bibnamefont {Nickel}}, \bibinfo {author}
  {\bibfnamefont {S.-T.}\ \bibnamefont {Kong}},\ and\ \bibinfo {author}
  {\bibfnamefont {C.}~\bibnamefont {Reiner}},\ }\bibfield  {title} {\bibinfo
  {title} {{Li$_7$PS$_6$} and {Li$_6$PS$_5X$} ({$X$}: {Cl}, {Br}, {I}):
  {Possible} three-dimensional diffusion pathways for lithium ions and
  temperature dependence of the ionic conductivity by impedance measurements},\
  }\href {https://doi.org/10.1002/zaac.201100158} {\bibfield  {journal}
  {\bibinfo  {journal} {Z. Anorg. Allg. Chem.}\ }\textbf {\bibinfo {volume}
  {637}},\ \bibinfo {pages} {1287} (\bibinfo {year} {2011})}\BibitemShut
  {NoStop}%
\bibitem [{\citenamefont {Wang}\ and\ \citenamefont {Shao}(2017)}]{Wang2017}%
  \BibitemOpen
  \bibfield  {author} {\bibinfo {author} {\bibfnamefont {Z.}~\bibnamefont
  {Wang}}\ and\ \bibinfo {author} {\bibfnamefont {G.}~\bibnamefont {Shao}},\
  }\bibfield  {title} {\bibinfo {title} {Theoretical design of solid
  electrolytes with superb ionic conductivity: {Alloying} effect on {Li}$^+$
  transportation in cubic {Li$_6$PA$_5X$} chalcogenides},\ }\href
  {https://doi.org/10.1039/C7TA06986C} {\bibfield  {journal} {\bibinfo
  {journal} {J. Mater. Chem. A}\ }\textbf {\bibinfo {volume} {5}},\ \bibinfo
  {pages} {21846} (\bibinfo {year} {2017})}\BibitemShut {NoStop}%
\bibitem [{\citenamefont {Golov}\ and\ \citenamefont
  {Carrasco}(2021)}]{Golov2021}%
  \BibitemOpen
  \bibfield  {author} {\bibinfo {author} {\bibfnamefont {A.}~\bibnamefont
  {Golov}}\ and\ \bibinfo {author} {\bibfnamefont {J.}~\bibnamefont
  {Carrasco}},\ }\bibfield  {title} {\bibinfo {title} {Molecular-level insight
  into the interfacial reactivity and ionic conductivity of a {Li}-argyrodite
  {Li$_6$PS$_5$Cl} solid electrolyte at bare and coated {Li}-metal anodes},\
  }\href {https://doi.org/10.1021/acsami.1c12753} {\bibfield  {journal}
  {\bibinfo  {journal} {ACS Appl. Mater. Interfaces}\ }\textbf {\bibinfo
  {volume} {13}},\ \bibinfo {pages} {43734} (\bibinfo {year}
  {2021})}\BibitemShut {NoStop}%
\bibitem [{\citenamefont {D'Amore}\ \emph {et~al.}(2022)\citenamefont
  {D'Amore}, \citenamefont {Daga}, \citenamefont {Rocca}, \citenamefont
  {Sgroi}, \citenamefont {Marana}, \citenamefont {Casassa}, \citenamefont
  {Maschio},\ and\ \citenamefont {Ferrari}}]{DAmore2022}%
  \BibitemOpen
  \bibfield  {author} {\bibinfo {author} {\bibfnamefont {M.}~\bibnamefont
  {D'Amore}}, \bibinfo {author} {\bibfnamefont {L.~E.}\ \bibnamefont {Daga}},
  \bibinfo {author} {\bibfnamefont {R.}~\bibnamefont {Rocca}}, \bibinfo
  {author} {\bibfnamefont {M.~F.}\ \bibnamefont {Sgroi}}, \bibinfo {author}
  {\bibfnamefont {N.~L.}\ \bibnamefont {Marana}}, \bibinfo {author}
  {\bibfnamefont {S.~M.}\ \bibnamefont {Casassa}}, \bibinfo {author}
  {\bibfnamefont {L.}~\bibnamefont {Maschio}},\ and\ \bibinfo {author}
  {\bibfnamefont {A.~M.}\ \bibnamefont {Ferrari}},\ }\bibfield  {title}
  {\bibinfo {title} {From symmetry breaking in the bulk to phase transitions at
  the surface: {A} quantum-mechanical exploration of {Li$_6$PS$_5$Cl}
  argyrodite superionic conductor},\ }\href
  {https://doi.org/10.1039/D2CP03599E} {\bibfield  {journal} {\bibinfo
  {journal} {Phys. Chem. Chem. Phys.}\ }\textbf {\bibinfo {volume} {24}},\
  \bibinfo {pages} {22978} (\bibinfo {year} {2022})}\BibitemShut {NoStop}%
\bibitem [{\citenamefont {Ou}\ \emph {et~al.}(2022)\citenamefont {Ou},
  \citenamefont {Ikeda}, \citenamefont {Clemens},\ and\ \citenamefont
  {Grabowski}}]{Ou2022}%
  \BibitemOpen
  \bibfield  {author} {\bibinfo {author} {\bibfnamefont {Y.}~\bibnamefont
  {Ou}}, \bibinfo {author} {\bibfnamefont {Y.}~\bibnamefont {Ikeda}}, \bibinfo
  {author} {\bibfnamefont {O.}~\bibnamefont {Clemens}},\ and\ \bibinfo {author}
  {\bibfnamefont {B.}~\bibnamefont {Grabowski}},\ }\bibfield  {title} {\bibinfo
  {title} {Dynamic stabilization of perovskites at elevated temperatures: {A}
  comparison between cubic {BaFeO$_3$} and vacancy-ordered monoclinic
  {BaFeO$_{2.67}$}},\ }\href {https://doi.org/10.1103/PhysRevB.106.064308}
  {\bibfield  {journal} {\bibinfo  {journal} {Phys. Rev. B}\ }\textbf {\bibinfo
  {volume} {106}},\ \bibinfo {pages} {064308} (\bibinfo {year}
  {2022})}\BibitemShut {NoStop}%
\bibitem [{\citenamefont {Momma}\ and\ \citenamefont
  {Izumi}(2011)}]{Momma2011}%
  \BibitemOpen
  \bibfield  {author} {\bibinfo {author} {\bibfnamefont {K.}~\bibnamefont
  {Momma}}\ and\ \bibinfo {author} {\bibfnamefont {F.}~\bibnamefont {Izumi}},\
  }\bibfield  {title} {\bibinfo {title} {{VESTA~\!3} for three-dimensional
  visualization of crystal, volumetric and morphology data},\ }\href
  {https://doi.org/10.1107/S0021889811038970} {\bibfield  {journal} {\bibinfo
  {journal} {J. Appl. Crystallogr.}\ }\textbf {\bibinfo {volume} {44}},\
  \bibinfo {pages} {1272} (\bibinfo {year} {2011})}\BibitemShut {NoStop}%
\bibitem [{\citenamefont {Cheng}\ \emph {et~al.}(2018)\citenamefont {Cheng},
  \citenamefont {Luo},\ and\ \citenamefont {Yang}}]{Cheng2018}%
  \BibitemOpen
  \bibfield  {author} {\bibinfo {author} {\bibfnamefont {J.}~\bibnamefont
  {Cheng}}, \bibinfo {author} {\bibfnamefont {J.}~\bibnamefont {Luo}},\ and\
  \bibinfo {author} {\bibfnamefont {K.}~\bibnamefont {Yang}},\ }\bibfield
  {title} {\bibinfo {title} {Aimsgb: {An} algorithm and open-source {Python}
  library to generate periodic grain boundary structures},\ }\href
  {https://doi.org/10.1016/j.commatsci.2018.08.029} {\bibfield  {journal}
  {\bibinfo  {journal} {Comput. Mater. Sci.}\ }\textbf {\bibinfo {volume}
  {155}},\ \bibinfo {pages} {92} (\bibinfo {year} {2018})}\BibitemShut
  {NoStop}%
\bibitem [{\citenamefont {Metzler}\ and\ \citenamefont
  {Klafter}(2000)}]{Metzler2000}%
  \BibitemOpen
  \bibfield  {author} {\bibinfo {author} {\bibfnamefont {R.}~\bibnamefont
  {Metzler}}\ and\ \bibinfo {author} {\bibfnamefont {J.}~\bibnamefont
  {Klafter}},\ }\bibfield  {title} {\bibinfo {title} {The random walk's guide
  to anomalous diffusion: {A} fractional dynamics approach},\ }\href
  {https://doi.org/10.1016/S0370-1573(00)00070-3} {\bibfield  {journal}
  {\bibinfo  {journal} {Phys. Rep.}\ }\textbf {\bibinfo {volume} {339}},\
  \bibinfo {pages} {1} (\bibinfo {year} {2000})}\BibitemShut {NoStop}%
\bibitem [{\citenamefont {Einstein}(1905)}]{Einstein1905}%
  \BibitemOpen
  \bibfield  {author} {\bibinfo {author} {\bibfnamefont {A.}~\bibnamefont
  {Einstein}},\ }\bibfield  {title} {\bibinfo {title} {Über die von der
  molekularkinetischen {Theorie} der {Wärme} geforderte {Bewegung} von in
  ruhenden {Flüssigkeiten} suspendierten {Teilchen}},\ }\href
  {https://doi.org/10.1002/andp.19053220806} {\bibfield  {journal} {\bibinfo
  {journal} {Ann. Phys.}\ }\textbf {\bibinfo {volume} {322}},\ \bibinfo {pages}
  {549} (\bibinfo {year} {1905})}\BibitemShut {NoStop}%
\bibitem [{\citenamefont {Dawson}\ \emph {et~al.}(2018)\citenamefont {Dawson},
  \citenamefont {Canepa}, \citenamefont {Famprikis}, \citenamefont
  {Masquelier},\ and\ \citenamefont {Islam}}]{Dawson2018}%
  \BibitemOpen
  \bibfield  {author} {\bibinfo {author} {\bibfnamefont {J.~A.}\ \bibnamefont
  {Dawson}}, \bibinfo {author} {\bibfnamefont {P.}~\bibnamefont {Canepa}},
  \bibinfo {author} {\bibfnamefont {T.}~\bibnamefont {Famprikis}}, \bibinfo
  {author} {\bibfnamefont {C.}~\bibnamefont {Masquelier}},\ and\ \bibinfo
  {author} {\bibfnamefont {M.~S.}\ \bibnamefont {Islam}},\ }\bibfield  {title}
  {\bibinfo {title} {Atomic-scale influence of grain boundaries on {Li}-ion
  conduction in solid electrolytes for all-solid-state batteries},\ }\href
  {https://doi.org/10.1021/jacs.7b10593} {\bibfield  {journal} {\bibinfo
  {journal} {J. Am. Chem. Soc.}\ }\textbf {\bibinfo {volume} {140}},\ \bibinfo
  {pages} {362} (\bibinfo {year} {2018})}\BibitemShut {NoStop}%
\bibitem [{\citenamefont {Lee}\ \emph {et~al.}(2023)\citenamefont {Lee},
  \citenamefont {Qi}, \citenamefont {Gadre}, \citenamefont {Huyan},
  \citenamefont {Ko}, \citenamefont {Zuo}, \citenamefont {Du}, \citenamefont
  {Li}, \citenamefont {Aoki}, \citenamefont {Wu}, \citenamefont {Luo},
  \citenamefont {Ong},\ and\ \citenamefont {Pan}}]{Lee2023}%
  \BibitemOpen
  \bibfield  {author} {\bibinfo {author} {\bibfnamefont {T.}~\bibnamefont
  {Lee}}, \bibinfo {author} {\bibfnamefont {J.}~\bibnamefont {Qi}}, \bibinfo
  {author} {\bibfnamefont {C.~A.}\ \bibnamefont {Gadre}}, \bibinfo {author}
  {\bibfnamefont {H.}~\bibnamefont {Huyan}}, \bibinfo {author} {\bibfnamefont
  {S.-T.}\ \bibnamefont {Ko}}, \bibinfo {author} {\bibfnamefont
  {Y.}~\bibnamefont {Zuo}}, \bibinfo {author} {\bibfnamefont {C.}~\bibnamefont
  {Du}}, \bibinfo {author} {\bibfnamefont {J.}~\bibnamefont {Li}}, \bibinfo
  {author} {\bibfnamefont {T.}~\bibnamefont {Aoki}}, \bibinfo {author}
  {\bibfnamefont {R.}~\bibnamefont {Wu}}, \bibinfo {author} {\bibfnamefont
  {J.}~\bibnamefont {Luo}}, \bibinfo {author} {\bibfnamefont {S.~P.}\
  \bibnamefont {Ong}},\ and\ \bibinfo {author} {\bibfnamefont {X.}~\bibnamefont
  {Pan}},\ }\bibfield  {title} {\bibinfo {title} {Atomic-scale origin of the
  low grain-boundary resistance in perovskite solid electrolyte
  {Li$_{0.375}$Sr$_{0.4375}$Ta$_{0.75}$Zr$_{0.25}$O$_{3}$}},\ }\href
  {https://doi.org/10.1038/s41467-023-37115-6} {\bibfield  {journal} {\bibinfo
  {journal} {Nat. Commun.}\ }\textbf {\bibinfo {volume} {14}},\ \bibinfo
  {pages} {1940} (\bibinfo {year} {2023})}\BibitemShut {NoStop}%
\bibitem [{\citenamefont {Jalem}\ \emph {et~al.}(2023)\citenamefont {Jalem},
  \citenamefont {Chandrappa}, \citenamefont {Qi}, \citenamefont {Tateyama},\
  and\ \citenamefont {Ong}}]{Jalem2023}%
  \BibitemOpen
  \bibfield  {author} {\bibinfo {author} {\bibfnamefont {R.}~\bibnamefont
  {Jalem}}, \bibinfo {author} {\bibfnamefont {M.~L.~H.}\ \bibnamefont
  {Chandrappa}}, \bibinfo {author} {\bibfnamefont {J.}~\bibnamefont {Qi}},
  \bibinfo {author} {\bibfnamefont {Y.}~\bibnamefont {Tateyama}},\ and\
  \bibinfo {author} {\bibfnamefont {S.~P.}\ \bibnamefont {Ong}},\ }\bibfield
  {title} {\bibinfo {title} {Lithium dynamics at grain boundaries of
  $\beta$-{Li$_3$PS$_4$} solid electrolyte},\ }\href
  {https://doi.org/10.1039/D3YA00234A} {\bibfield  {journal} {\bibinfo
  {journal} {Energy Adv.}\ }\textbf {\bibinfo {volume} {2}},\ \bibinfo {pages}
  {2029} (\bibinfo {year} {2023})}\BibitemShut {NoStop}%
\bibitem [{\citenamefont {Hart}(1957)}]{Hart1957}%
  \BibitemOpen
  \bibfield  {author} {\bibinfo {author} {\bibfnamefont {E.~W.}\ \bibnamefont
  {Hart}},\ }\bibfield  {title} {\bibinfo {title} {On the role of dislocations
  in bulk diffusion},\ }\href {https://doi.org/10.1016/0001-6160(57)90127-X}
  {\bibfield  {journal} {\bibinfo  {journal} {Acta Metall.}\ }\textbf {\bibinfo
  {volume} {5}},\ \bibinfo {pages} {597} (\bibinfo {year} {1957})}\BibitemShut
  {NoStop}%
\bibitem [{\citenamefont {Belova}\ and\ \citenamefont
  {Murch}(2003)}]{Belova2003}%
  \BibitemOpen
  \bibfield  {author} {\bibinfo {author} {\bibfnamefont {I.~V.}\ \bibnamefont
  {Belova}}\ and\ \bibinfo {author} {\bibfnamefont {G.~E.}\ \bibnamefont
  {Murch}},\ }\bibfield  {title} {\bibinfo {title} {Diffusion in
  nanocrystalline materials},\ }\href
  {https://doi.org/10.1016/S0022-3697(02)00421-3} {\bibfield  {journal}
  {\bibinfo  {journal} {J. Phys. Chem. Solids}\ }\textbf {\bibinfo {volume}
  {64}},\ \bibinfo {pages} {873} (\bibinfo {year} {2003})}\BibitemShut
  {NoStop}%
\bibitem [{\citenamefont {Kaur}\ \emph {et~al.}(1995)\citenamefont {Kaur},
  \citenamefont {Mishin},\ and\ \citenamefont {Gust}}]{Kaur1995}%
  \BibitemOpen
  \bibfield  {author} {\bibinfo {author} {\bibfnamefont {I.}~\bibnamefont
  {Kaur}}, \bibinfo {author} {\bibfnamefont {Y.}~\bibnamefont {Mishin}},\ and\
  \bibinfo {author} {\bibfnamefont {W.}~\bibnamefont {Gust}},\ }\href@noop {}
  {\emph {\bibinfo {title} {Fundamentals of grain and interface boundary
  diffusion}}}\ (\bibinfo  {publisher} {Wiley \& Sons LTD},\ \bibinfo {year}
  {1995})\BibitemShut {NoStop}%
\bibitem [{\citenamefont {Blöchl}(1994)}]{Bloechl1994}%
  \BibitemOpen
  \bibfield  {author} {\bibinfo {author} {\bibfnamefont {P.~E.}\ \bibnamefont
  {Blöchl}},\ }\bibfield  {title} {\bibinfo {title} {Projector augmented-wave
  method},\ }\href {https://doi.org/10.1103/PhysRevB.50.17953} {\bibfield
  {journal} {\bibinfo  {journal} {Phys. Rev. B}\ }\textbf {\bibinfo {volume}
  {50}},\ \bibinfo {pages} {17953} (\bibinfo {year} {1994})}\BibitemShut
  {NoStop}%
\bibitem [{\citenamefont {Perdew}\ \emph {et~al.}(1996)\citenamefont {Perdew},
  \citenamefont {Burke},\ and\ \citenamefont {Ernzerhof}}]{Perdew1996}%
  \BibitemOpen
  \bibfield  {author} {\bibinfo {author} {\bibfnamefont {J.~P.}\ \bibnamefont
  {Perdew}}, \bibinfo {author} {\bibfnamefont {K.}~\bibnamefont {Burke}},\ and\
  \bibinfo {author} {\bibfnamefont {M.}~\bibnamefont {Ernzerhof}},\ }\bibfield
  {title} {\bibinfo {title} {Generalized gradient approximation made simple},\
  }\href {https://doi.org/10.1103/PhysRevLett.77.3865} {\bibfield  {journal}
  {\bibinfo  {journal} {Phys. Rev. Lett.}\ }\textbf {\bibinfo {volume} {77}},\
  \bibinfo {pages} {3865} (\bibinfo {year} {1996})}\BibitemShut {NoStop}%
\bibitem [{\citenamefont {Kresse}(1995)}]{Kresse1995}%
  \BibitemOpen
  \bibfield  {author} {\bibinfo {author} {\bibfnamefont {G.}~\bibnamefont
  {Kresse}},\ }\bibfield  {title} {\bibinfo {title} {\textit{Ab initio}
  molecular dynamics for liquid metals},\ }\href
  {https://doi.org/10.1016/0022-3093(95)00355-X} {\bibfield  {journal}
  {\bibinfo  {journal} {J. Non-Cryst. Solids}\ }\textbf {\bibinfo {volume}
  {192-193}},\ \bibinfo {pages} {222} (\bibinfo {year} {1995})}\BibitemShut
  {NoStop}%
\bibitem [{\citenamefont {Kresse}\ and\ \citenamefont
  {Furthmüller}(1996)}]{Kresse1996}%
  \BibitemOpen
  \bibfield  {author} {\bibinfo {author} {\bibfnamefont {G.}~\bibnamefont
  {Kresse}}\ and\ \bibinfo {author} {\bibfnamefont {J.}~\bibnamefont
  {Furthmüller}},\ }\bibfield  {title} {\bibinfo {title} {Efficient iterative
  schemes for \textit{ab initio} total-energy calculations using a plane-wave
  basis set},\ }\href {https://doi.org/10.1103/PhysRevB.54.11169} {\bibfield
  {journal} {\bibinfo  {journal} {Phys. Rev. B}\ }\textbf {\bibinfo {volume}
  {54}},\ \bibinfo {pages} {11169} (\bibinfo {year} {1996})}\BibitemShut
  {NoStop}%
\bibitem [{\citenamefont {Kresse}\ and\ \citenamefont
  {Joubert}(1999)}]{Kresse1999}%
  \BibitemOpen
  \bibfield  {author} {\bibinfo {author} {\bibfnamefont {G.}~\bibnamefont
  {Kresse}}\ and\ \bibinfo {author} {\bibfnamefont {D.}~\bibnamefont
  {Joubert}},\ }\bibfield  {title} {\bibinfo {title} {From ultrasoft
  pseudopotentials to the projector augmented-wave method},\ }\href
  {https://doi.org/10.1103/PhysRevB.59.1758} {\bibfield  {journal} {\bibinfo
  {journal} {Phys. Rev. B}\ }\textbf {\bibinfo {volume} {59}},\ \bibinfo
  {pages} {1758} (\bibinfo {year} {1999})}\BibitemShut {NoStop}%
\bibitem [{\citenamefont {Blöchl}\ \emph {et~al.}(1994)\citenamefont
  {Blöchl}, \citenamefont {Jepsen},\ and\ \citenamefont
  {Andersen}}]{Bloechl1994a}%
  \BibitemOpen
  \bibfield  {author} {\bibinfo {author} {\bibfnamefont {P.~E.}\ \bibnamefont
  {Blöchl}}, \bibinfo {author} {\bibfnamefont {O.}~\bibnamefont {Jepsen}},\
  and\ \bibinfo {author} {\bibfnamefont {O.~K.}\ \bibnamefont {Andersen}},\
  }\bibfield  {title} {\bibinfo {title} {Improved tetrahedron method for
  {Brillouin-zone} integrations},\ }\href
  {https://doi.org/10.1103/PhysRevB.49.16223} {\bibfield  {journal} {\bibinfo
  {journal} {Phys. Rev. B}\ }\textbf {\bibinfo {volume} {49}},\ \bibinfo
  {pages} {16223} (\bibinfo {year} {1994})}\BibitemShut {NoStop}%
\bibitem [{\citenamefont {Thompson}\ \emph {et~al.}(2022)\citenamefont
  {Thompson}, \citenamefont {Aktulga}, \citenamefont {Berger}, \citenamefont
  {Bolintineanu}, \citenamefont {Brown}, \citenamefont {Crozier}, \citenamefont
  {in~'t Veld}, \citenamefont {Kohlmeyer}, \citenamefont {Moore}, \citenamefont
  {Nguyen}, \citenamefont {Shan}, \citenamefont {Stevens}, \citenamefont
  {Tranchida}, \citenamefont {Trott},\ and\ \citenamefont
  {Plimpton}}]{Thompson2022}%
  \BibitemOpen
  \bibfield  {author} {\bibinfo {author} {\bibfnamefont {A.~P.}\ \bibnamefont
  {Thompson}}, \bibinfo {author} {\bibfnamefont {H.~M.}\ \bibnamefont
  {Aktulga}}, \bibinfo {author} {\bibfnamefont {R.}~\bibnamefont {Berger}},
  \bibinfo {author} {\bibfnamefont {D.~S.}\ \bibnamefont {Bolintineanu}},
  \bibinfo {author} {\bibfnamefont {W.~M.}\ \bibnamefont {Brown}}, \bibinfo
  {author} {\bibfnamefont {P.~S.}\ \bibnamefont {Crozier}}, \bibinfo {author}
  {\bibfnamefont {P.~J.}\ \bibnamefont {in~'t Veld}}, \bibinfo {author}
  {\bibfnamefont {A.}~\bibnamefont {Kohlmeyer}}, \bibinfo {author}
  {\bibfnamefont {S.~G.}\ \bibnamefont {Moore}}, \bibinfo {author}
  {\bibfnamefont {T.~D.}\ \bibnamefont {Nguyen}}, \bibinfo {author}
  {\bibfnamefont {R.}~\bibnamefont {Shan}}, \bibinfo {author} {\bibfnamefont
  {M.~J.}\ \bibnamefont {Stevens}}, \bibinfo {author} {\bibfnamefont
  {J.}~\bibnamefont {Tranchida}}, \bibinfo {author} {\bibfnamefont
  {C.}~\bibnamefont {Trott}},\ and\ \bibinfo {author} {\bibfnamefont {S.~J.}\
  \bibnamefont {Plimpton}},\ }\bibfield  {title} {\bibinfo {title} {{LAMMPS} --
  a flexible simulation tool for particle-based materials modeling at the
  atomic, meso, and continuum scales},\ }\href
  {https://doi.org/10.1016/j.cpc.2021.108171} {\bibfield  {journal} {\bibinfo
  {journal} {Comput. Phys. Commun.}\ }\textbf {\bibinfo {volume} {271}},\
  \bibinfo {pages} {108171} (\bibinfo {year} {2022})}\BibitemShut {NoStop}%
\bibitem [{\citenamefont {Wagih}\ and\ \citenamefont
  {Schuh}(2023)}]{Wagih2023Dec}%
  \BibitemOpen
  \bibfield  {author} {\bibinfo {author} {\bibfnamefont {M.}~\bibnamefont
  {Wagih}}\ and\ \bibinfo {author} {\bibfnamefont {C.~A.}\ \bibnamefont
  {Schuh}},\ }\bibfield  {title} {\bibinfo {title} {{Viewpoint: Can symmetric
  tilt grain boundaries represent polycrystals?}},\ }\href
  {https://doi.org/10.1016/j.scriptamat.2023.115716} {\bibfield  {journal}
  {\bibinfo  {journal} {Scr. Mater.}\ }\textbf {\bibinfo {volume} {237}},\
  \bibinfo {pages} {115716} (\bibinfo {year} {2023})}\BibitemShut {NoStop}%
\bibitem [{\citenamefont {Novikov}\ \emph {et~al.}(2020)\citenamefont
  {Novikov}, \citenamefont {Gubaev}, \citenamefont {Podryabinkin},\ and\
  \citenamefont {Shapeev}}]{Novikov2020}%
  \BibitemOpen
  \bibfield  {author} {\bibinfo {author} {\bibfnamefont {I.~S.}\ \bibnamefont
  {Novikov}}, \bibinfo {author} {\bibfnamefont {K.}~\bibnamefont {Gubaev}},
  \bibinfo {author} {\bibfnamefont {E.~V.}\ \bibnamefont {Podryabinkin}},\ and\
  \bibinfo {author} {\bibfnamefont {A.~V.}\ \bibnamefont {Shapeev}},\
  }\bibfield  {title} {\bibinfo {title} {The {MLIP} package: {Moment} tensor
  potentials with {MPI} and active learning},\ }\href
  {https://doi.org/10.1088/2632-2153/abc9fe} {\bibfield  {journal} {\bibinfo
  {journal} {Mach. Learn.: Sci. Technol.}\ }\textbf {\bibinfo {volume} {2}},\
  \bibinfo {pages} {025002} (\bibinfo {year} {2020})}\BibitemShut {NoStop}%
\bibitem [{\citenamefont {Gubaev}\ \emph {et~al.}(2021)\citenamefont {Gubaev},
  \citenamefont {Ikeda}, \citenamefont {Tasnádi}, \citenamefont {Neugebauer},
  \citenamefont {Shapeev}, \citenamefont {Grabowski},\ and\ \citenamefont
  {Körmann}}]{Gubaev2021}%
  \BibitemOpen
  \bibfield  {author} {\bibinfo {author} {\bibfnamefont {K.}~\bibnamefont
  {Gubaev}}, \bibinfo {author} {\bibfnamefont {Y.}~\bibnamefont {Ikeda}},
  \bibinfo {author} {\bibfnamefont {F.}~\bibnamefont {Tasnádi}}, \bibinfo
  {author} {\bibfnamefont {J.}~\bibnamefont {Neugebauer}}, \bibinfo {author}
  {\bibfnamefont {A.~V.}\ \bibnamefont {Shapeev}}, \bibinfo {author}
  {\bibfnamefont {B.}~\bibnamefont {Grabowski}},\ and\ \bibinfo {author}
  {\bibfnamefont {F.}~\bibnamefont {Körmann}},\ }\bibfield  {title} {\bibinfo
  {title} {Finite-temperature interplay of structural stability, chemical
  complexity, and elastic properties of bcc multicomponent alloys from
  \textit{ab initio} trained machine-learning potentials},\ }\href
  {https://doi.org/10.1103/PhysRevMaterials.5.073801} {\bibfield  {journal}
  {\bibinfo  {journal} {Phys. Rev. Mater.}\ }\textbf {\bibinfo {volume} {5}},\
  \bibinfo {pages} {073801} (\bibinfo {year} {2021})}\BibitemShut {NoStop}%
\bibitem [{\citenamefont {Stukowski}(2009)}]{Stukowski2009}%
  \BibitemOpen
  \bibfield  {author} {\bibinfo {author} {\bibfnamefont {A.}~\bibnamefont
  {Stukowski}},\ }\bibfield  {title} {\bibinfo {title} {Visualization and
  analysis of atomistic simulation data with {OVITO} – the open visualization
  tool},\ }\href {https://doi.org/10.1088/0965-0393/18/1/015012} {\bibfield
  {journal} {\bibinfo  {journal} {Model. Simul. Mater. Sci. Eng.}\ }\textbf
  {\bibinfo {volume} {18}},\ \bibinfo {pages} {015012} (\bibinfo {year}
  {2009})}\BibitemShut {NoStop}%
\bibitem [{\citenamefont {Larsen}\ \emph {et~al.}(2017)\citenamefont {Larsen},
  \citenamefont {Mortensen}, \citenamefont {Blomqvist}, \citenamefont
  {Castelli}, \citenamefont {Christensen}, \citenamefont {Dułak},
  \citenamefont {Friis}, \citenamefont {Groves}, \citenamefont {Hammer},
  \citenamefont {Hargus}, \citenamefont {Hermes}, \citenamefont {Jennings},
  \citenamefont {Jensen}, \citenamefont {Kermode}, \citenamefont {Kitchin},
  \citenamefont {Kolsbjerg}, \citenamefont {Kubal}, \citenamefont {Kaasbjerg},
  \citenamefont {Lysgaard}, \citenamefont {Maronsson}, \citenamefont {Maxson},
  \citenamefont {Olsen}, \citenamefont {Pastewka}, \citenamefont {Peterson},
  \citenamefont {Rostgaard}, \citenamefont {Schiøtz}, \citenamefont {Schütt},
  \citenamefont {Strange}, \citenamefont {Thygesen}, \citenamefont {Vegge},
  \citenamefont {Vilhelmsen}, \citenamefont {Walter}, \citenamefont {Zeng},\
  and\ \citenamefont {Jacobsen}}]{ase-paper}%
  \BibitemOpen
  \bibfield  {author} {\bibinfo {author} {\bibfnamefont {A.~H.}\ \bibnamefont
  {Larsen}}, \bibinfo {author} {\bibfnamefont {J.~J.}\ \bibnamefont
  {Mortensen}}, \bibinfo {author} {\bibfnamefont {J.}~\bibnamefont
  {Blomqvist}}, \bibinfo {author} {\bibfnamefont {I.~E.}\ \bibnamefont
  {Castelli}}, \bibinfo {author} {\bibfnamefont {R.}~\bibnamefont
  {Christensen}}, \bibinfo {author} {\bibfnamefont {M.}~\bibnamefont {Dułak}},
  \bibinfo {author} {\bibfnamefont {J.}~\bibnamefont {Friis}}, \bibinfo
  {author} {\bibfnamefont {M.~N.}\ \bibnamefont {Groves}}, \bibinfo {author}
  {\bibfnamefont {B.}~\bibnamefont {Hammer}}, \bibinfo {author} {\bibfnamefont
  {C.}~\bibnamefont {Hargus}}, \bibinfo {author} {\bibfnamefont {E.~D.}\
  \bibnamefont {Hermes}}, \bibinfo {author} {\bibfnamefont {P.~C.}\
  \bibnamefont {Jennings}}, \bibinfo {author} {\bibfnamefont {P.~B.}\
  \bibnamefont {Jensen}}, \bibinfo {author} {\bibfnamefont {J.}~\bibnamefont
  {Kermode}}, \bibinfo {author} {\bibfnamefont {J.~R.}\ \bibnamefont
  {Kitchin}}, \bibinfo {author} {\bibfnamefont {E.~L.}\ \bibnamefont
  {Kolsbjerg}}, \bibinfo {author} {\bibfnamefont {J.}~\bibnamefont {Kubal}},
  \bibinfo {author} {\bibfnamefont {K.}~\bibnamefont {Kaasbjerg}}, \bibinfo
  {author} {\bibfnamefont {S.}~\bibnamefont {Lysgaard}}, \bibinfo {author}
  {\bibfnamefont {J.~B.}\ \bibnamefont {Maronsson}}, \bibinfo {author}
  {\bibfnamefont {T.}~\bibnamefont {Maxson}}, \bibinfo {author} {\bibfnamefont
  {T.}~\bibnamefont {Olsen}}, \bibinfo {author} {\bibfnamefont
  {L.}~\bibnamefont {Pastewka}}, \bibinfo {author} {\bibfnamefont
  {A.}~\bibnamefont {Peterson}}, \bibinfo {author} {\bibfnamefont
  {C.}~\bibnamefont {Rostgaard}}, \bibinfo {author} {\bibfnamefont
  {J.}~\bibnamefont {Schiøtz}}, \bibinfo {author} {\bibfnamefont
  {O.}~\bibnamefont {Schütt}}, \bibinfo {author} {\bibfnamefont
  {M.}~\bibnamefont {Strange}}, \bibinfo {author} {\bibfnamefont {K.~S.}\
  \bibnamefont {Thygesen}}, \bibinfo {author} {\bibfnamefont {T.}~\bibnamefont
  {Vegge}}, \bibinfo {author} {\bibfnamefont {L.}~\bibnamefont {Vilhelmsen}},
  \bibinfo {author} {\bibfnamefont {M.}~\bibnamefont {Walter}}, \bibinfo
  {author} {\bibfnamefont {Z.}~\bibnamefont {Zeng}},\ and\ \bibinfo {author}
  {\bibfnamefont {K.~W.}\ \bibnamefont {Jacobsen}},\ }\bibfield  {title}
  {\bibinfo {title} {The atomic simulation environment -- a {Python} library
  for working with atoms},\ }\href
  {http://stacks.iop.org/0953-8984/29/i=27/a=273002} {\bibfield  {journal}
  {\bibinfo  {journal} {J. Phys. Condens. Matter.}\ }\textbf {\bibinfo {volume}
  {29}},\ \bibinfo {pages} {273002} (\bibinfo {year} {2017})}\BibitemShut
  {NoStop}%
\bibitem [{\citenamefont {Glass}\ \emph {et~al.}(2006)\citenamefont {Glass},
  \citenamefont {Oganov},\ and\ \citenamefont {Hansen}}]{Glass2006}%
  \BibitemOpen
  \bibfield  {author} {\bibinfo {author} {\bibfnamefont {C.~W.}\ \bibnamefont
  {Glass}}, \bibinfo {author} {\bibfnamefont {A.~R.}\ \bibnamefont {Oganov}},\
  and\ \bibinfo {author} {\bibfnamefont {N.}~\bibnamefont {Hansen}},\
  }\bibfield  {title} {\bibinfo {title} {{USPEX} -- {Evolutionary} crystal
  structure prediction},\ }\href {https://doi.org/10.1016/j.cpc.2006.07.020}
  {\bibfield  {journal} {\bibinfo  {journal} {Comput. Phys. Commun.}\ }\textbf
  {\bibinfo {volume} {175}},\ \bibinfo {pages} {713} (\bibinfo {year}
  {2006})}\BibitemShut {NoStop}%
\bibitem [{\citenamefont {Guo}\ \emph {et~al.}(2017)\citenamefont {Guo},
  \citenamefont {Wang},\ and\ \citenamefont {Saidi}}]{Guo2017}%
  \BibitemOpen
  \bibfield  {author} {\bibinfo {author} {\bibfnamefont {Y.}~\bibnamefont
  {Guo}}, \bibinfo {author} {\bibfnamefont {Q.}~\bibnamefont {Wang}},\ and\
  \bibinfo {author} {\bibfnamefont {W.~A.}\ \bibnamefont {Saidi}},\ }\bibfield
  {title} {\bibinfo {title} {Structural stabilities and electronic properties
  of high-angle grain boundaries in perovskite cesium lead halides},\ }\href
  {https://doi.org/10.1021/acs.jpcc.6b11434} {\bibfield  {journal} {\bibinfo
  {journal} {J. Phys. Chem. C}\ }\textbf {\bibinfo {volume} {121}},\ \bibinfo
  {pages} {1715} (\bibinfo {year} {2017})}\BibitemShut {NoStop}%
\bibitem [{\citenamefont {Koju}\ and\ \citenamefont {Mishin}(2020)}]{Koju2020}%
  \BibitemOpen
  \bibfield  {author} {\bibinfo {author} {\bibfnamefont {R.~K.}\ \bibnamefont
  {Koju}}\ and\ \bibinfo {author} {\bibfnamefont {Y.}~\bibnamefont {Mishin}},\
  }\bibfield  {title} {\bibinfo {title} {Relationship between grain boundary
  segregation and grain boundary diffusion in {Cu}-{Ag} alloys},\ }\href
  {https://doi.org/10.1103/PhysRevMaterials.4.073403} {\bibfield  {journal}
  {\bibinfo  {journal} {Phys. Rev. Mater.}\ }\textbf {\bibinfo {volume} {4}},\
  \bibinfo {pages} {073403} (\bibinfo {year} {2020})}\BibitemShut {NoStop}%
\bibitem [{\citenamefont {Hanghofer}\ \emph {et~al.}(2019)\citenamefont
  {Hanghofer}, \citenamefont {Brinek}, \citenamefont {Eisbacher}, \citenamefont
  {Bitschnau}, \citenamefont {Volck}, \citenamefont {Hennige}, \citenamefont
  {Hanzu}, \citenamefont {Rettenwander},\ and\ \citenamefont
  {Wilkening}}]{Hanghofer2019}%
  \BibitemOpen
  \bibfield  {author} {\bibinfo {author} {\bibfnamefont {I.}~\bibnamefont
  {Hanghofer}}, \bibinfo {author} {\bibfnamefont {M.}~\bibnamefont {Brinek}},
  \bibinfo {author} {\bibfnamefont {S.~L.}\ \bibnamefont {Eisbacher}}, \bibinfo
  {author} {\bibfnamefont {B.}~\bibnamefont {Bitschnau}}, \bibinfo {author}
  {\bibfnamefont {M.}~\bibnamefont {Volck}}, \bibinfo {author} {\bibfnamefont
  {V.}~\bibnamefont {Hennige}}, \bibinfo {author} {\bibfnamefont
  {I.}~\bibnamefont {Hanzu}}, \bibinfo {author} {\bibfnamefont
  {D.}~\bibnamefont {Rettenwander}},\ and\ \bibinfo {author} {\bibfnamefont
  {H.~M.~R.}\ \bibnamefont {Wilkening}},\ }\bibfield  {title} {\bibinfo {title}
  {Substitutional disorder: {Structure} and ion dynamics of the argyrodites
  {Li$_6$PS$_5$Cl}, {Li$_6$PS$_5$Br} and {Li$_6$PS$_5$I}},\ }\href
  {https://doi.org/10.1039/C9CP00664H} {\bibfield  {journal} {\bibinfo
  {journal} {Phys. Chem. Chem. Phys.}\ }\textbf {\bibinfo {volume} {21}},\
  \bibinfo {pages} {8489} (\bibinfo {year} {2019})}\BibitemShut {NoStop}%
\bibitem [{\citenamefont {Luo}\ \emph {et~al.}(2023)\citenamefont {Luo},
  \citenamefont {Meziere}, \citenamefont {Samolyuk}, \citenamefont {Hart},
  \citenamefont {Daymond},\ and\ \citenamefont {Béland}}]{Luo2023}%
  \BibitemOpen
  \bibfield  {author} {\bibinfo {author} {\bibfnamefont {Y.}~\bibnamefont
  {Luo}}, \bibinfo {author} {\bibfnamefont {J.~A.}\ \bibnamefont {Meziere}},
  \bibinfo {author} {\bibfnamefont {G.~D.}\ \bibnamefont {Samolyuk}}, \bibinfo
  {author} {\bibfnamefont {G.~L.~W.}\ \bibnamefont {Hart}}, \bibinfo {author}
  {\bibfnamefont {M.~R.}\ \bibnamefont {Daymond}},\ and\ \bibinfo {author}
  {\bibfnamefont {L.~K.}\ \bibnamefont {Béland}},\ }\bibfield  {title}
  {\bibinfo {title} {A set of moment tensor potentials for zirconium with
  increasing complexity},\ }\href {https://doi.org/10.1021/acs.jctc.3c00488}
  {\bibfield  {journal} {\bibinfo  {journal} {J. Chem. Theory Comput.}\
  }\textbf {\bibinfo {volume} {19}},\ \bibinfo {pages} {6848} (\bibinfo {year}
  {2023})}\BibitemShut {NoStop}%
\bibitem [{\citenamefont {Bock}\ \emph {et~al.}(2024)\citenamefont {Bock},
  \citenamefont {Tasnádi},\ and\ \citenamefont {Abrikosov}}]{Bock2024}%
  \BibitemOpen
  \bibfield  {author} {\bibinfo {author} {\bibfnamefont {F.}~\bibnamefont
  {Bock}}, \bibinfo {author} {\bibfnamefont {F.}~\bibnamefont {Tasnádi}},\
  and\ \bibinfo {author} {\bibfnamefont {I.~A.}\ \bibnamefont {Abrikosov}},\
  }\bibfield  {title} {\bibinfo {title} {Active learning with moment tensor
  potentials to predict material properties: {Ti$_{0.5}$Al$_{0.5}$N} at
  elevated temperature},\ }\href {https://doi.org/10.1116/6.0003260} {\bibfield
   {journal} {\bibinfo  {journal} {J. Vac. Sci. Technol. A}\ }\textbf {\bibinfo
  {volume} {42}},\ \bibinfo {pages} {013412} (\bibinfo {year}
  {2024})}\BibitemShut {NoStop}%
\bibitem [{\citenamefont {Xu}\ \emph {et~al.}(2023)\citenamefont {Xu},
  \citenamefont {Zhang}, \citenamefont {Ruban}, \citenamefont {Schmauder},\
  and\ \citenamefont {Grabowski}}]{Xu2023}%
  \BibitemOpen
  \bibfield  {author} {\bibinfo {author} {\bibfnamefont {X.}~\bibnamefont
  {Xu}}, \bibinfo {author} {\bibfnamefont {X.}~\bibnamefont {Zhang}}, \bibinfo
  {author} {\bibfnamefont {A.}~\bibnamefont {Ruban}}, \bibinfo {author}
  {\bibfnamefont {S.}~\bibnamefont {Schmauder}},\ and\ \bibinfo {author}
  {\bibfnamefont {B.}~\bibnamefont {Grabowski}},\ }\bibfield  {title} {\bibinfo
  {title} {Strong impact of spin fluctuations on the antiphase boundaries of
  weak itinerant ferromagnetic {Ni$_3$Al}},\ }\href
  {https://doi.org/10.1016/j.actamat.2023.118986} {\bibfield  {journal}
  {\bibinfo  {journal} {Acta Mater.}\ }\textbf {\bibinfo {volume} {255}},\
  \bibinfo {pages} {118986} (\bibinfo {year} {2023})}\BibitemShut {NoStop}%
\bibitem [{\citenamefont {Xu}\ \emph {et~al.}(2024)\citenamefont {Xu},
  \citenamefont {Zhang}, \citenamefont {Ruban}, \citenamefont {Schmauder},\
  and\ \citenamefont {Grabowski}}]{Xu2024}%
  \BibitemOpen
  \bibfield  {author} {\bibinfo {author} {\bibfnamefont {X.}~\bibnamefont
  {Xu}}, \bibinfo {author} {\bibfnamefont {X.}~\bibnamefont {Zhang}}, \bibinfo
  {author} {\bibfnamefont {A.}~\bibnamefont {Ruban}}, \bibinfo {author}
  {\bibfnamefont {S.}~\bibnamefont {Schmauder}},\ and\ \bibinfo {author}
  {\bibfnamefont {B.}~\bibnamefont {Grabowski}},\ }\bibfield  {title} {\bibinfo
  {title} {Accurate complex-stacking-fault {Gibbs} energy in {Ni$_3$Al} at high
  temperatures},\ }\href {https://doi.org/10.1016/j.scriptamat.2023.115934}
  {\bibfield  {journal} {\bibinfo  {journal} {Scr. Mater.}\ }\textbf {\bibinfo
  {volume} {242}},\ \bibinfo {pages} {115934} (\bibinfo {year}
  {2024})}\BibitemShut {NoStop}%
\bibitem [{\citenamefont {Zotov}\ \emph {et~al.}(2024)\citenamefont {Zotov},
  \citenamefont {Gubaev}, \citenamefont {Wörner},\ and\ \citenamefont
  {Grabowski}}]{Zotov2024}%
  \BibitemOpen
  \bibfield  {author} {\bibinfo {author} {\bibfnamefont {N.}~\bibnamefont
  {Zotov}}, \bibinfo {author} {\bibfnamefont {K.}~\bibnamefont {Gubaev}},
  \bibinfo {author} {\bibfnamefont {J.}~\bibnamefont {Wörner}},\ and\ \bibinfo
  {author} {\bibfnamefont {B.}~\bibnamefont {Grabowski}},\ }\bibfield  {title}
  {\bibinfo {title} {Moment tensor potential for static and dynamic
  investigations of screw dislocations in bcc {Nb}},\ }\href
  {https://doi.org/10.1088/1361-651X/ad2d68} {\bibfield  {journal} {\bibinfo
  {journal} {Model. Simul. Mat. Sci. Eng.}\ }\textbf {\bibinfo {volume} {32}},\
  \bibinfo {pages} {035032} (\bibinfo {year} {2024})}\BibitemShut {NoStop}%
\bibitem [{\citenamefont {Grabowski}\ \emph {et~al.}(2019)\citenamefont
  {Grabowski}, \citenamefont {Ikeda}, \citenamefont {Srinivasan}, \citenamefont
  {Körmann}, \citenamefont {Freysoldt}, \citenamefont {Duff}, \citenamefont
  {Shapeev},\ and\ \citenamefont {Neugebauer}}]{Grabowski2019}%
  \BibitemOpen
  \bibfield  {author} {\bibinfo {author} {\bibfnamefont {B.}~\bibnamefont
  {Grabowski}}, \bibinfo {author} {\bibfnamefont {Y.}~\bibnamefont {Ikeda}},
  \bibinfo {author} {\bibfnamefont {P.}~\bibnamefont {Srinivasan}}, \bibinfo
  {author} {\bibfnamefont {F.}~\bibnamefont {Körmann}}, \bibinfo {author}
  {\bibfnamefont {C.}~\bibnamefont {Freysoldt}}, \bibinfo {author}
  {\bibfnamefont {A.~I.}\ \bibnamefont {Duff}}, \bibinfo {author}
  {\bibfnamefont {A.}~\bibnamefont {Shapeev}},\ and\ \bibinfo {author}
  {\bibfnamefont {J.}~\bibnamefont {Neugebauer}},\ }\bibfield  {title}
  {\bibinfo {title} {\textit{Ab initio} vibrational free energies including
  anharmonicity for multicomponent alloys},\ }\href
  {https://doi.org/10.1038/s41524-019-0218-8} {\bibfield  {journal} {\bibinfo
  {journal} {Npj Comput. Mater.}\ }\textbf {\bibinfo {volume} {5}},\ \bibinfo
  {pages} {1} (\bibinfo {year} {2019})}\BibitemShut {NoStop}%
\bibitem [{\citenamefont {Jung}\ \emph
  {et~al.}(2023{\natexlab{a}})\citenamefont {Jung}, \citenamefont {Srinivasan},
  \citenamefont {Forslund},\ and\ \citenamefont {Grabowski}}]{Jung2023}%
  \BibitemOpen
  \bibfield  {author} {\bibinfo {author} {\bibfnamefont {J.~H.}\ \bibnamefont
  {Jung}}, \bibinfo {author} {\bibfnamefont {P.}~\bibnamefont {Srinivasan}},
  \bibinfo {author} {\bibfnamefont {A.}~\bibnamefont {Forslund}},\ and\
  \bibinfo {author} {\bibfnamefont {B.}~\bibnamefont {Grabowski}},\ }\bibfield
  {title} {\bibinfo {title} {High-accuracy thermodynamic properties to the
  melting point from \textit{ab initio} calculations aided by machine-learning
  potentials},\ }\href {https://doi.org/10.1038/s41524-022-00956-8} {\bibfield
  {journal} {\bibinfo  {journal} {Npj Comput. Mater.}\ }\textbf {\bibinfo
  {volume} {9}},\ \bibinfo {pages} {1} (\bibinfo {year}
  {2023}{\natexlab{a}})}\BibitemShut {NoStop}%
\bibitem [{\citenamefont {Jung}\ \emph
  {et~al.}(2023{\natexlab{b}})\citenamefont {Jung}, \citenamefont {Forslund},
  \citenamefont {Srinivasan},\ and\ \citenamefont {Grabowski}}]{Jung2023a}%
  \BibitemOpen
  \bibfield  {author} {\bibinfo {author} {\bibfnamefont {J.~H.}\ \bibnamefont
  {Jung}}, \bibinfo {author} {\bibfnamefont {A.}~\bibnamefont {Forslund}},
  \bibinfo {author} {\bibfnamefont {P.}~\bibnamefont {Srinivasan}},\ and\
  \bibinfo {author} {\bibfnamefont {B.}~\bibnamefont {Grabowski}},\ }\bibfield
  {title} {\bibinfo {title} {Dynamically stabilized phases with full \textit{ab
  initio} accuracy: {Thermodynamics} of {Ti}, {Zr}, {Hf} with a focus on the
  hcp-bcc transition},\ }\href {https://doi.org/10.1103/PhysRevB.108.184107}
  {\bibfield  {journal} {\bibinfo  {journal} {Phys. Rev. B}\ }\textbf {\bibinfo
  {volume} {108}},\ \bibinfo {pages} {184107} (\bibinfo {year}
  {2023}{\natexlab{b}})}\BibitemShut {NoStop}%
\bibitem [{\citenamefont {Forslund}\ \emph {et~al.}(2023)\citenamefont
  {Forslund}, \citenamefont {Jung}, \citenamefont {Srinivasan},\ and\
  \citenamefont {Grabowski}}]{Forslund2023}%
  \BibitemOpen
  \bibfield  {author} {\bibinfo {author} {\bibfnamefont {A.}~\bibnamefont
  {Forslund}}, \bibinfo {author} {\bibfnamefont {J.~H.}\ \bibnamefont {Jung}},
  \bibinfo {author} {\bibfnamefont {P.}~\bibnamefont {Srinivasan}},\ and\
  \bibinfo {author} {\bibfnamefont {B.}~\bibnamefont {Grabowski}},\ }\bibfield
  {title} {\bibinfo {title} {Thermodynamic properties on the homologous
  temperature scale from direct upsampling: {Understanding} electron-vibration
  coupling and thermal vacancies in bcc refractory metals},\ }\href
  {https://doi.org/10.1103/PhysRevB.107.174309} {\bibfield  {journal} {\bibinfo
   {journal} {Phys. Rev. B}\ }\textbf {\bibinfo {volume} {107}},\ \bibinfo
  {pages} {174309} (\bibinfo {year} {2023})}\BibitemShut {NoStop}%
\bibitem [{\citenamefont {Zhou}\ \emph {et~al.}(2022)\citenamefont {Zhou},
  \citenamefont {Srinivasan}, \citenamefont {Körmann}, \citenamefont
  {Grabowski}, \citenamefont {Smith}, \citenamefont {Goddard},\ and\
  \citenamefont {Duff}}]{Zhou2022}%
  \BibitemOpen
  \bibfield  {author} {\bibinfo {author} {\bibfnamefont {Y.}~\bibnamefont
  {Zhou}}, \bibinfo {author} {\bibfnamefont {P.}~\bibnamefont {Srinivasan}},
  \bibinfo {author} {\bibfnamefont {F.}~\bibnamefont {Körmann}}, \bibinfo
  {author} {\bibfnamefont {B.}~\bibnamefont {Grabowski}}, \bibinfo {author}
  {\bibfnamefont {R.}~\bibnamefont {Smith}}, \bibinfo {author} {\bibfnamefont
  {P.}~\bibnamefont {Goddard}},\ and\ \bibinfo {author} {\bibfnamefont {A.~I.}\
  \bibnamefont {Duff}},\ }\bibfield  {title} {\bibinfo {title} {Thermodynamics
  up to the melting point in a {TaVCrW} high entropy alloy: {Systematic}
  \textit{ab initio} study aided by machine learning potentials},\ }\href
  {https://doi.org/10.1103/PhysRevB.105.214302} {\bibfield  {journal} {\bibinfo
   {journal} {Phys. Rev. B}\ }\textbf {\bibinfo {volume} {105}},\ \bibinfo
  {pages} {214302} (\bibinfo {year} {2022})}\BibitemShut {NoStop}%
\bibitem [{\citenamefont {Batzner}\ \emph {et~al.}(2022)\citenamefont
  {Batzner}, \citenamefont {Musaelian}, \citenamefont {Sun}, \citenamefont
  {Geiger}, \citenamefont {Mailoa}, \citenamefont {Kornbluth}, \citenamefont
  {Molinari}, \citenamefont {Smidt},\ and\ \citenamefont
  {Kozinsky}}]{Batzner2022}%
  \BibitemOpen
  \bibfield  {author} {\bibinfo {author} {\bibfnamefont {S.}~\bibnamefont
  {Batzner}}, \bibinfo {author} {\bibfnamefont {A.}~\bibnamefont {Musaelian}},
  \bibinfo {author} {\bibfnamefont {L.}~\bibnamefont {Sun}}, \bibinfo {author}
  {\bibfnamefont {M.}~\bibnamefont {Geiger}}, \bibinfo {author} {\bibfnamefont
  {J.~P.}\ \bibnamefont {Mailoa}}, \bibinfo {author} {\bibfnamefont
  {M.}~\bibnamefont {Kornbluth}}, \bibinfo {author} {\bibfnamefont
  {N.}~\bibnamefont {Molinari}}, \bibinfo {author} {\bibfnamefont {T.~E.}\
  \bibnamefont {Smidt}},\ and\ \bibinfo {author} {\bibfnamefont
  {B.}~\bibnamefont {Kozinsky}},\ }\bibfield  {title} {\bibinfo {title}
  {{E(3)}-equivariant graph neural networks for data-efficient and accurate
  interatomic potentials},\ }\href {https://doi.org/10.1038/s41467-022-29939-5}
  {\bibfield  {journal} {\bibinfo  {journal} {Nat. Commun.}\ }\textbf {\bibinfo
  {volume} {13}},\ \bibinfo {pages} {2453} (\bibinfo {year}
  {2022})}\BibitemShut {NoStop}%
\bibitem [{\citenamefont {Batatia}\ \emph {et~al.}(2022)\citenamefont
  {Batatia}, \citenamefont {Kovacs}, \citenamefont {Simm}, \citenamefont
  {Ortner},\ and\ \citenamefont {Csanyi}}]{NEURIPS2022_4a36c3c5}%
  \BibitemOpen
  \bibfield  {author} {\bibinfo {author} {\bibfnamefont {I.}~\bibnamefont
  {Batatia}}, \bibinfo {author} {\bibfnamefont {D.~P.}\ \bibnamefont {Kovacs}},
  \bibinfo {author} {\bibfnamefont {G.}~\bibnamefont {Simm}}, \bibinfo {author}
  {\bibfnamefont {C.}~\bibnamefont {Ortner}},\ and\ \bibinfo {author}
  {\bibfnamefont {G.}~\bibnamefont {Csanyi}},\ }\bibfield  {title} {\bibinfo
  {title} {{MACE}: {Higher} order equivariant message passing neural networks
  for fast and accurate force fields},\ }in\ \href
  {https://proceedings.neurips.cc/paper_files/paper/2022/file/4a36c3c51af11ed9f34615b81edb5bbc-Paper-Conference.pdf}
  {\emph {\bibinfo {booktitle} {Adv. Neural Inf. Process Syst.}}},\
  Vol.~\bibinfo {volume} {35},\ \bibinfo {editor} {edited by\ \bibinfo {editor}
  {\bibfnamefont {S.}~\bibnamefont {Koyejo}}, \bibinfo {editor} {\bibfnamefont
  {S.}~\bibnamefont {Mohamed}}, \bibinfo {editor} {\bibfnamefont
  {A.}~\bibnamefont {Agarwal}}, \bibinfo {editor} {\bibfnamefont
  {D.}~\bibnamefont {Belgrave}}, \bibinfo {editor} {\bibfnamefont
  {K.}~\bibnamefont {Cho}},\ and\ \bibinfo {editor} {\bibfnamefont
  {A.}~\bibnamefont {Oh}}}\ (\bibinfo  {publisher} {Curran Associates, Inc.},\
  \bibinfo {year} {2022})\ pp.\ \bibinfo {pages} {11423--11436}\BibitemShut
  {NoStop}%
\bibitem [{\citenamefont {Xie}\ \emph {et~al.}(2024)\citenamefont {Xie},
  \citenamefont {Deng}, \citenamefont {Liu}, \citenamefont {Famprikis},
  \citenamefont {Butler},\ and\ \citenamefont {Canepa}}]{Xie2024}%
  \BibitemOpen
  \bibfield  {author} {\bibinfo {author} {\bibfnamefont {W.}~\bibnamefont
  {Xie}}, \bibinfo {author} {\bibfnamefont {Z.}~\bibnamefont {Deng}}, \bibinfo
  {author} {\bibfnamefont {Z.}~\bibnamefont {Liu}}, \bibinfo {author}
  {\bibfnamefont {T.}~\bibnamefont {Famprikis}}, \bibinfo {author}
  {\bibfnamefont {K.~T.}\ \bibnamefont {Butler}},\ and\ \bibinfo {author}
  {\bibfnamefont {P.}~\bibnamefont {Canepa}},\ }\bibfield  {title} {\bibinfo
  {title} {Effects of grain boundaries and surfaces on electronic and
  mechanical properties of solid electrolytes},\ }\href
  {https://doi.org/10.1002/aenm.202304230} {\bibfield  {journal} {\bibinfo
  {journal} {Adv. Energy Mater.}\ }\textbf {\bibinfo {volume} {14}},\ \bibinfo
  {pages} {2304230} (\bibinfo {year} {2024})}\BibitemShut {NoStop}%
\bibitem [{\citenamefont {Symington}\ \emph {et~al.}(2021)\citenamefont
  {Symington}, \citenamefont {Molinari}, \citenamefont {Dawson}, \citenamefont
  {Statham}, \citenamefont {Purton}, \citenamefont {Canepa},\ and\
  \citenamefont {Parker}}]{Symington2021}%
  \BibitemOpen
  \bibfield  {author} {\bibinfo {author} {\bibfnamefont {A.~R.}\ \bibnamefont
  {Symington}}, \bibinfo {author} {\bibfnamefont {M.}~\bibnamefont {Molinari}},
  \bibinfo {author} {\bibfnamefont {J.~A.}\ \bibnamefont {Dawson}}, \bibinfo
  {author} {\bibfnamefont {J.~M.}\ \bibnamefont {Statham}}, \bibinfo {author}
  {\bibfnamefont {J.}~\bibnamefont {Purton}}, \bibinfo {author} {\bibfnamefont
  {P.}~\bibnamefont {Canepa}},\ and\ \bibinfo {author} {\bibfnamefont {S.~C.}\
  \bibnamefont {Parker}},\ }\bibfield  {title} {\bibinfo {title} {Elucidating
  the nature of grain boundary resistance in lithium lanthanum titanate},\
  }\href {https://doi.org/10.1039/D0TA11539H} {\bibfield  {journal} {\bibinfo
  {journal} {J. Mater. Chem. A}\ }\textbf {\bibinfo {volume} {9}},\ \bibinfo
  {pages} {6487} (\bibinfo {year} {2021})}\BibitemShut {NoStop}%
\bibitem [{\citenamefont {Quirk}\ and\ \citenamefont
  {Dawson}(2023)}]{Quirk2023}%
  \BibitemOpen
  \bibfield  {author} {\bibinfo {author} {\bibfnamefont {J.~A.}\ \bibnamefont
  {Quirk}}\ and\ \bibinfo {author} {\bibfnamefont {J.~A.}\ \bibnamefont
  {Dawson}},\ }\bibfield  {title} {\bibinfo {title} {Design principles for
  grain boundaries in solid-state lithium-ion conductors},\ }\href
  {https://doi.org/10.1002/aenm.202301114} {\bibfield  {journal} {\bibinfo
  {journal} {Adv. Energy Mater.}\ }\textbf {\bibinfo {volume} {13}},\ \bibinfo
  {pages} {2301114} (\bibinfo {year} {2023})}\BibitemShut {NoStop}%
\bibitem [{\citenamefont {Hayamizu}\ \emph {et~al.}(2016)\citenamefont
  {Hayamizu}, \citenamefont {Aihara}, \citenamefont {Watanabe}, \citenamefont
  {Yamada}, \citenamefont {Ito},\ and\ \citenamefont {Machida}}]{Hayamizu2016}%
  \BibitemOpen
  \bibfield  {author} {\bibinfo {author} {\bibfnamefont {K.}~\bibnamefont
  {Hayamizu}}, \bibinfo {author} {\bibfnamefont {Y.}~\bibnamefont {Aihara}},
  \bibinfo {author} {\bibfnamefont {T.}~\bibnamefont {Watanabe}}, \bibinfo
  {author} {\bibfnamefont {T.}~\bibnamefont {Yamada}}, \bibinfo {author}
  {\bibfnamefont {S.}~\bibnamefont {Ito}},\ and\ \bibinfo {author}
  {\bibfnamefont {N.}~\bibnamefont {Machida}},\ }\bibfield  {title} {\bibinfo
  {title} {{NMR} studies on lithium ion migration in sulfide-based conductors,
  amorphous and crystalline {Li$_3$PS$_4$}},\ }\href
  {https://doi.org/10.1016/j.ssi.2015.06.016} {\bibfield  {journal} {\bibinfo
  {journal} {Solid State Ion.}\ }\textbf {\bibinfo {volume} {285}},\ \bibinfo
  {pages} {51} (\bibinfo {year} {2016})}\BibitemShut {NoStop}%
\bibitem [{\citenamefont {Zhou}\ \emph {et~al.}(2021)\citenamefont {Zhou},
  \citenamefont {Minafra}, \citenamefont {Zeier},\ and\ \citenamefont
  {Nazar}}]{Zhou2021}%
  \BibitemOpen
  \bibfield  {author} {\bibinfo {author} {\bibfnamefont {L.}~\bibnamefont
  {Zhou}}, \bibinfo {author} {\bibfnamefont {N.}~\bibnamefont {Minafra}},
  \bibinfo {author} {\bibfnamefont {W.~G.}\ \bibnamefont {Zeier}},\ and\
  \bibinfo {author} {\bibfnamefont {L.~F.}\ \bibnamefont {Nazar}},\ }\bibfield
  {title} {\bibinfo {title} {Innovative approaches to {Li}-argyrodite solid
  electrolytes for all-solid-state lithium batteries},\ }\href
  {https://doi.org/10.1021/acs.accounts.0c00874} {\bibfield  {journal}
  {\bibinfo  {journal} {Acc. Chem. Res.}\ }\textbf {\bibinfo {volume} {54}},\
  \bibinfo {pages} {2717} (\bibinfo {year} {2021})}\BibitemShut {NoStop}%
\bibitem [{\citenamefont {Borisov}\ \emph {et~al.}(1964)\citenamefont
  {Borisov}, \citenamefont {Golikov},\ and\ \citenamefont
  {Scherbedinskiy}}]{Borisov1964}%
  \BibitemOpen
  \bibfield  {author} {\bibinfo {author} {\bibfnamefont {V.}~\bibnamefont
  {Borisov}}, \bibinfo {author} {\bibfnamefont {V.}~\bibnamefont {Golikov}},\
  and\ \bibinfo {author} {\bibfnamefont {G.}~\bibnamefont {Scherbedinskiy}},\
  }\bibfield  {title} {\bibinfo {title} {Relation between diffusion
  coefficients and grain boundary energy},\ }\href@noop {} {\bibfield
  {journal} {\bibinfo  {journal} {Phys. Met. Metall.}\ }\textbf {\bibinfo
  {volume} {17}},\ \bibinfo {pages} {881} (\bibinfo {year} {1964})}\BibitemShut
  {NoStop}%
\bibitem [{\citenamefont {Pelleg}(1966)}]{Pelleg1966}%
  \BibitemOpen
  \bibfield  {author} {\bibinfo {author} {\bibfnamefont {J.}~\bibnamefont
  {Pelleg}},\ }\bibfield  {title} {\bibinfo {title} {On the relation between
  diffusion coefficients and grain boundary energy},\ }\href
  {https://doi.org/10.1080/14786436608211954} {\bibfield  {journal} {\bibinfo
  {journal} {Philos. Mag.: J. Theor. Exp. Appl. Phys.}\ }\textbf {\bibinfo
  {volume} {14}},\ \bibinfo {pages} {595} (\bibinfo {year} {1966})}\BibitemShut
  {NoStop}%
\bibitem [{\citenamefont {Page}\ \emph {et~al.}(2021)\citenamefont {Page},
  \citenamefont {Varela}, \citenamefont {Johnson}, \citenamefont {Fullwood},\
  and\ \citenamefont {Homer}}]{Page2021}%
  \BibitemOpen
  \bibfield  {author} {\bibinfo {author} {\bibfnamefont {D.~E.}\ \bibnamefont
  {Page}}, \bibinfo {author} {\bibfnamefont {K.~F.}\ \bibnamefont {Varela}},
  \bibinfo {author} {\bibfnamefont {O.~K.}\ \bibnamefont {Johnson}}, \bibinfo
  {author} {\bibfnamefont {D.~T.}\ \bibnamefont {Fullwood}},\ and\ \bibinfo
  {author} {\bibfnamefont {E.~R.}\ \bibnamefont {Homer}},\ }\bibfield  {title}
  {\bibinfo {title} {Measuring simulated hydrogen diffusion in symmetric tilt
  nickel grain boundaries and examining the relevance of the {Borisov}
  relationship for individual boundary diffusion},\ }\href
  {https://doi.org/10.1016/j.actamat.2021.116882} {\bibfield  {journal}
  {\bibinfo  {journal} {Acta Mater.}\ }\textbf {\bibinfo {volume} {212}},\
  \bibinfo {pages} {116882} (\bibinfo {year} {2021})}\BibitemShut {NoStop}%
\bibitem [{\citenamefont {Divinski}\ \emph {et~al.}(2010)\citenamefont
  {Divinski}, \citenamefont {Reglitz},\ and\ \citenamefont
  {Wilde}}]{Divinski2010}%
  \BibitemOpen
  \bibfield  {author} {\bibinfo {author} {\bibfnamefont {S.~V.}\ \bibnamefont
  {Divinski}}, \bibinfo {author} {\bibfnamefont {G.}~\bibnamefont {Reglitz}},\
  and\ \bibinfo {author} {\bibfnamefont {G.}~\bibnamefont {Wilde}},\ }\bibfield
   {title} {\bibinfo {title} {Grain boundary self-diffusion in polycrystalline
  nickel of different purity levels},\ }\href
  {https://doi.org/10.1016/j.actamat.2009.09.015} {\bibfield  {journal}
  {\bibinfo  {journal} {Acta Mater.}\ }\textbf {\bibinfo {volume} {58}},\
  \bibinfo {pages} {386} (\bibinfo {year} {2010})}\BibitemShut {NoStop}%
\bibitem [{\citenamefont {Prokoshkina}\ \emph {et~al.}(2013)\citenamefont
  {Prokoshkina}, \citenamefont {Esin}, \citenamefont {Wilde},\ and\
  \citenamefont {Divinski}}]{Prokoshkina2013}%
  \BibitemOpen
  \bibfield  {author} {\bibinfo {author} {\bibfnamefont {D.}~\bibnamefont
  {Prokoshkina}}, \bibinfo {author} {\bibfnamefont {V.~A.}\ \bibnamefont
  {Esin}}, \bibinfo {author} {\bibfnamefont {G.}~\bibnamefont {Wilde}},\ and\
  \bibinfo {author} {\bibfnamefont {S.~V.}\ \bibnamefont {Divinski}},\
  }\bibfield  {title} {\bibinfo {title} {Grain boundary width, energy and
  self-diffusion in nickel: {Effect} of material purity},\ }\href
  {https://doi.org/10.1016/j.actamat.2013.05.010} {\bibfield  {journal}
  {\bibinfo  {journal} {Acta Mater.}\ }\textbf {\bibinfo {volume} {61}},\
  \bibinfo {pages} {5188} (\bibinfo {year} {2013})}\BibitemShut {NoStop}%
\bibitem [{\citenamefont {Li}\ \emph {et~al.}(2023)\citenamefont {Li},
  \citenamefont {Lu}, \citenamefont {Divinski},\ and\ \citenamefont
  {Vitos}}]{Li2023}%
  \BibitemOpen
  \bibfield  {author} {\bibinfo {author} {\bibfnamefont {C.}~\bibnamefont
  {Li}}, \bibinfo {author} {\bibfnamefont {S.}~\bibnamefont {Lu}}, \bibinfo
  {author} {\bibfnamefont {S.}~\bibnamefont {Divinski}},\ and\ \bibinfo
  {author} {\bibfnamefont {L.}~\bibnamefont {Vitos}},\ }\bibfield  {title}
  {\bibinfo {title} {Theoretical and experimental grain boundary energies in
  body-centered cubic metals},\ }\href
  {https://doi.org/10.1016/j.actamat.2023.119074} {\bibfield  {journal}
  {\bibinfo  {journal} {Acta Mater.}\ }\textbf {\bibinfo {volume} {255}},\
  \bibinfo {pages} {119074} (\bibinfo {year} {2023})}\BibitemShut {NoStop}%
\bibitem [{\citenamefont {Karkkainen}\ \emph {et~al.}(2000)\citenamefont
  {Karkkainen}, \citenamefont {Sihvola},\ and\ \citenamefont
  {Nikoskinen}}]{Karkkainen2000}%
  \BibitemOpen
  \bibfield  {author} {\bibinfo {author} {\bibfnamefont {K.}~\bibnamefont
  {Karkkainen}}, \bibinfo {author} {\bibfnamefont {A.}~\bibnamefont
  {Sihvola}},\ and\ \bibinfo {author} {\bibfnamefont {K.}~\bibnamefont
  {Nikoskinen}},\ }\bibfield  {title} {\bibinfo {title} {Effective permittivity
  of mixtures: {Numerical} validation by the {FDTD} method},\ }\href
  {https://doi.org/10.1109/36.843023} {\bibfield  {journal} {\bibinfo
  {journal} {IEEE Trans. Geosci. Remote Sens.}\ }\textbf {\bibinfo {volume}
  {38}},\ \bibinfo {pages} {1303} (\bibinfo {year} {2000})}\BibitemShut
  {NoStop}%
\bibitem [{\citenamefont {Chen}\ and\ \citenamefont {Schuh}(2006)}]{Chen2006}%
  \BibitemOpen
  \bibfield  {author} {\bibinfo {author} {\bibfnamefont {Y.}~\bibnamefont
  {Chen}}\ and\ \bibinfo {author} {\bibfnamefont {C.~A.}\ \bibnamefont
  {Schuh}},\ }\bibfield  {title} {\bibinfo {title} {Diffusion on grain boundary
  networks: {Percolation} theory and effective medium approximations},\ }\href
  {https://doi.org/10.1016/j.actamat.2006.06.011} {\bibfield  {journal}
  {\bibinfo  {journal} {Acta Mater.}\ }\textbf {\bibinfo {volume} {54}},\
  \bibinfo {pages} {4709} (\bibinfo {year} {2006})}\BibitemShut {NoStop}%
\bibitem [{\citenamefont {Heo}\ \emph {et~al.}(2021)\citenamefont {Heo},
  \citenamefont {Grieder}, \citenamefont {Wang}, \citenamefont {Wood},
  \citenamefont {Hsu}, \citenamefont {Akhade}, \citenamefont {Wan},
  \citenamefont {Chen}, \citenamefont {Adelstein},\ and\ \citenamefont
  {Wood}}]{Heo2021}%
  \BibitemOpen
  \bibfield  {author} {\bibinfo {author} {\bibfnamefont {T.~W.}\ \bibnamefont
  {Heo}}, \bibinfo {author} {\bibfnamefont {A.}~\bibnamefont {Grieder}},
  \bibinfo {author} {\bibfnamefont {B.}~\bibnamefont {Wang}}, \bibinfo {author}
  {\bibfnamefont {M.}~\bibnamefont {Wood}}, \bibinfo {author} {\bibfnamefont
  {T.}~\bibnamefont {Hsu}}, \bibinfo {author} {\bibfnamefont {S.~A.}\
  \bibnamefont {Akhade}}, \bibinfo {author} {\bibfnamefont {L.~F.}\
  \bibnamefont {Wan}}, \bibinfo {author} {\bibfnamefont {L.-Q.}\ \bibnamefont
  {Chen}}, \bibinfo {author} {\bibfnamefont {N.}~\bibnamefont {Adelstein}},\
  and\ \bibinfo {author} {\bibfnamefont {B.~C.}\ \bibnamefont {Wood}},\
  }\bibfield  {title} {\bibinfo {title} {Microstructural impacts on ionic
  conductivity of oxide solid electrolytes from a combined atomistic-mesoscale
  approach},\ }\href {https://doi.org/10.1038/s41524-021-00681-8} {\bibfield
  {journal} {\bibinfo  {journal} {Npj Comput. Mater.}\ }\textbf {\bibinfo
  {volume} {7}},\ \bibinfo {pages} {1} (\bibinfo {year} {2021})}\BibitemShut
  {NoStop}%
\bibitem [{\citenamefont {Faka}\ \emph {et~al.}(2024)\citenamefont {Faka},
  \citenamefont {Agne}, \citenamefont {Lange}, \citenamefont {Daisenberger},
  \citenamefont {Wankmiller}, \citenamefont {Schwarzmüller}, \citenamefont
  {Huppertz}, \citenamefont {Maus}, \citenamefont {Helm}, \citenamefont
  {Böger}, \citenamefont {Hartel}, \citenamefont {Gerdes}, \citenamefont
  {Molaison}, \citenamefont {Kieslich}, \citenamefont {Hansen},\ and\
  \citenamefont {Zeier}}]{Faka2024}%
  \BibitemOpen
  \bibfield  {author} {\bibinfo {author} {\bibfnamefont {V.}~\bibnamefont
  {Faka}}, \bibinfo {author} {\bibfnamefont {M.~T.}\ \bibnamefont {Agne}},
  \bibinfo {author} {\bibfnamefont {M.~A.}\ \bibnamefont {Lange}}, \bibinfo
  {author} {\bibfnamefont {D.}~\bibnamefont {Daisenberger}}, \bibinfo {author}
  {\bibfnamefont {B.}~\bibnamefont {Wankmiller}}, \bibinfo {author}
  {\bibfnamefont {S.}~\bibnamefont {Schwarzmüller}}, \bibinfo {author}
  {\bibfnamefont {H.}~\bibnamefont {Huppertz}}, \bibinfo {author}
  {\bibfnamefont {O.}~\bibnamefont {Maus}}, \bibinfo {author} {\bibfnamefont
  {B.}~\bibnamefont {Helm}}, \bibinfo {author} {\bibfnamefont {T.}~\bibnamefont
  {Böger}}, \bibinfo {author} {\bibfnamefont {J.}~\bibnamefont {Hartel}},
  \bibinfo {author} {\bibfnamefont {J.~M.}\ \bibnamefont {Gerdes}}, \bibinfo
  {author} {\bibfnamefont {J.~J.}\ \bibnamefont {Molaison}}, \bibinfo {author}
  {\bibfnamefont {G.}~\bibnamefont {Kieslich}}, \bibinfo {author}
  {\bibfnamefont {M.~R.}\ \bibnamefont {Hansen}},\ and\ \bibinfo {author}
  {\bibfnamefont {W.~G.}\ \bibnamefont {Zeier}},\ }\bibfield  {title} {\bibinfo
  {title} {Pressure-induced dislocations and their influence on ionic transport
  in {Li$^+$}-conducting argyrodites},\ }\href
  {https://doi.org/10.1021/jacs.3c12323} {\bibfield  {journal} {\bibinfo
  {journal} {J. Am. Chem. Soc.}\ }\textbf {\bibinfo {volume} {146}},\ \bibinfo
  {pages} {1710} (\bibinfo {year} {2024})}\BibitemShut {NoStop}%
\bibitem [{\citenamefont {Janek}\ and\ \citenamefont
  {Zeier}(2023)}]{Janek2023}%
  \BibitemOpen
  \bibfield  {author} {\bibinfo {author} {\bibfnamefont {J.}~\bibnamefont
  {Janek}}\ and\ \bibinfo {author} {\bibfnamefont {W.~G.}\ \bibnamefont
  {Zeier}},\ }\bibfield  {title} {\bibinfo {title} {Challenges in speeding up
  solid-state battery development},\ }\href
  {https://doi.org/10.1038/s41560-023-01208-9} {\bibfield  {journal} {\bibinfo
  {journal} {Nat. Energy}\ }\textbf {\bibinfo {volume} {8}},\ \bibinfo {pages}
  {230} (\bibinfo {year} {2023})}\BibitemShut {NoStop}%
\bibitem [{\citenamefont {{\ifmmode\check{Z}\else\v{Z}\fi}guns}\ and\
  \citenamefont {Yildiz}(2022)}]{Zguns2022Jul}%
  \BibitemOpen
  \bibfield  {author} {\bibinfo {author} {\bibfnamefont {P.}~\bibnamefont
  {{\ifmmode\check{Z}\else\v{Z}\fi}guns}}\ and\ \bibinfo {author}
  {\bibfnamefont {B.}~\bibnamefont {Yildiz}},\ }\bibfield  {title} {\bibinfo
  {title} {Strain sensitivity of {Li}-ion conductivity in
  $\beta$-{Li}$_{3}${PS}$_{4}$ solid electrolyte},\ }\href
  {https://doi.org/10.1103/PRXEnergy.1.023003} {\bibfield  {journal} {\bibinfo
  {journal} {PRX Energy}\ }\textbf {\bibinfo {volume} {1}},\ \bibinfo {pages}
  {023003} (\bibinfo {year} {2022})}\BibitemShut {NoStop}%
\bibitem [{\citenamefont {Podryabinkin}\ and\ \citenamefont
  {Shapeev}(2017)}]{Podryabinkin2017}%
  \BibitemOpen
  \bibfield  {author} {\bibinfo {author} {\bibfnamefont {E.~V.}\ \bibnamefont
  {Podryabinkin}}\ and\ \bibinfo {author} {\bibfnamefont {A.~V.}\ \bibnamefont
  {Shapeev}},\ }\bibfield  {title} {\bibinfo {title} {Active learning of
  linearly parametrized interatomic potentials},\ }\href
  {https://doi.org/10.1016/j.commatsci.2017.08.031} {\bibfield  {journal}
  {\bibinfo  {journal} {Comput. Mater. Sci.}\ }\textbf {\bibinfo {volume}
  {140}},\ \bibinfo {pages} {171} (\bibinfo {year} {2017})}\BibitemShut
  {NoStop}%
\bibitem [{\citenamefont {Gubaev}\ \emph {et~al.}(2018)\citenamefont {Gubaev},
  \citenamefont {Podryabinkin},\ and\ \citenamefont {Shapeev}}]{Gubaev2018}%
  \BibitemOpen
  \bibfield  {author} {\bibinfo {author} {\bibfnamefont {K.}~\bibnamefont
  {Gubaev}}, \bibinfo {author} {\bibfnamefont {E.~V.}\ \bibnamefont
  {Podryabinkin}},\ and\ \bibinfo {author} {\bibfnamefont {A.~V.}\ \bibnamefont
  {Shapeev}},\ }\bibfield  {title} {\bibinfo {title} {Machine learning of
  molecular properties: {Locality} and active learning},\ }\href
  {https://doi.org/10.1063/1.5005095} {\bibfield  {journal} {\bibinfo
  {journal} {J. Chem. Phys.}\ }\textbf {\bibinfo {volume} {148}},\ \bibinfo
  {pages} {241727} (\bibinfo {year} {2018})}\BibitemShut {NoStop}%
\bibitem [{\citenamefont {Uitz}\ \emph {et~al.}(2017)\citenamefont {Uitz},
  \citenamefont {Epp}, \citenamefont {Bottke},\ and\ \citenamefont
  {Wilkening}}]{Uitz2017}%
  \BibitemOpen
  \bibfield  {author} {\bibinfo {author} {\bibfnamefont {M.}~\bibnamefont
  {Uitz}}, \bibinfo {author} {\bibfnamefont {V.}~\bibnamefont {Epp}}, \bibinfo
  {author} {\bibfnamefont {P.}~\bibnamefont {Bottke}},\ and\ \bibinfo {author}
  {\bibfnamefont {M.}~\bibnamefont {Wilkening}},\ }\bibfield  {title} {\bibinfo
  {title} {Ion dynamics in solid electrolytes for lithium batteries},\ }\href
  {https://doi.org/10.1007/s10832-017-0071-4} {\bibfield  {journal} {\bibinfo
  {journal} {J. Electroceram.}\ }\textbf {\bibinfo {volume} {38}},\ \bibinfo
  {pages} {142} (\bibinfo {year} {2017})}\BibitemShut {NoStop}%
\bibitem [{\citenamefont {Kizilyalli}\ \emph {et~al.}(1999)\citenamefont
  {Kizilyalli}, \citenamefont {Corish},\ and\ \citenamefont
  {Metselaar}}]{Kizilyalli1999}%
  \BibitemOpen
  \bibfield  {author} {\bibinfo {author} {\bibfnamefont {M.}~\bibnamefont
  {Kizilyalli}}, \bibinfo {author} {\bibfnamefont {J.}~\bibnamefont {Corish}},\
  and\ \bibinfo {author} {\bibfnamefont {R.}~\bibnamefont {Metselaar}},\
  }\bibfield  {title} {\bibinfo {title} {Definitions of terms for diffusion in
  the solid state},\ }\href {https://doi.org/10.1351/pac199971071307}
  {\bibfield  {journal} {\bibinfo  {journal} {Pure Appl. Chem.}\ }\textbf
  {\bibinfo {volume} {71}},\ \bibinfo {pages} {1307} (\bibinfo {year}
  {1999})}\BibitemShut {NoStop}%
\bibitem [{\citenamefont {Marcolongo}\ and\ \citenamefont
  {Marzari}(2017)}]{Marcolongo2017}%
  \BibitemOpen
  \bibfield  {author} {\bibinfo {author} {\bibfnamefont {A.}~\bibnamefont
  {Marcolongo}}\ and\ \bibinfo {author} {\bibfnamefont {N.}~\bibnamefont
  {Marzari}},\ }\bibfield  {title} {\bibinfo {title} {Ionic correlations and
  failure of {Nernst}-{Einstein} relation in solid-state electrolytes},\ }\href
  {https://doi.org/10.1103/PhysRevMaterials.1.025402} {\bibfield  {journal}
  {\bibinfo  {journal} {Phys. Rev. Mater.}\ }\textbf {\bibinfo {volume} {1}},\
  \bibinfo {pages} {025402} (\bibinfo {year} {2017})}\BibitemShut {NoStop}%
\bibitem [{\citenamefont {Garnett}\ and\ \citenamefont
  {Larmor}(1997)}]{Garnett1997}%
  \BibitemOpen
  \bibfield  {author} {\bibinfo {author} {\bibfnamefont {J.~C.~M.}\
  \bibnamefont {Garnett}}\ and\ \bibinfo {author} {\bibfnamefont
  {J.}~\bibnamefont {Larmor}},\ }\bibfield  {title} {\bibinfo {title} {{XII}.
  {Colours} in metal glasses and in metallic films},\ }\href
  {https://doi.org/10.1098/rsta.1904.0024} {\bibfield  {journal} {\bibinfo
  {journal} {Philos. Trans. R. Soc. A}\ }\textbf {\bibinfo {volume} {203}},\
  \bibinfo {pages} {385} (\bibinfo {year} {1997})}\BibitemShut {NoStop}%
\bibitem [{\citenamefont {Kalnin}\ \emph {et~al.}(2002)\citenamefont {Kalnin},
  \citenamefont {Kotomin},\ and\ \citenamefont {Maier}}]{Kalnin2002}%
  \BibitemOpen
  \bibfield  {author} {\bibinfo {author} {\bibfnamefont {J.~R.}\ \bibnamefont
  {Kalnin}}, \bibinfo {author} {\bibfnamefont {E.~A.}\ \bibnamefont
  {Kotomin}},\ and\ \bibinfo {author} {\bibfnamefont {J.}~\bibnamefont
  {Maier}},\ }\bibfield  {title} {\bibinfo {title} {Calculations of the
  effective diffusion coefficient for inhomogeneous media},\ }\href
  {https://doi.org/10.1016/S0022-3697(01)00159-7} {\bibfield  {journal}
  {\bibinfo  {journal} {J. Phys. Chem. Solids}\ }\textbf {\bibinfo {volume}
  {63}},\ \bibinfo {pages} {449} (\bibinfo {year} {2002})}\BibitemShut
  {NoStop}%
\bibitem [{\citenamefont {Chen}\ and\ \citenamefont {Schuh}(2007)}]{Chen2007}%
  \BibitemOpen
  \bibfield  {author} {\bibinfo {author} {\bibfnamefont {Y.}~\bibnamefont
  {Chen}}\ and\ \bibinfo {author} {\bibfnamefont {C.~A.}\ \bibnamefont
  {Schuh}},\ }\bibfield  {title} {\bibinfo {title} {Geometric considerations
  for diffusion in polycrystalline solids},\ }\href
  {https://doi.org/10.1063/1.2711820} {\bibfield  {journal} {\bibinfo
  {journal} {J. Appl. Phys.}\ }\textbf {\bibinfo {volume} {101}},\ \bibinfo
  {pages} {063524} (\bibinfo {year} {2007})}\BibitemShut {NoStop}%
\bibitem [{\citenamefont {Voigt}(1889)}]{Voigt1889}%
  \BibitemOpen
  \bibfield  {author} {\bibinfo {author} {\bibfnamefont {W.}~\bibnamefont
  {Voigt}},\ }\bibfield  {title} {\bibinfo {title} {Über die {Beziehung}
  zwischen den beiden {Elasticitätsconstanten} isotroper {Körper}},\ }\href
  {https://doi.org/10.1002/andp.18892741206} {\bibfield  {journal} {\bibinfo
  {journal} {Ann. Phys.}\ }\textbf {\bibinfo {volume} {274}},\ \bibinfo {pages}
  {573} (\bibinfo {year} {1889})}\BibitemShut {NoStop}%
\bibitem [{\citenamefont {Reuss}(1929)}]{Reuss1929}%
  \BibitemOpen
  \bibfield  {author} {\bibinfo {author} {\bibfnamefont {A.}~\bibnamefont
  {Reuss}},\ }\bibfield  {title} {\bibinfo {title} {Berechnung der
  {Fließgrenze} von {Mischkristallen} auf {Grund} der {Plastizitätsbedingung}
  für {Einkristalle}},\ }\href {https://doi.org/10.1002/zamm.19290090104}
  {\bibfield  {journal} {\bibinfo  {journal} {J. Appl. Math. Mech.}\ }\textbf
  {\bibinfo {volume} {9}},\ \bibinfo {pages} {49} (\bibinfo {year}
  {1929})}\BibitemShut {NoStop}%
\bibitem [{\citenamefont {Hill}(1963)}]{Hill1963}%
  \BibitemOpen
  \bibfield  {author} {\bibinfo {author} {\bibfnamefont {R.}~\bibnamefont
  {Hill}},\ }\bibfield  {title} {\bibinfo {title} {Elastic properties of
  reinforced solids: {Some} theoretical principles},\ }\href
  {https://doi.org/10.1016/0022-5096(63)90036-X} {\bibfield  {journal}
  {\bibinfo  {journal} {J. Mech. Phys. Solids}\ }\textbf {\bibinfo {volume}
  {11}},\ \bibinfo {pages} {357} (\bibinfo {year} {1963})}\BibitemShut
  {NoStop}%
\bibitem [{\citenamefont {Yoon}\ \emph {et~al.}(2023)\citenamefont {Yoon},
  \citenamefont {Sulaimon},\ and\ \citenamefont {Siegel}}]{Yoon2023}%
  \BibitemOpen
  \bibfield  {author} {\bibinfo {author} {\bibfnamefont {J.~S.}\ \bibnamefont
  {Yoon}}, \bibinfo {author} {\bibfnamefont {H.}~\bibnamefont {Sulaimon}},\
  and\ \bibinfo {author} {\bibfnamefont {D.~J.}\ \bibnamefont {Siegel}},\
  }\bibfield  {title} {\bibinfo {title} {Exploiting grain boundary diffusion to
  minimize dendrite formation in lithium metal-solid state batteries},\ }\href
  {https://doi.org/10.1039/D3TA03814A} {\bibfield  {journal} {\bibinfo
  {journal} {J. Mater. Chem. A}\ }\textbf {\bibinfo {volume} {11}},\ \bibinfo
  {pages} {23288} (\bibinfo {year} {2023})}\BibitemShut {NoStop}%
\bibitem [{\citenamefont {Thorvaldsen}(1997)}]{Thorvaldsen1997}%
  \BibitemOpen
  \bibfield  {author} {\bibinfo {author} {\bibfnamefont {A.}~\bibnamefont
  {Thorvaldsen}},\ }\bibfield  {title} {\bibinfo {title} {The intercept method
  -- 1. {Evaluation} of grain shape},\ }\href
  {https://doi.org/10.1016/S1359-6454(96)00197-8} {\bibfield  {journal}
  {\bibinfo  {journal} {Acta Mater.}\ }\textbf {\bibinfo {volume} {45}},\
  \bibinfo {pages} {587} (\bibinfo {year} {1997})}\BibitemShut {NoStop}%
\bibitem [{\citenamefont {Zhang}\ and\ \citenamefont
  {Liu}(2012)}]{Zhang2012-partitioning}%
  \BibitemOpen
  \bibfield  {author} {\bibinfo {author} {\bibfnamefont {Y.}~\bibnamefont
  {Zhang}}\ and\ \bibinfo {author} {\bibfnamefont {L.}~\bibnamefont {Liu}},\
  }\bibfield  {title} {\bibinfo {title} {On diffusion in heterogeneous media},\
  }\href {https://doi.org/10.2475/09.2012.03} {\bibfield  {journal} {\bibinfo
  {journal} {Am. J. Sci.}\ }\textbf {\bibinfo {volume} {312}},\ \bibinfo
  {pages} {1028} (\bibinfo {year} {2012})}\BibitemShut {NoStop}%
\bibitem [{\citenamefont {Harrison}(1961)}]{Harrison1961}%
  \BibitemOpen
  \bibfield  {author} {\bibinfo {author} {\bibfnamefont {L.~G.}\ \bibnamefont
  {Harrison}},\ }\bibfield  {title} {\bibinfo {title} {Influence of
  dislocations on diffusion kinetics in solids with particular reference to the
  alkali halides},\ }\href {https://doi.org/10.1039/TF9615701191} {\bibfield
  {journal} {\bibinfo  {journal} {Trans. Faraday Soc.}\ }\textbf {\bibinfo
  {volume} {57}},\ \bibinfo {pages} {1191} (\bibinfo {year}
  {1961})}\BibitemShut {NoStop}%
\bibitem [{\citenamefont {Paul}\ \emph {et~al.}(2014)\citenamefont {Paul},
  \citenamefont {Laurila}, \citenamefont {Vuorinen},\ and\ \citenamefont
  {Divinski}}]{Paul2014}%
  \BibitemOpen
  \bibfield  {author} {\bibinfo {author} {\bibfnamefont {A.}~\bibnamefont
  {Paul}}, \bibinfo {author} {\bibfnamefont {T.}~\bibnamefont {Laurila}},
  \bibinfo {author} {\bibfnamefont {V.}~\bibnamefont {Vuorinen}},\ and\
  \bibinfo {author} {\bibfnamefont {S.~V.}\ \bibnamefont {Divinski}},\
  }\href@noop {} {\emph {\bibinfo {title} {{Thermodynamics, diffusion and the
  Kirkendall effect in solids}}}}\ (\bibinfo  {publisher} {Springer
  International Publishing},\ \bibinfo {address} {Switzerland},\ \bibinfo
  {year} {2014})\BibitemShut {NoStop}%
\bibitem [{\citenamefont {Divinski}\ \emph
  {et~al.}(2004{\natexlab{a}})\citenamefont {Divinski}, \citenamefont {Lee},\
  and\ \citenamefont {Herzig}}]{Divinski2004a}%
  \BibitemOpen
  \bibfield  {author} {\bibinfo {author} {\bibfnamefont {S.~V.}\ \bibnamefont
  {Divinski}}, \bibinfo {author} {\bibfnamefont {J.~S.}\ \bibnamefont {Lee}},\
  and\ \bibinfo {author} {\bibfnamefont {C.}~\bibnamefont {Herzig}},\
  }\bibfield  {title} {\bibinfo {title} {Grain boundary diffusion and
  segregation in compacted and sintered nanocrystalline alloys},\ }\href
  {https://doi.org/10.4028/www.scientific.net/JMNM.19.55} {\bibfield  {journal}
  {\bibinfo  {journal} {J. Metastable Nanocryst. Mater.}\ }\textbf {\bibinfo
  {volume} {19}},\ \bibinfo {pages} {55} (\bibinfo {year}
  {2004}{\natexlab{a}})}\BibitemShut {NoStop}%
\bibitem [{\citenamefont {Divinski}\ \emph
  {et~al.}(2004{\natexlab{b}})\citenamefont {Divinski}, \citenamefont {Hisker},
  \citenamefont {Kang}, \citenamefont {Lee},\ and\ \citenamefont
  {Herzig}}]{DIVINSKI2004b}%
  \BibitemOpen
  \bibfield  {author} {\bibinfo {author} {\bibfnamefont {S.}~\bibnamefont
  {Divinski}}, \bibinfo {author} {\bibfnamefont {F.}~\bibnamefont {Hisker}},
  \bibinfo {author} {\bibfnamefont {Y.-S.}\ \bibnamefont {Kang}}, \bibinfo
  {author} {\bibfnamefont {J.-S.}\ \bibnamefont {Lee}},\ and\ \bibinfo {author}
  {\bibfnamefont {C.}~\bibnamefont {Herzig}},\ }\bibfield  {title} {\bibinfo
  {title} {{Ag diffusion and interface segregation in nanocrystalline
  $\gamma$-FeNi alloy with a two-scale microstructure}},\ }\href
  {https://doi.org/https://doi.org/10.1016/j.actamat.2003.09.045} {\bibfield
  {journal} {\bibinfo  {journal} {Acta Mater.}\ }\textbf {\bibinfo {volume}
  {52}},\ \bibinfo {pages} {631} (\bibinfo {year}
  {2004}{\natexlab{b}})}\BibitemShut {NoStop}%
\end{thebibliography}
\end{document}